\documentclass[aps,prb,twocolumn,floats,epsfig,pdflatex,longbibliography]{revtex4-2}
\usepackage{color,ulem,verbatim,float}
\usepackage{amssymb,amsbsy,amsmath,mathrsfs}
\usepackage{amssymb}
\usepackage{amsbsy}
\usepackage{amsmath}
\usepackage{float}
\usepackage{graphicx}
\usepackage{xcolor}
\usepackage{bm}
\usepackage{natbib}
\usepackage[plainpages=false,pdfpagelabels,colorlinks=true,linkcolor=red,urlcolor=blue,citecolor=blue,pdftitle={},pdfauthor={},pdfdisplaydoctitle={},pdfduplex=DuplexFlipLongEdge]{hyperref}

\newcommand{\XX}[1] {\textcolor{black}{#1}}

\newcommand{\drim}[1] {\textcolor{black}{#1}}
\newcommand{\cyan}[1] {\textcolor{black}{#1}}
\newcommand\bea{\begin{eqnarray}}
\newcommand\eea{\end{eqnarray}}
\newcommand\beq{\begin{equation}}
\newcommand\eeq{\end{equation}}

\newcommand{\noi}{\noindent}
\newcommand{\non}{\nonumber}
\newcommand{\al}{\alpha}
\newcommand{\de}{\delta}

\newcommand{\ga}{\gamma}

\newcommand{\ka}{\kappa}

\newcommand{\si}{\sigma}
\newcommand{\ta}{\theta}
\newcommand{\y}{\psi}
\newcommand{\om}{\omega}
\newcommand{\da}{\dagger}
\newcommand{\pa}{\partial}
\newcommand{\la}{\langle}
\newcommand{\ra}{\rangle}
\newcommand{\St}{\mathcal{S}}
\newcommand{\hd}{h_{D}^{x}}
\newcommand{\Jxij}{J^{x}_{ij}}
\newcommand{\Jyij}{J^{y}_{ij}}
\newcommand{\Jzij}{J^{z}_{ij}}
\newcommand{\Sgn}[0]{\mathrm{Sgn}}
\newcommand{\newold}[2]{\textcolor{black}{#1}}
\newcommand{\newoldB}[2]{\textcolor{black}{#1}}
\newcommand{\newoldA}[2]{\textcolor{black}{#1}}
\newcommand{\newoldC}[2] {\textcolor{black}{#1}}

\begin{document}

\title{Dynamical Freezing and Scar Points in Strongly Driven Floquet Matter: Resonance vs Emergent 
Conservation Laws}

\author{Asmi Haldar$^{1,3}$, Diptiman Sen$^2$, Roderich Moessner$^3$, and 
Arnab Das$^1$}

\affiliation{$^1$Indian Association for the Cultivation of Science, \\
2A \& 2B Raja S. C. Mullick Road, Kolkata 700032, India \\
$^2$Centre for High Energy Physics and Department of Physics, 
Indian Institute of Science, Bengaluru 560012, India \\
$^3$Max Planck Institute for the Physics of Complex Systems, Dresden, Germany}

\date\today

\begin{abstract}
We consider a clean quantum system subject to strong periodic driving. 
The existence of a dominant energy scale, $h_D^x$, can generate 
considerable structure in an effective description of a system which,
in the absence of the drive, is non-integrable, interacting, 
and does not host localization. In particular, we 
uncover points of freezing in the space of drive 
parameters (frequency and amplitude). At those points, the dynamics 
is severely constrained due to the emergence of a\newold{n almost exact}{} local conserved quantity, 
which \newoldC{scars the {\it entire} Floquet spectrum by preventing}{prevents} the system from heating up ergodically,
starting from any generic state, even though it delocalizes over an
appropriate subspace. At large drive frequencies, where a na\"ive 
Magnus expansion would predict a vanishing effective (average) drive, we 
devise instead a strong-drive Magnus expansion in a moving frame. There, the
emergent conservation law is reflected in the appearance of an `integrability' 
of an effective Hamiltonian. These results hold for a wide variety 
of Hamiltonians, including the Ising model in a transverse field in {\it 
any dimension} and for {\it any form of Ising interactions}. 
\newoldB{The phenomenon is also shown to be robust in the presence of 
\drim{{\it two-body Heisenberg interactions with any arbitrary choice of 
couplings}}}{}.
Further, we construct a real-time perturbation theory which 
captures resonance phenomena where the conservation breaks down, giving way to 
unbounded heating. This opens a window on the low-frequency regime where 
the Magnus expansion fails. 
\end{abstract}

\maketitle

\section{Introduction} 

For closed systems with time-independent Hamiltonians, the notion of ergodicity 
has been formulated at the level of eigenstates as the eigenstate 
thermalization hypothesis (ETH)~\cite{Srednicki, Rigol_Nature}. According 
to ETH, the expectation value of a local observable in a single energy 
eigenstate of a complex (disorder-free) many-body quantum system is equal to 
the thermal expectation value of the observable at a temperature corresponding 
to the energy density of that eigenstate. The implication of the ergodicity 
hypothesis in the context of time-dependent (`driven') closed quantum systems 
is an open question of fundamental importance. 

Relatively recent progress along this line has occurred for 
systems subjected to a periodic drive (Floquet 
systems)~\cite{LDM_PRE, Rigol_Infinite_T}, which are perhaps conceptually 
closest to a static system. These studies indicate that a quantum system 
that satisfies ETH, when subjected to a periodic drive, approaches a state 
which locally looks like an entirely featureless `infinite-temperature' state. 
This is in accordance 
with the ergodicity hypothesis -- in systems which satisfy ETH (we will 
call them generic), energy is the only local conserved quantity, and any 
time-dependence breaks this conservation, allowing the system to explore 
the entire Hilbert space. 

The breakdown of ETH in interacting systems due to the presence of 
localized states -- either due to disorder (many-body 
localization)~\cite{Altshuler_MBL,ADP_MBL} or other mechanisms (like many-body 
Wannier-Stark localization)~\cite{Kris_Subir_Tilted_Mott,2019PNAS..116.9269V, 
Moessner_Stark} is well-known within the equilibrium set-up, and their 
persistence under periodic perturbations has also been
observed~\cite{FlqMBL1,FlqMBL2,Bhaskar_Scar}. \XX{Absolute stability bestowed upon
such Floquet systems by disorder even allows for interesting spatio-temporal
phases and long-range orderd in those systems}~\cite{Vedika_Stability,Norm_Chetan_Rev}.
But the common intuition is
that a translationally invariant, interacting, non-integrable many-body system
will be ergodic. However, this intuition has encountered a number of 
remarkable counterexamples recently within the static setting. It has been 
shown that in such systems there can be highly excited energy eigenstates, 
dubbed as scars, which do not satisfy ETH~\cite{Scar_Lukin_Nature,
Scar_Abanin_NatPhys, Scar_Shiraishi-Mori, Scar_Abanin_PRB, Scar_Vedika_PRB, 
Scar_Lukin_PRL, Scar_Serbyn_PRL, Scar_Bernevig_1, Scar_Vedika_Rahul,Scar_Bernevig_2}. 
Most of these examples (see, however, Ref.~\cite{Rigol_Comment}) indicate the 
non-trivial (weak) breaking of ergodicity by certain eigenstates. 

On the non-equilibrium side, stable Floquet states are seen in finite-size 
closed interacting Floquet systems which are not localized in the absence of 
a drive~\cite{Luca_Polku,Bukov_Polku_Huse,
Anatoli_Rev,Adhip_Diptiman,Sayak_Utsa_Amit,Bordia_Knap_Bloch,Sreejith_Mahesh, 
Qin_Hofstetter,gil-fss,Prosen_prl_98,Prosen_Tilted_NoHeat,AL_DL_PRB}.
In particular, it has recently been shown that ergodicity is broken 
in disorder-free generic systems under a periodic drive if the drive strength 
is greater than a threshold value (compared with the interaction strength) -- 
a KAM-like scenario~\cite{Onset}. 

The emergence of constraints on dynamics and approximately (stroboscopically) conserved quantities 
under strong periodic driving is known for non-interacting systems: for strongly driven 
\newold{spin chains that can be mapped to}{} free fermions, 
there exist special points in the space of the drive parameters, where 
{any arbitrary initial state}, for any 
(including infinite) 
system size, is 
frozen~\cite{AD-DMF,SB_AD_SDG,Mahesh_Freezing,Kris-Periodic,Russomanno_JStatMech}. 
This is surprising since the appropriate description for such a system is a 
periodic generalized Gibbs' ensemble (PGE)~\cite{PGE}. Such an ensemble,
though much less ergodic than a thermal one due to the presence of an extensive
number of (periodically) conserved quantities, still leaves ample space for 
substantial dynamics. \newold{In particular, the emergent (approximate) conserved quantity
is not one of the exact stroboscopically conserved quantities that characterizes the PGE,
and hence integrability does not necessarily assure its approximate conservation in any
trivial way.}{} Hence, in addition to the integrability, other constraints emerge at those 
special freezing points. 

\newold{Here, we  extend the reach of this phenomenology to generic {\it interacting} Floquet 
systems, far from integrability. Crucially, we also provide}{Here we uncover and similar scar phenomenology in {\it interacting}
Floquet systems, and provide}
a physical mechanism and an analytical 
understanding of the resulting non-ergodicity.
Concretely, we demonstrate that generic, interacting, translationally invariant 
Ising \newold{as well as Heisenberg}{} systems can exhibit 
non-ergodic behavior under a strong periodic drive \newold{and the non-ergodicity
is due to emergence of a new approximate (stroboscopic) conservation law not present in the
undriven quantum chaotic system.}{} For certain isolated sets 
of values of the drive parameters -- the scar/freezing points in the drive parameter space --
\newold{the conservation is most accurate, leading to almost perfect freezing of the conserved 
quantity for any generic initial state. 
}{a local quasi-conserved quantity (that exhibits only small fluctuations about
its initial) emerges}

The Floquet Hamiltonian is then no longer 
ergodic, i.e., its eigenstates (Floquet states) do not look like the otherwise 
expected infinite temperature states~\cite{LDM_PRE,Rigol_Infinite_T},
but instead are characterized by eigenvalues of the quasi-conserved quantity. 
This is because the dynamics does not mix different eigenstates with different 
eigenvalues of the quasi-conserved quantity. This, however, does not mean 
that there is no dynamics. Indeed, even at the scar points, we
see pronounced dynamics evidenced by substantial growth in sub-system 
entanglement entropy as delocalization takes place within each eigenvalue 
sector. A finite-size analysis of the numerical results indicates 
the stability of the scars under an increase in the system size. This 
is quite distinct from emergence of non-thermal Floquet states only at 
finite sizes, where the trend towards thermalization is clearly visible 
with increasing system size~\cite{Kolodrubetz}. 

\newoldB{We emphasize that -- unlike the conventional scars, which are manifested as a few (measure zero)
exceptional eigenstates -- here, near the scar point, the Hilbert space is fractured into dynamically disjoint
sectors, and hence the dynamics of {\it any initial state} is constrained.}{}

At high driving frequencies, the conventional Magnus expansion -- controlled 
by the driving frequency as the largest energy scale -- fails, as the average 
Hamiltonian generally does not exhibit the conservation law in question. To 
remedy this, we present a strong-drive Magnus expansion, 
constructed in a `moving' frame incorporating the strong driving term. Here, 
the conservation law is manifest at low order in the expansion. For a general 
class of Hamiltonians, including the Ising model in a transverse field in 
{\it any dimension} and {\it any form} of the Ising interactions, we find that 
the effective Hamiltonian satisfies the conservation law up to the two leading 
orders for our example, capturing the freezing (observed from exact numerics)
to a good approximation away from the resonances. This suggests that the 
expansion is either convergent or asymptotic. 

For lower drive frequencies, controlled approximation schemes for Floquet 
systems are sparse (see, however, Refs.~\cite{Babak_1,Babak_2,Babak_3}). 
Here, we formulate a novel perturbation theory, called Floquet-Dyson 
perturbation theory (FDPT), which again uses the fact that the drive 
amplitude is large. We find that this works best in the low-frequency regime, 
where we benchmark it for simple systems against an exact solution, and against exact numerics.
The advantage of our FDPT is that it enables us to account for 
isolated first-order resonances, which are of particular interest as their sparseness implies 
stable non-thermal states at this order. The stability is maintained in 
the thermodynamic limit if our perturbation expansion is an asymptotic one, 
which is indicated by the finite-size analysis of our numerical results -- the 
freezing is insensitive to an increase in the system-size 
(see the finite-size results in Sec.~\ref{subsec:Ldep}).
In particular, the FDPT is remarkably accurate in predicting the resonances 
(obtained from exact numerics) close to integrability, and hence at the scars.
This opens up a recipe to construct stable Floquet state with 
desired properties by choosing suitable drive terms.

We organize this paper as follows. After briefly introducing Floquet physics and 
our notations, we first present the phenomenology of scarring. We then develop 
the high-field Magnus expansion, and the FDPT. \newoldA{Our presentation 
focuses on
one particular quantum chaotic many-body model; we then go on to demonstrate
the key feature, namely, the emergence of the local conservation law and 
resulting absence of thermalization in diverse other models, namely, ones with
three-spin interactions, long-range (power-law) interactions, as well as general Heisenberg interactions, establishing the
 generality of the non-ergodic behavior that we have uncovered.}{} 
We conclude with a summary and an outlook. 

\section{Floquet in a Nutshell}

\begin{figure*}[ht]
\begin{center}
\includegraphics[width=0.31\linewidth]{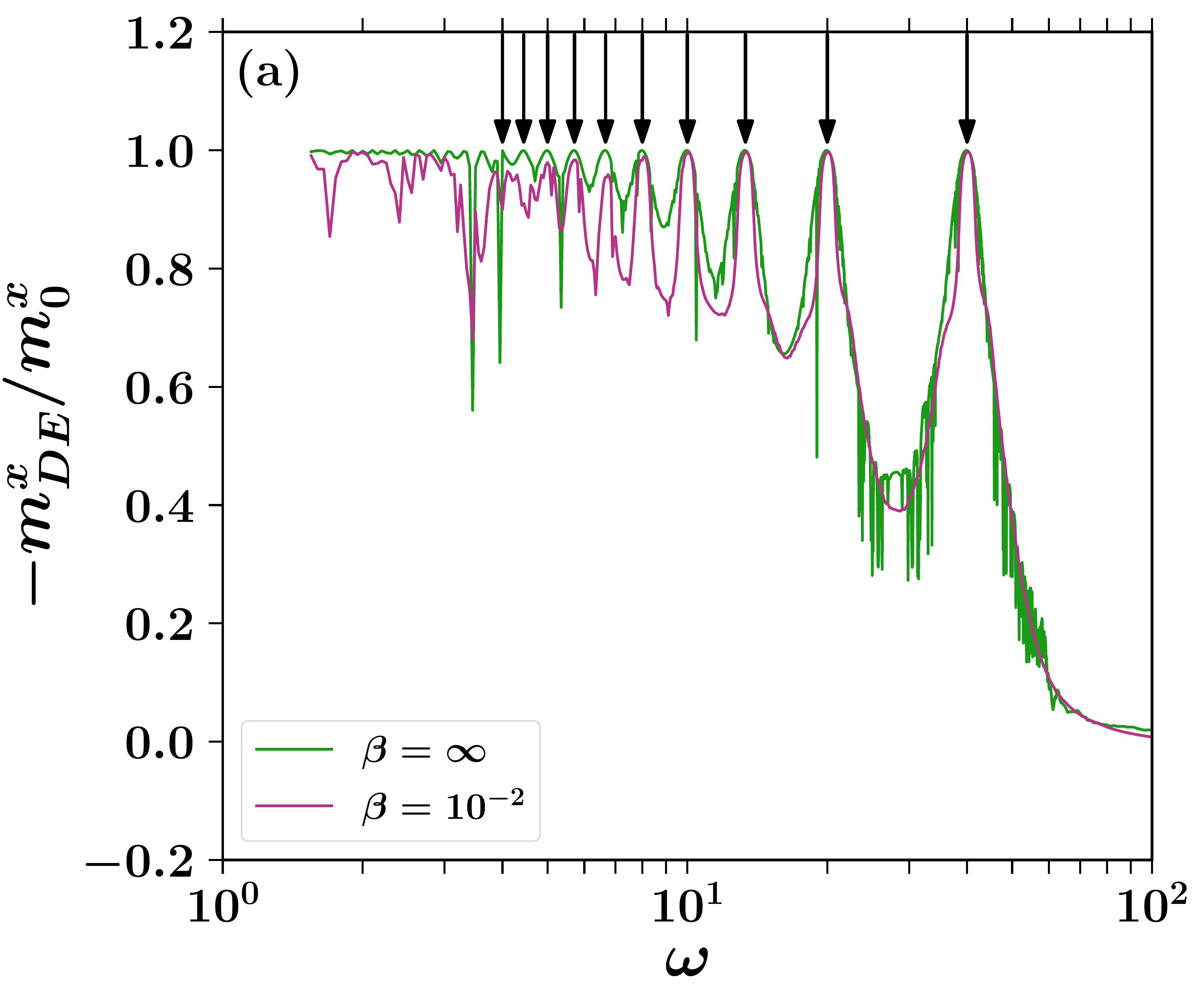}
\includegraphics[width=0.31\linewidth]{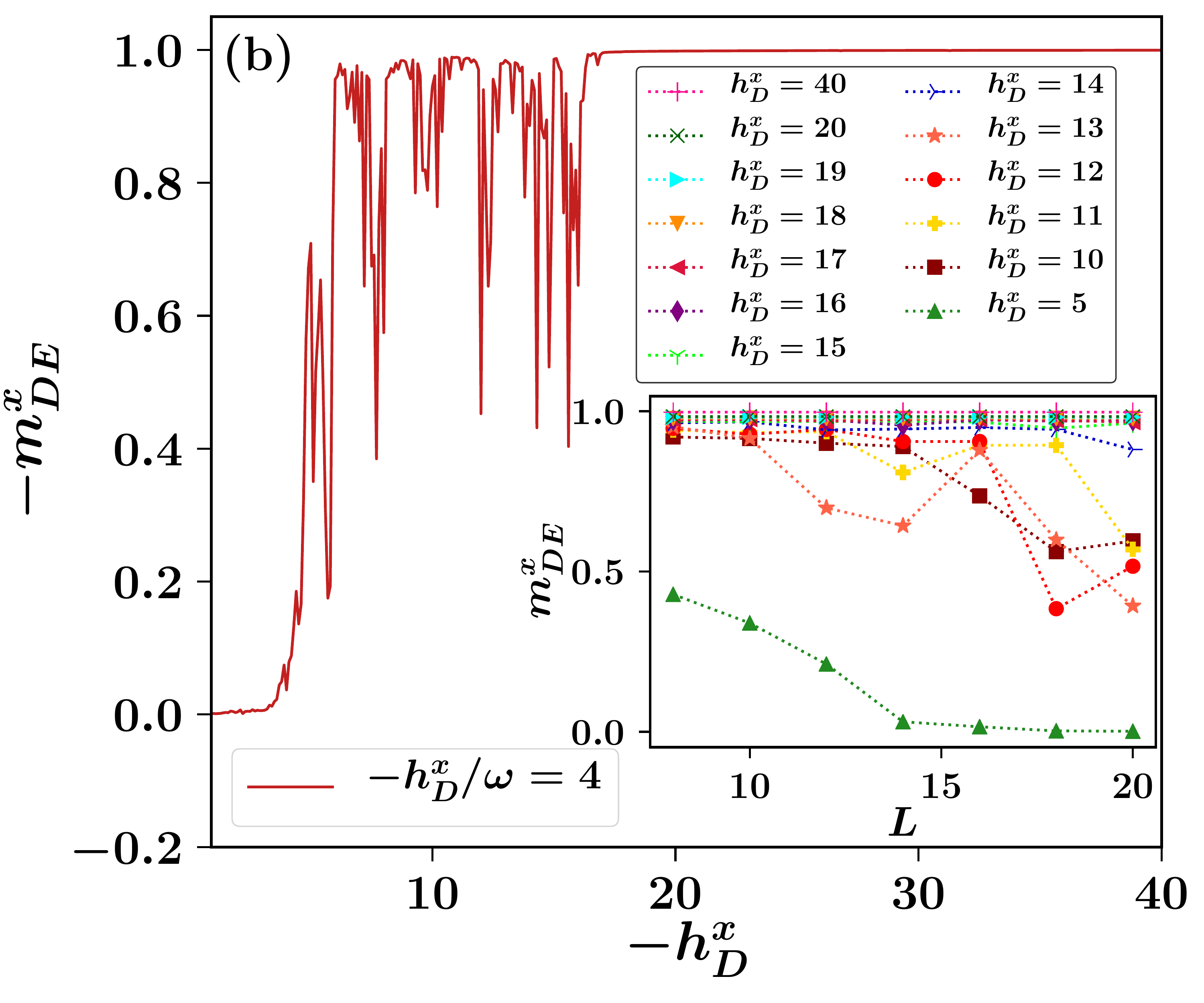}
\includegraphics[width=0.31\linewidth]{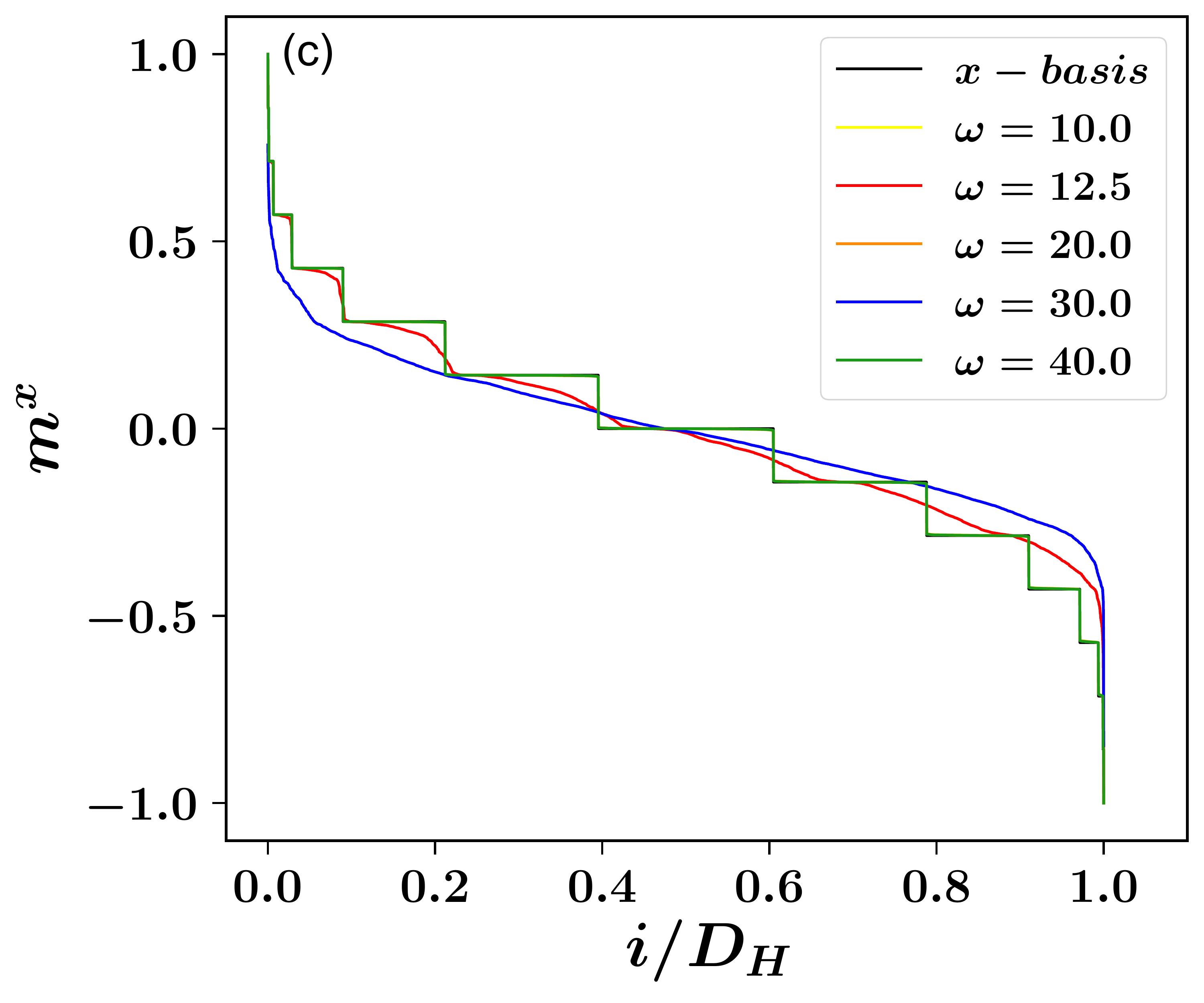}
\end{center}
\caption{Scars, resonances and emergent conservation law. 
{\bf(a):} $m_{DE}^{x}/m^x_0$, the ratio of magnetizations after 
infinite (diagonal ensemble average) and 0 (initial state)
cycles versus drive frequency $\om$. Freezing, reflected in a large value of 
this ratio, occurs over a broad range of $\om$, and is strongest at 
particular `scar' points (marked with arrows) $h^{x}_{D} = k\om,$ where $k$ 
is an integer (for $h^{x}_{D}= - 40$ here, the ten arrows mark $\omega 
= 40/k;~ k=1,2,..,10$). Results are shown for zero and high-temperature 
initial states: the former the 
ground state of $H(0)$ (which gives an initial magnetization $m^{x}(0) 
\alt 1$), and the latter the Gibbs state with $\beta = 10^{-2}$ ($m^{x}(0) 
\approx 0.05$) for $H_{I}$ of the form $H(0)$, but with $h_{D}^{x}=5$; all 
other parameters are the same as the driven Hamiltonian, namely, $J=1, ~\kappa 
= 0.7\pi /3, ~h^{x}_{0}= e/ 10, ~h_{D}^{x}=40, ~h^{z}=1.2, ~L=14$. The sharp 
dips in the green lines
represent resonances, discussed in detail in the main text on Floquet-Dyson 
perturbation theory. Parameters are chosen to avoid these resonances.
	\XX{{\bf (b):} The figure shows $m^x$ as a function of $h^{x}_{D}$ for a fixed ratio $|\hd/\om| = 4,$ showing 
the freezing cutoff  ($h^{x}_{D} \approx 18)$ above which a stable regime of freezing sets in.
	{\bf Inset:} Finite-size behaviour of the freezing. For intermediate strengths of the driving field, higher-order resonances lead to non-monotonic behaviour of $m^x_{DE}$ on $L$, which is then sensitively dependent on small variations of $h^x_D$.} 
{\bf(c):} $\la m^x \ra$ of the Floquet states plotted against 
the serial number (normalized by the Hilbert space dimension $D_{H}$) of the 
Floquet states, arranged in decreasing order of $\la m^x\ra.$ At the scar 
points ($\om = 10, ~20,$ and $40$) the $\la m^x\ra$ values form steps 
coinciding with the eigenvalues of $m^x$ arranged and plotted in the same 
order: $m^x$ emerges as a quasi-conserved quantity, hence the freezing of 
$m^x$ for {\it any} generic initial state.
}
\label{Peak_Valley_Steps} 
\end{figure*}

The Floquet states $|\mu_{n}\ra$ are elements of a complete orthonormal set 
of eigenstates of the time-evolution operator $U(T,0)$ for time 
evolution from $t=0$ to $t=T,$ for a system governed by a time-periodic 
Hamiltonian with a period $T = 2\pi/\om.$ The Floquet formalism is 
particularly useful for following the dynamics stroboscopically at 
discrete time instants $t = nT.$ From the above definition it follows that
\beq U(T,0)|\mu_{n}\ra ~=~ e^{-i\mu_{n}}|\mu_{n}\ra, 
\label{Floquet_EigEq} \eeq
where the $\mu_{n}$'s are real. It is customary to define an effective 
Floquet Hamiltonian $H_{eff}$ as 
\beq U(T,0) = e^{-iH_{eff} T}. \label{Heff:def} \eeq
(We will set $\hbar = 1$ in this paper).
When observed stroboscopically at times $t = nT,$ the dynamics can be thought 
of as being governed by the time-independent Hamiltonian $H_{eff},$ which has 
eigenvalues $\mu_{n}/T$ (modulo integer multiples of $2\pi /T$) 
and eigenvectors $|\mu_{n}\ra.$ In the infinite time limit, the 
expectation values of a local operator ${\cal O}$ can be
written in terms of the expectation values in the Floquet eigenstates as
\beq \lim_{N\to\infty} \la\y(NT)|{\cal O}|\y(NT)\ra = \sum_{n} |c_{n}|^{2}
\la\mu_{n}|{\cal O}|\mu_{n}\ra = {\cal O}_{_{DE}}, \label{DEA_formal} \eeq
where $|\psi(0)\ra = \sum_{n}c_{n}|\mu_{n}\ra,$ and the subscript 
``${DE}$" denotes the diagonal ensemble 
average~\cite{Srednicki_DE,Rigol_GGE_1,Rigol_GGE_2,
Rigol_Nature} as defined above. \newoldA{The DE description has been shown 
to be a very accurate description for generic interacting systems
after a quench at long times~\cite{Reimann}. A Floquet system under stroboscopic observation is
equivalent to a quench with $H_{eff}$ in conjugation with stroboscopic observations. 
We mainly focus on the longitudinal magnetization (polarization in the $x$ direction), given by 
\beq m^x ~=~ \frac{1}{L} ~\sum_{i}^{L} ~\si_{i}^{x}. 
\label{mx:def} 
\eeq
The diagonal
ensemble average}{} is equivalent to a ``classical" average over the properties of the 
Floquet eigenstates $\{|\mu_{n}\ra\}.$ The diagonal ensemble average of 
$m^{x}$ given by
\beq m^{x}_{_{DE}} ~=~ \sum_{n} ~|c_{n}|^{2} ~\la\mu_{n}|m^{x}|\mu_{n}\ra.
\label{DEAx} \eeq 
The absence of interference between the Floquet states in a DE average 
ensures that it is sufficient to study the properties of individual Floquet 
states (and their spectrum average) in order to characterize the gross 
behavior of the driven system in the infinite-time limit. In the following 
we will therefore mostly concentrate on DE averages and the properties of 
the Floquet states.

\section{The Scar Phenomenology}

\subsection{Freezing and Quasi-Conservation}
 
This section discusses the scar phenomenology for a 
periodically driven, interacting, non-integrable Ising chain described by
\bea H(t) &=& H_{0}(t) ~+~ V, ~~ {\rm where} \non \\
H_{0} (t) &=& H_{0}^{x} ~+~ \Sgn (\sin (\om t)) ~H_{D}, ~~ {\rm with} 
\non \\
H_{0}^{x} = &-& ~\sum_{n=1}^L ~J \si_n^x \si_{n+1}^x + \sum_{n=1}^L ~\ka 
\si_n^x \si_{n+2}^x - h_{0}^{x}~\sum_{n=1}^L\si_n^x, \non \\
H_{D} = &-& ~ h_D^x ~\sum_{n=1}^L \si_n^x, ~~ {\rm and} \non \\
V = &-& ~ h^z \sum_{n=1}^L \si_n^z, 
\label{ham3} 
\eea
where $\si_{n}^{x/y/z}$ are the Pauli matrices. 
\newoldA{Note that $H_{0}^{x}$ is,
by definition, the sum of all the terms that commute with $H_{D}(t),$ and $V$ is the sum of {\it all the
remaining terms} in the time-independent part of the Hamiltonian (while $V$ is
termed ``perturbation" later, there is no implied distinction between the relative strengths of $H^{x}_{0}$ and $V$). This partition of the static part into $H_{0}^{x}$ and $V$ is for computational 
bookkeeping convenience for our analytical derivations.}{}

The main result is that at large drive amplitude $h_{D}^{x},$ the longitudinal 
magnetization $m^x$ 
emerges as a quasi-conserved quantity under the drive condition (`scar
points' in the drive parameter space) given by
\beq h^{x}_{D} ~=~ k\om, \label{scar:def} \eeq
where $k$ are integers. 
Fig.~\ref{Peak_Valley_Steps}(a), main frame, shows that at the scar points 
(marked with arrows), the diagonal ensemble average $m^x_{DE}$ 
(Eq.~\eqref{DEA_formal}) for $m^x$ is equal to the initial value $m^x(0),$ to 
very high accuracy, indicating that $m^x$ remains frozen at its 
initial value for arbitrarily long times. 
As seen from the figure, this happens for a very broad range of $\om.$ 

\XX{
However, as $\hd$ is reduced, the stable frozen regime eventually gives way to 
Floquet thermalized regime. Below a clear  freezing cutoff (around $h^{D}_{x} \approx 18$), $m^x_{DE}$ exhibits strong fluctuations as a function of $h_{D}^{x},$ (short frozen stretches punctuated by 
higher order resonance; see Sec~\ref{Reso_Scar}) followed by a subsequent sharp decline
to almost zero below a thermalization threshold (around $h_{D}^{x}\approx 5$). A locally infinite-temperature like Floquet thermalized regime is observed below this threshold, as shown in Fig.~\ref{Peak_Valley_Steps}(b).
This threshold does not exhibit any perceptible shift with system-size~\cite{Onset}. 
The inset of Fig.~\ref{Peak_Valley_Steps}(b) shows that there is no perceptible $L$-dependence in 
the freezing of $m^{x}_{DE}$ as long as $\hd$ is above the freezing cutoff.
}
The phenomenon is reminiscent 
of the non-monotonic peak-valley structure of freezing observed in integrable 
Floquet systems in the thermodynamic limit~\cite{AD-DMF,SB_AD_SDG}.

The figure shows that freezing happens for two very different kinds of initial 
states, namely, the highly polarized initial {\it ground} state of $H(0)$ 
as well as a {\it high-temperature} thermal state. The initial thermal density 
matrix is of the form
\beq \rho_{_{Th}}(t=0) ~=~ \sum_{j=1}^{2^L} ~
\frac{e^{-\beta\varepsilon_{j}}}{\mathcal Z}~
|\varepsilon_{j}\ra\la \varepsilon_{j}|, \label{rhoTh_Init} \eeq
where $|\varepsilon_{j}\ra$ is the $j$-th eigenstate of an initial 
Hamiltonian $H_{I},$ with eigenvalue $\varepsilon_{j}.$ We have chosen $H_{I} = 
H(t=0,h^{x}_{D}=5.0, h^{x}_{0}=0.1, J=1,\kappa=0.7),$ with $H(t)$ from 
(Eq.~\eqref{ham3}), and ${\mathcal Z} = \sum_{j}e^{-\beta\varepsilon_{j}}$ is 
the partition function. Eq.~\eqref{rhoTh_Init} represents a mixture of 
eigenstates $|\varepsilon _j\ra.$ Hence we obtain the final diagonal ensemble 
density matrix by taking the diagonal ensemble density matrix for each 
$|\varepsilon_j\ra$, weighted by its Boltzmann weight in $\rho_{_{Th}}(0),$ 
i.e., 
\begin{eqnarray}
\rho_{_{DE}}(t\to\infty) ~=~ \sum_{j}\frac{e^{-\beta\varepsilon_{j}}}{\mathcal 
Z} \left( \sum_{k} ~|\la\varepsilon_{j}|\mu_{k}\ra|^{2} ~|\mu_{k}\ra
\la \mu_{k}|\right) &~& \non \\
= \sum_{k}\left(\sum_{j} ~\frac{e^{-\beta\varepsilon_{j}}}{\mathcal Z} ~
|\la\varepsilon_{j}|\mu_{k}\ra|^{2} ~\right)|\mu_{k}\ra\la \mu_{k}|. ~~~ &~& 
\label{rhoTh_DE}
\end{eqnarray}

The quasi-conservation of $m^x$ for a generic thermal state suggests that all 
the Floquet states must be organized according to the emergent conservation 
law. This is shown to be true in Fig.~\ref{Peak_Valley_Steps}~(c), which displays 
the expectation value $\la m^x \ra$ in the Floquet eigenstates 
(corresponding to the drive in Fig.~\ref{Peak_Valley_Steps}~(c)), plotted 
against their serial number (normalized by the dimension $D_{H}$ of the 
Hilbert space), arranged in decreasing order of their $\la m^x \ra$
values. For the scar points, for $h_{D}^{x}=40$ at $\om = 10, 20, 40,$
the values of $\la m^x \ra$ of the Floquet states coincide 
 with the eigenvalues of $m^x,$ indicating that all the 
eigenstates of $m^x$ which participate in a given Floquet state 
have the same $m^x$ eigenvalues. This explains conservation/freezing
of $m^x$ for dynamics starting with any generic initial state. As we will see 
later, the condition \newoldC{for encountering such a scar point}{of the scar} 
(Eq.~\eqref{scar:def}) can be deduced both 
from the FDPT and a Magnus expansion in a time-dependent frame, and the 
latter confirms the effect over the entire spectrum and explains the steps 
in $\la m^x \ra$ to the leading orders.

\begin{figure*}[ht]
\begin{center}
\includegraphics[width=0.8\linewidth]{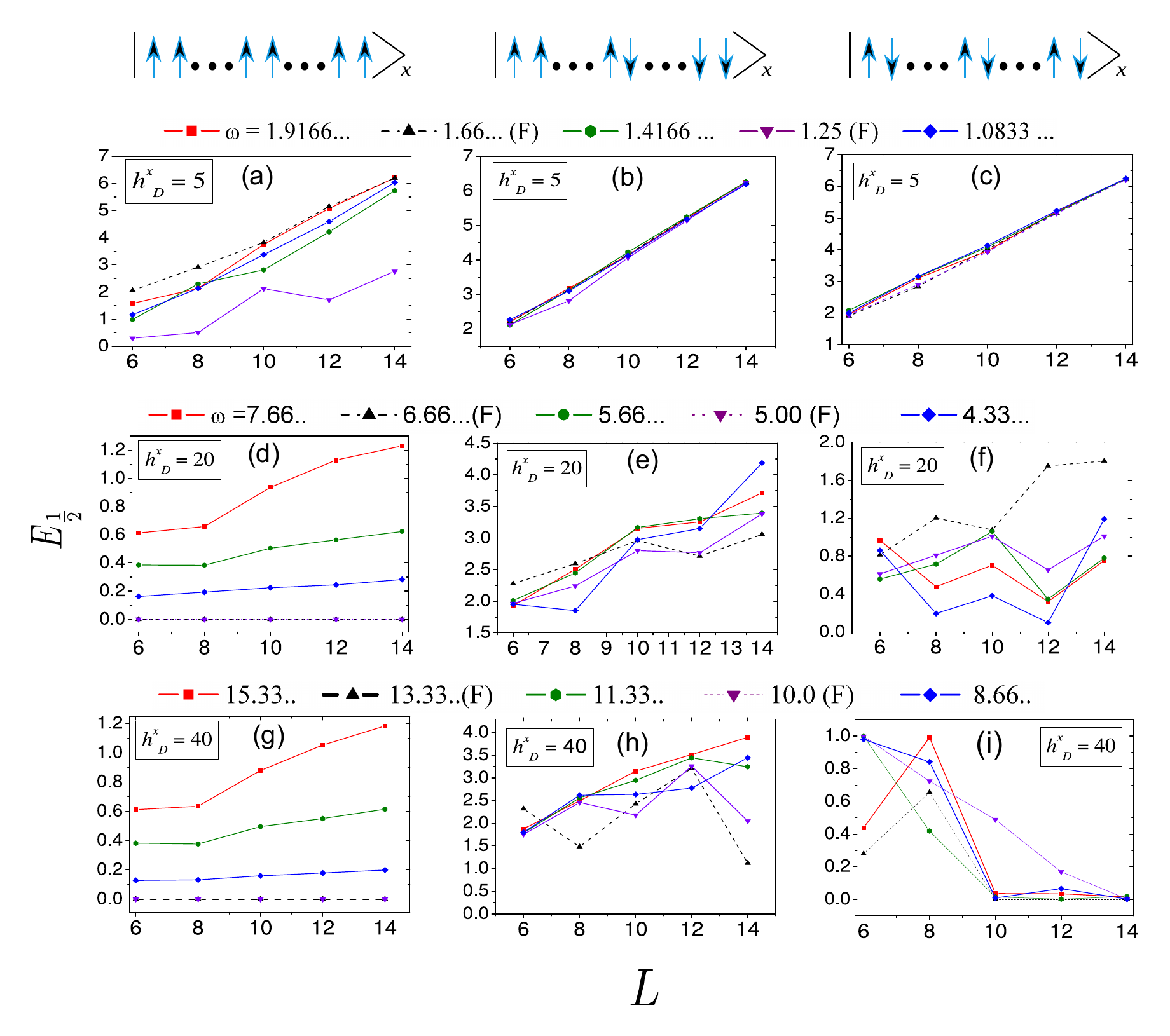}
\end{center}
\caption{(De)localization of the wave function over the $x$-basis 
(simultaneous eigenstates of all of the $\sigma_{i}^{x}$'s) 
as evidenced by the half-chain entanglement entropy ($E_{\frac{1}{2}}$)
versus system size $L,$ for different driving strengths $h_{D}^{x}$ (rows) 
and initial states (columns: left: maximally $m^x$ polarized; middle: 
$L/2$-domain-pair state with vanishing total $m^x$; right: N\'eel state). 
Top row (small $h_{D}^{x} = 5$): $E_{\frac{1}{2}}$ entropy grows linearly 
with system size for all initial states, signaling ergodicity. 
For stronger drives ($h_{D}^{x} = 20,40$ in middle, bottom row, respectively), 
scars appear, and $E_{\frac{1}{2}}$ depends strongly on the initial states, 
reflecting the size of the emergent magnetization sectors: for the fully 
polarized initial states (left column), 
$E_{\frac{1}{2}}$ does not grow at all for the freezing/scar points (marked 
as (F) in the figure legends and represented by almost indistinguishably 
coincidental black and violet triangles), while for the N\'{e}el and the 
$L/2$-domain-pair initial states, there is considerable growth in 
$E_{\frac{1}{2}}$ even at the scar points, reflecting (at least partial) 
delocalization over the large concomitant magnetization sectors. The results 
are for $J=1, ~\kappa = 0.7, ~h^{x}_{0}=e/ 10, ~h^{z}=1.2, ~L=14$, averaged 
over $10^4$ cycles after driving for $10^{10}$ cycles.} \label{EE} \end{figure*}
\subsection{Dynamics of the Unentangled Eigenstates of $m^x$: Growth of Entanglement Entropy}
We define an unentangled, complete, orthonormal set 
of eigenstates of $m^x,$ which we call the $x-$basis. Each element of 
the $x-$basis is a simultaneous eigenstate of all the $\si_{i}^{x}$ 
operators. The non-triviality of the dynamics at the scar points and the 
consequence of the quasi-conservation is manifested in the growth of the 
half-chain entanglement entropy $E_{\frac{1}{2}}$ at the scar points, 
especially with different $x$-basis eigenstates of $m^x$ as initial states. 
We study the half-chain entanglement entropy 
\beq E_{\frac{1}{2}} ~=~ - ~Tr [\rho_{_\frac{1}{2}} \log_{2}
{\rho_{_\frac{1}{2}}}], \label{E_hlf} \eeq
where $\rho_{_\frac{1}{2}}$ is the density matrix of one half of the chain, 
obtained by tracing out the other half. 

The results are shown in Fig.~\ref{EE}. These highlight that, even though $m^x$ 
is conserved for large enough $h_{D}^{x}$ at the scar points, there is 
substantial dynamics even at those points. 
For large enough $h_{D}^{x},$ Figs.~\ref{EE} (d-i), we see that different 
eigenstates of $m^x$ evolve quite differently even at the scar points, at 
which $m^x$ is conserved to a very good approximation for all initial states. 
For example, for the fully polarized initial state entanglement does not grow 
even after $10^{10}$ drive cycles, but for the N\'{e}el and the 
$L/2$-domain-pair initial states, it does. This reflects the respective sizes 
of the $m^x$ subspaces with maximal and zero magnetization. 

The growth of $E_{\frac{1}{2}}$ also reflects the role of interactions in the 
dynamics even at the scar points, without which we would not see such a 
substantial growth of entanglement.

In App.~\ref{app:varfield}, we show that the suppression of entanglement growth
is robust in that it is observed for other patterns of the drive field, as 
long as the concomitant emergent conservation law gives rise to well-defined 
sectors which contain only a small number of states. 

\section{Strong-drive Magnus Expansion}
\label{sec:Magnus}

We next provide a modified Magnus expansion which
incorporates the large size of the drive from the start, 
using the inverse of the driving field as a small parameter. 
This makes the emergence of a conserved quantity manifest, for a wide range 
of Hamiltonians -- the terms in the time-independent part of the Hamiltonian 
that commute with the time-dependent part of the Hamiltonian ($H^{x}_{0}$ 
here) can have any form. This is because the factor pre-multiplying
the terms involving $H_{0}^{x}$, vanishes to second order regardless of the 
form of $H_{0}^{x}$.
For example, it applies to transverse field Ising models in {\it any} 
dimension, with {\it any} Ising interaction. From this, one can immediately 
read off the scars found above.

The conventional Magnus expansion uses the inverse of a large frequency as a 
small parameter (see, e.g., \cite{Anatoli_Rev,Andre_Anisimovas_Rev}) for obtaining the Floquet 
Hamiltonian $H_{eff}$ (Eq.~\eqref{Heff:def}) as given below.
\bea H_{eff} &=& \sum_{n=0}^{\infty} H^{(n)}_{F}, ~ ~ {\rm where} \non \\
H^{(0)}_{F} &=& \frac{1}{T} \int_{0}^{T} dt ~H(t), \non \\
H^{(1)}_{F} &=& \frac{1}{2 ! (i) T}\int_{0}^{T}dt_{1}\int_{0}^{t_{1}} dt_2
[H(t_1), H(t_2)], 
\label{Magnus} \eea
and so on. In our case, we have $h_{D}^{x} > \om,$ making the 
series non-convergent even when $\om$ is greater than all other couplings in 
the Hamiltonian, so the na\"ive Magnus expansion is qualitatively wrong even at leading order: 
the first-order term $H^{(0)}$ is the time average over one period of $H(t)$ 
(Eq.~\eqref{ham3}), an interacting generic Hamiltonian which 
does not conserve $m^x.$ Hence we would have no hint of the scars from even 
the first-order term. 

This problem can be remedied when the strong drive modulates the strength of 
a fixed field/potential (this is a very natural way of applying a periodic 
drive). The largest coupling ($h_{D}^{x}$ here) can be eliminated from 
the Hamiltonian by switching to a time-dependent frame as 
follows~\cite{Anatoli_Rev}. We introduce a unitary transformation 
\bea |\psi_{mov}(t)\ra &=& W(t)^{\da}|\psi(t)\ra, \non \\
\hat{\cal O}_{mov} &=& W(t)^{\da}\hat{\cal O}W(t), \label{Moving_Frame} \eea
where $|\psi(t)\ra$ is the wave function and $\hat{\cal O}$ is any predefined 
operator (the subscript $_{mov}$ marks the quantities in the moving frame).

The crux of the expansion is then apparent for a $W(t)$ of the following form,
\bea W(t) &=& \exp{\left[-i\int_{0}^{t} dt^{\prime} ~r(t^{\prime}) ~H_D
\right]}, \label{Urot} \eea
\noindent where $r(t)$ is $T$-periodic parameter. 
If the total Hamiltonian were {\it constant up to the time-dependent prefactor 
$r(t)$}, i.e. $H(t)=r(t) H(0)$, 
the above would just give the solution of the static Schr\"odinger equation, 
but with a rate of phase accumulation for each (time-independent) eigenstate 
given by the integrand of the variable prefactor. In particular, any 
conservation law of $H(0)$ would be bequeathed to the time-dependent problem. 
Now, if the drive is not the only, but still the dominant part, of the 
Hamiltonian, there will be corrections to this picture, but it suggests 
the eigenbasis of the drive and its conservation law(s) should remain 
perturbatively useful starting points. \\

Given the form of $H_{D}(t)$ in Eq.~\eqref{ham3}, the transformed Hamiltonian 
reads
\beq H_{mov} ~=~ W(t)^{\da}H(t)W(t) ~-~ i W(t)^{\da}\partial_{t}W \ , 
\label{Ham_Mov} \eeq 
\noindent where the second term exactly cancels the part from the first 
term which
has $h^{x}_{D}$ as its coupling, and hence $H_{mov}$ is free from any coupling 
of order $h_{D}^{x}$~(see App.~\ref{app:Integrals} for details).

\subsection{Scars in the Driven Interacting Ising Chain}

In the case of Eq.~(\ref{ham3}), we have 
\beq H(t) = H_{0}^{x} + V - \Sgn (\sin{(\omega t)}) ~h_{D}^{x} \sum_{i} 
\sigma_{i}^{x}. \eeq
\noindent
Switching to the moving frame by using the transformation in Eq.~(\ref{Urot}) 
gives
\bea H_{mov} &=& H_{0}^{x} -h^{z}\sum_{i}\left[ \cos{(2\theta)}\si_{i}^{z} 
+ \sin{(2\theta)}\si_{i}^{y} \right], ~ {\rm where} \non \\
\theta(t) &=& - ~h_{D}^{x}\int_{0}^{t} dt' ~\Sgn (\sin{\om t'}).
\label{Hmov_Ising} \eea
After some algebra, we find the Magnus expansion of $H_{mov}$ to have the 
following leading terms:
\bea 
H_{F}^{(0)} = H_{0}^{x} &-& \frac{h^z}{\hd ~ T}\left[ 
\sin{(h_{D}^{x}T)}\sum_{i}\si_{i}^{z} \right. \non \\
&~&-(1 - \cos{(h_{D}^{x}T)}  \left. )\sum_{i}\si_{i}^{y} 
\right]. \label{HF_0} 
\eea
Note that this is useful for $h_D^x\gg 1/T$, the regime we are interested in. 
The next-order term is given by :
\bea
H_{F}^{(1)} &=&  \frac{1}{2! Ti} \int_0^T dt_1 \int_0^{t_1} dt_2 \left[H^{\mathrm{mov}}(t_1),H^{\mathrm{mov}}(t_2)\right]
\label{HF2_Modules}
\eea
Calling $\ta(t_1) = \ta_1$; $\ta(t_2) = \ta_2$ and $\sum_i \si_i^{z/y} = S^{z/y}$ and using the form of $H_{mov}$
from Eq.~(\ref{Hmov_Ising}), we get
%
\begin{widetext}
\bea
H_{F}^{(1)} &=&
\left[S^z,H^x_0 \right] h^z (\cos 2\ta_2 - \cos 2 \ta_1) +
\left[S^y,H^x_0 \right] h^z (\sin 2 \ta_2 - \sin 2 \ta_1)  + \left[S^y,S^z \right] (h^z)^2 \sin (2\ta_1 - 2\ta_2).
\eea
\end{widetext}
\noindent
Upon integration (see App.~\ref{app:Ising} and ~\ref{app:Integrals}
for details), this {\it identically} gives
\beq
H_{F}^{(1)} = 0.
\label{HF2_S1_S2_S3}
\eeq
The end result -- a homogeneous expansion in the small parameters 
$1/h^x_D$ and $1/T$ from the two initial orders -- given in Eqs.~\eqref{HF_0}, 
is quite remarkable.
First, for $\hd T = 2\pi k$ (where $k$ can be any integer), $H_{F}^{(0)} = 
H_{0}^{x}$; this is precisely the condition for 
scars observed numerically (Eq.~\eqref{scar:def}) and also from the FDPT (see 
Eq.~\eqref{mx3}). 
Clearly, to this approximation, $H_{eff}$ not only has 
a conservation law, but is also integrable; indeed it is classical, with all 
terms commuting. 
Numerical results suggest that the above expansion (unlike the Magnus expansion
in the static frame) is an asymptotic one, at least in the neighborhood of 
the scar points, since the leading order terms represent the exact numerical 
results accurately. 

Secondly, it is clear from the forms of $H_{F}^{(0)}$ and $H_{F}^{(1)}$ that 
the results hold independently of the form of $H_{0}^{x}$; this could be in any 
spatial dimension, and can incorporate any form of Ising interactions!
This wide generality implies that stable
quasi-conservation laws and constraints (in keeping with the
possible asymptotic nature of the expansion) may emerge in
generic interacting Floquet systems in the thermodynamic limit.
\newoldA{Since $H^{0}_{x}$ is by definition the portion
of the static part of the Hamiltonian that commutes with $H_{D},$ the statement
of generality obtained from the above analysis stands as follows: while the 
nature of the whole static part can be tuned over a wide variety of many-body 
Hamiltonians depending on the form of $H_{0}^{x}$ 
(ranging from non-interacting to interacting, integrable to non-integrable, low to high dimensional), the emergence of the conservation law and the resultant 
scarring do not depend on the form of $H_0^x$. In Sec.~\ref{Sec:Gen} we 
support this statement by considering various kinds of Ising interactions, 
and going beyond, we demonstrate the freezing in the presence of anisotropic 
Heisenberg interactions.}{}

\section{Floquet-Dyson Perturbation Theory}
\label{sec:pt}

In this section, we develop a theory which opens up a window on the otherwise 
difficult-to-access~\cite{Babak_1,Babak_2,Babak_3} low-frequency regime. 
We first test it for an exactly soluble problem, and then apply it to 
the Ising chain studied in the previous section. 

\begin{figure*}[t]
\begin{center}
\includegraphics[width=0.88\linewidth]{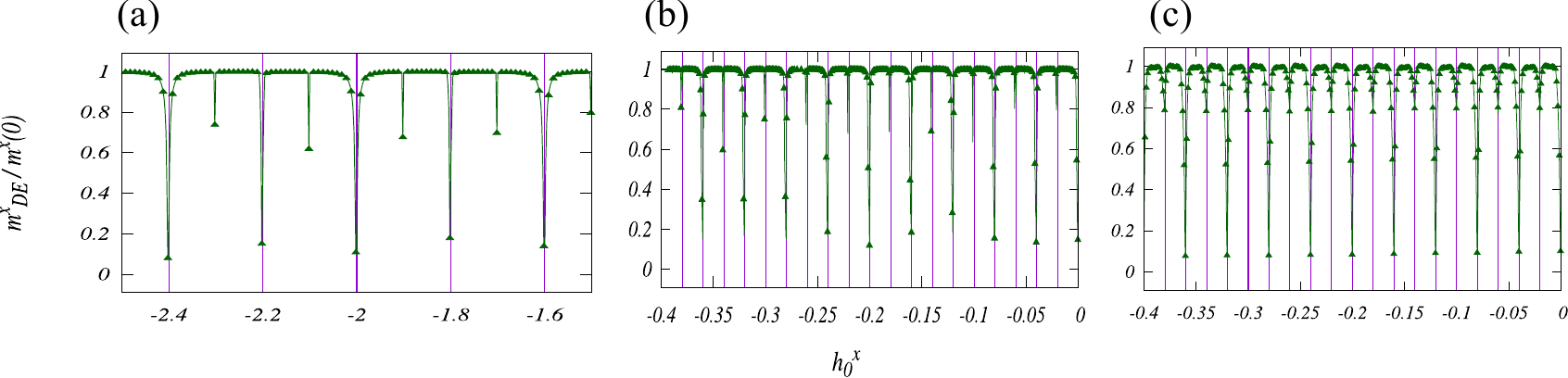}
\end{center}
\caption{{\bf The Interacting Case:} Freezing and resonances in the 
magnetization ratio $m_{DE}^{x}/m^x_0$ versus $h_0^x$. The 
observable, initial states at zero (panels a, b) and high temperature 
(inverse temperature $\beta = 10^{-2}$) (panel c), and other parameters are
as described in 
Fig.~\ref{Peak_Valley_Steps} (a). Results shown for slow (a: $\om = 0.4$) and
very slow (b, c: $\om = 0.04$) drives (green line-points). The resonances 
obtained from first-order FDPT, Eq.~\eqref{Gen_Res}, (purple vertical lines)
show a remarkable match with the numerical values of dips in $m^x.$
(Some higher order resonances are also visible at $\om=0.4$ in panel a). The 
other parameters are the same as in Fig.~\ref{Peak_Valley_Steps}.} 
\label{Resonances} \end{figure*}

We find the theory provides valuable insights for both systems. In particular, 
it identifies a resonance condition corresponding to the dips, as well as a 
freezing condition corresponding to the maxima in the response plotted in 
Figs.~\ref{Resonances} and \ref{Peak_Valley_Steps}, respectively.
A coincidence of the two accounts for the varying dip depths in that figure. 
While a comprehensive treatment of the general many-body problem is not yet 
possible, we believe that these items capture ingredients central for its 
understanding.

We first present the general formulation of the FDPT. 
The goal is to construct 
the Floquet states $|\mu_{n}\ra.$ We resort to a setting where 
the unperturbed Hamiltonian is time-dependent and the perturbation is static~\cite{soori}.
\cyan{The central idea is to construct the 
Floquet states in presence of the small static
perturbation from the known unperturbed Floquet states
by applying time-dependent perturbation theory
(a Dyson-like series for the wave-function). For this, one needs to know the 
unperturbed Floquet states, which comes from the
solution of the time-dependent Schr\"{o}dinger equation 
with only the time-dependent part in the Hamiltonian (including the static 
parts that commute with it at all time). 
In our case, this is naturally achieved as follows. The central ingredient is that the driven 
Hamiltonian
\beq H(t) ~=~ H_0 (t) ~+~ V \label{ham1} \eeq
contains a large time-dependent term $H_0 (t)$ which has {\it a 
time-independent set of eigenstates} and a perturbation $V$ that is time-independent. 
Those states then serve as the unperturbed Floquet states,
and $V$ can then be treated as a small (compared to the drive amplitude) perturbation.} \\

We work in the basis of eigenstates of $H_0 (t)$ (these are the unperturbed Floquet states), 
denoted as $| n \ra$, 
so that
\beq H_0 (t) | n \ra ~=~ E_n (t) | n \ra, \label{eig1} \eeq
and $\la m | n \ra = \de_{mn}$. 

Next, we assume without loss of generality that $V$ is 
completely off-diagonal in this basis, namely,
\beq \la n | V | n \ra ~=~ 0 \label{nvn} \eeq
for all $n$. We will now find solutions of the time-dependent Schr\"{o}dinger 
equation
\beq i \frac{\pa |\psi_n\ra}{\pa t} ~=~ H(t) |\psi_n (t)\ra, \label{sch1} \eeq
which satisfy 
\beq |\psi_n (T)\ra ~=~ e^{- i \mu_n} ~|\psi_n (0)\ra. \label{floeig1} \eeq

For $V=0$, each eigenstate $|n\ra$ of $H_{0}(t)$ is a Floquet state, with 
Floquet quasienergy $\mu_{n}^{(0)} = \int_{0}^{T} dt E_{n}(t)$ (defined 
modulo $2\pi$).

For $V$ non-zero but small, we develop a Dyson-like series for the 
wave function to first order in $V$. 
Clearly $V$ is a small perturbation as long as $|V /h^{x}_{D}| \ll 1,$ though it 
can otherwise be comparable to or larger than the other couplings of the undriven Hamiltonian.
In our ansatz, the $n$-th eigenstate is written as 
\beq 
|\psi_n (t)\ra ~=~ \sum_m ~c_m (t) ~e^{-i \int_0^t dt' E_m (t')} ~ |m \ra 
\label{nine}, 
\eeq
where $c_n (t) \simeq 1$ for all $t$ while $c_m (t)$ is of order $V$ (and
therefore small) for all $m \ne n$ and all $t$. \\

\cyan{We then
substitute the form for the wave-function in Eq.~\ref{nine}
in the time-dependent Schr\"{o}dinger equation, and then 
apply the key condition of the method, namely, we demand $|\psi_{n}(0)\ra = |\mu_{n}\ra,$ i.e., 
\beq
|\psi_n(T)\ra = e^{i\mu_{n}}|\psi_{n}(0)\rangle. 
\label{Floquet_Cond}
\eeq
\noi
Then taking
the overlaps with the basis states $|m\ra,$  we find}
(for details of the algebra, see App.~\ref{app:FDPT}):
\beq c_m (0) ~=~ - i ~\la m | V | n \ra ~\frac{\int_0^T dt ~
e^{i \int_0^t dt' [E_m (t') - E_n (t')]}}{e^{i \int_0^T dt [E_m (t) - E_n (t)]}
~-~ 1}. \label{cmt2} \eeq
We see that $c_m (t)$ is indeed of order $V$ provided that the denominator on
the right hand side of Eq.~\eqref{cmt2} does not vanish; we call this case 
non-degenerate. If
\beq e^{i \int_0^T dt [E_m (t) - E_n (t)]} ~=~ 1, \label{res1} \eeq
we have a resonance between states $| m \ra$ and $|n \ra$, and the above 
analysis breaks down. Now, if there are several states 
which are connected to $|n\ra$ by the perturbation $V$, Eq.~\eqref{cmt2} 
describes the amplitude to go to each of them from $|n\ra$.
Up to order $V^2$, the total probability of excitation away from $|n\ra$
is given by $\sum_{m \ne n} |c_m (0)|^2$ at time $t=0$.


\subsection{Single Large Spin: An Exactly Soluble Test-bed}
\label{sec:singlespin}
\begin{figure}[h!]
\begin{center}
\includegraphics[width=0.49\linewidth]{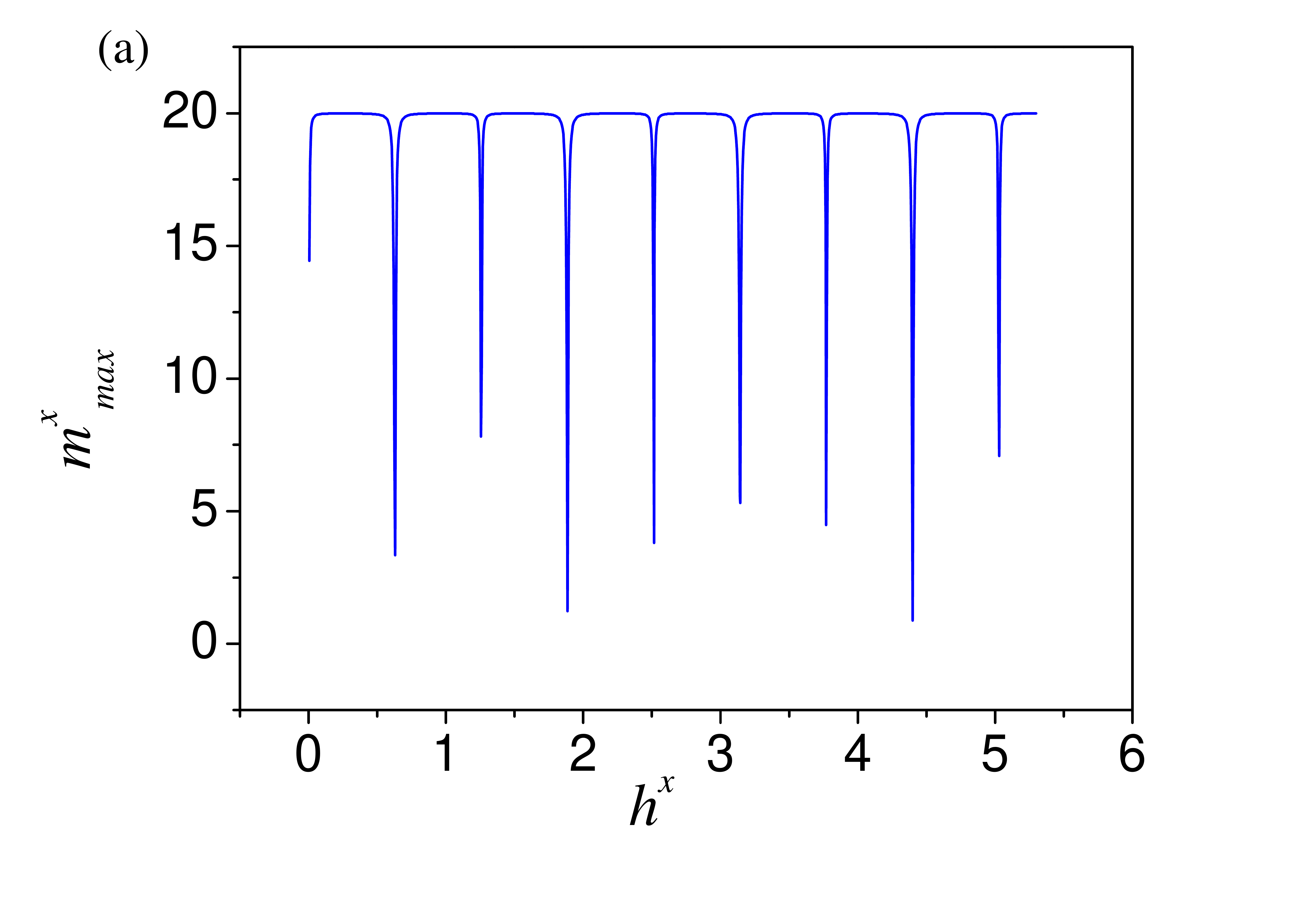}
\includegraphics[width=0.49\linewidth]{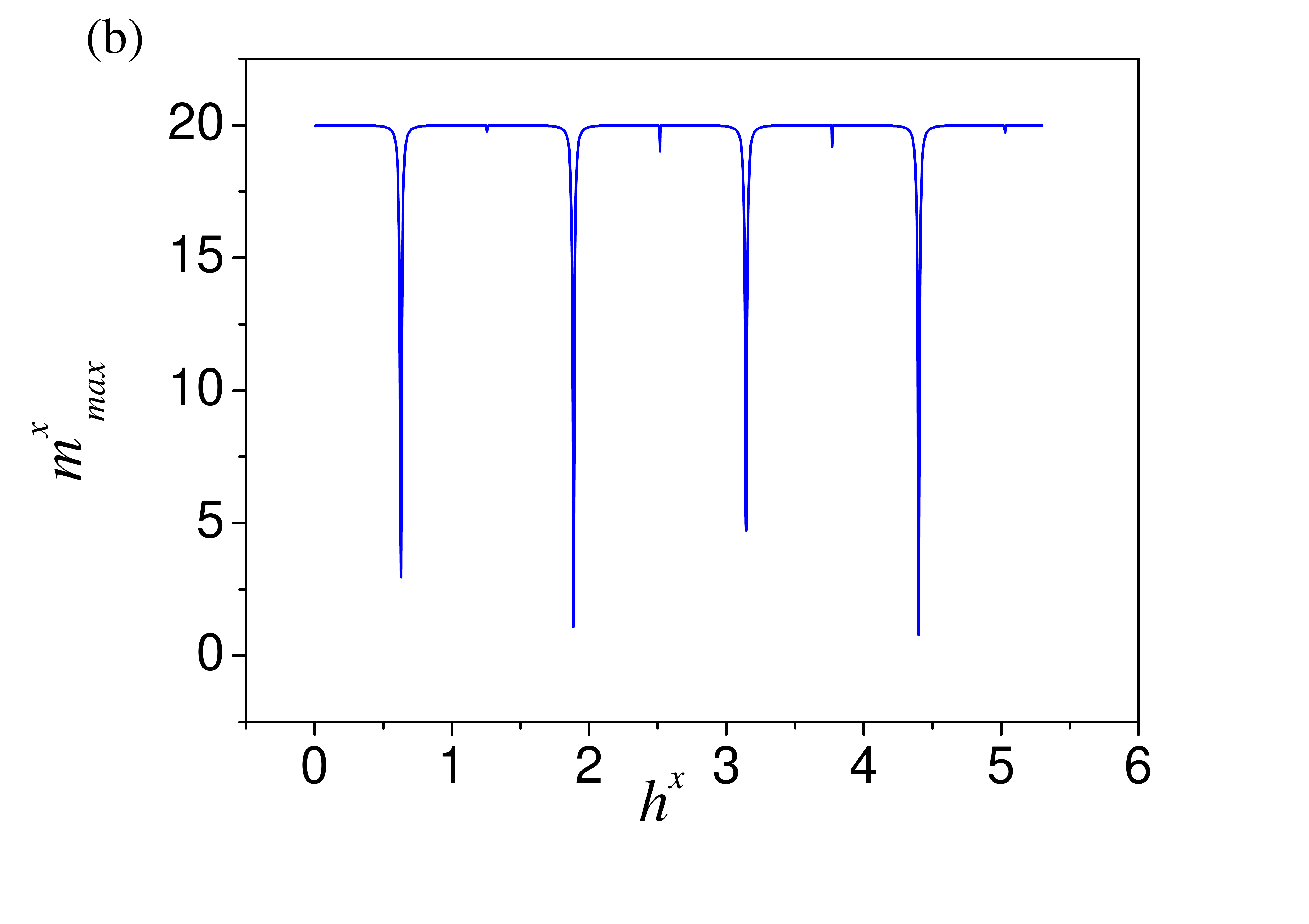}
\end{center}
\caption{Plots of the maximum expectation value of $S^x$ versus $h^x$, for
$S=20, ~T=10, ~h^z =1$, and (a) $h_D^x = 40$ and (b) $h_D^x = 12.8 \pi \simeq 
40.212$. In figure (a) we see pronounced dips for $h^x$ equal to all integer 
multiples of $2\pi/T$, while in figure (b) we see pronounced dips only when 
$h^x$ is equal to odd integer multiples of $2\pi/T$, as predicted by the 
FDPT result, Eq.~\eqref{cmt10}.} 
\label{mmaxvshx1} 
\end{figure}

As a simple illustration of the FDPT, we discuss a system with a single 
spin governed by a time-dependent Hamiltonian. We briefly discuss some 
results obtained from the FDPT (which give the conditions for perfect
freezing and resonances), numerical results, and exact results for the 
Floquet operator. The details are presented in App.~\ref{app:FDPT}. \\

{\it Model:} We consider a single spin $\vec S$, with ${\vec S}^2 = S(S+1)$, 
which is governed by a Hamiltonian of the form
\beq H (t) ~=~ -~ h^x S^x ~-~ h^z S^z ~-~ h_D^x ~\Sgn (\sin (\om t)) ~S^x.
\label{ham4} \eeq
The time period is $T=2\pi/\om$. Since $\sin (\om t)$ is positive
for $0 < t < T/2$ and negative for $T/2 < t < T$, the Floquet operator is 
given by
\bea U &=& e^{(iT/2)~[(h^x - h_D^x) S^x ~+~ h^z S^z]} \non \\
&& \times ~e^{(iT/2)~[(h^x + h_D^x) S^x ~+~ h^z S^z]}. \label{u1} \eea
It is clear from the group properties of matrices of the form $e^{i{\vec a} 
\cdot {\vec S}}$, that $U$ in Eq.~\eqref{u1} must be of the same form and
can be written as
\bea U &=& e^{i\ga {\hat k} \cdot {\vec S}}, \non \\
{\rm where} ~~~ {\hat k} &=& (\cos \ta, \sin \ta \cos \phi, \sin \ta \sin 
\phi). \label{u2} \eea
We work in the basis in which $S^x$ is diagonal. Since the eigenstates of 
$U$ in Eq.~\eqref{u2} are the same as the eigenstates of the matrix $M = 
{\hat k} \cdot {\vec S},$ the expectation values of $S^x$ in the different
eigenstates take the values $\cos \ta$ times $S, S-1, \cdots, -S$.
The maximum expectation value is given by $m^x_{max} = S \cos \ta$. \\

{\it Analytical results from FDPT:}
We can use the FDPT to derive the correction to $m^x_{max}$ to first 
order in the small parameter $h^z / h^x_D$. Namely, we find how the state 
given by $|0 \ra \equiv |S^x = S\ra$ mixes with the state $|1 \ra \equiv | S^x 
= S-1 \ra$. We discover that 
\beq c_1 (0) ~=~ \frac{\sqrt{2S} ~h^z}{h_D^x} ~\frac{e^{ih^x T/2} ~
[e^{ih_D^x T/2} - \cos (h^x T/2)]}{e^{ih^x T} ~-~ 1}, \label{cmt10} \eeq
	Three possibilities arise at this stage. \\
(i) The denominator of Eq.~\eqref{cmt10} is not zero. Then the expectation 
value of $S^x$ in this state will be close to $S$ since $h^z/h_D^x$ is small. 
In addition, if the numerator of Eq.~\eqref{cmt10} vanishes, we get perfect
freezing, namely, $\langle S^x \rangle = S$. \\
(ii) The denominator of Eq.~\eqref{cmt10} vanishes, i.e., $h^x$ is an 
integer multiple of $2\pi/T$, but the numerator does not vanish.
This is called the resonance condition. Clearly, the perturbative result
for $c_1 (0)$ breaks down in this case, and we have to either develop a 
degenerate perturbation theory or do an exact calculation. \\
(iii) Both the numerator and the denominator of Eq.~\eqref{cmt10} vanish.
Once again the perturbative result breaks down and we have to do a more 
careful calculation.

We would like to make a comment on the dependence of the result 
in Eq.~\eqref{cmt10} on the value of $S$. At $t=0$, the probability of state 
$| 1 \ra$ is $|c_1(0)|^2$ and the probability of state $| 0 \ra$ is 
$1- |c_1(0)|^2$. Hence the expectation value of $S^x/S$ is given by 
\bea && \frac{m^x_{max}}{S} ~=~ \frac{1}{S} ~\Bigl[ S ~(1 ~-~ |c_1(0)|^2) ~+~ 
(S ~-~ 1) ~ |c_1 (0)|^2 \Bigr] \non \\
&& = 1 ~-~ 2 \left(\frac{h^z}{h_D^x} \right)^2\times ~~~~~~~~~~~~~~~~~~~~
~~~~~~~~~~~~~~~~ \non \\
&& \frac{1 ~+~ \cos^2{(h^x T/2)} ~-~ 2 \cos{(h^x T/2)}\cos{(h_D^x T/2)}}{4 
\sin^2{(h^x T/2)}}. \label{mx5} \eea
We expect Eq.~\eqref{cmt10} to break down at a sufficiently large value of $S$
since it was derived using first-order perturbation theory which is accurate
only if $|c_1(0)| \ll 1$. However, we observe that the value of $m^x_{max}/S$ 
in Eq.~\eqref{mx5} is independent of $S$.
We therefore have the striking result in this model that we
can use first-order perturbation theory for values of $S$ which are not large
to derive an expression like Eq.~\eqref{mx5} which is then found to hold for 
arbitrarily large values of $S$. \\

{\it Numerical results:} Given the values of 
the parameters $S, ~T, ~h^x, ~h^z$ and $h_D^x$, we can
numerically compute $U$ and its eigenstates. From the eigenstates, we can
calculate $m^x_{max}$ which is the maximum value of the expectation value of
$\la S^x \ra$. In Fig.~\ref{mmaxvshx1}, we plot $m^x_{max}$ versus $h^x$, for 
$S=20, ~T=10, ~h^z =1$, and (a) $h_D^x =40$ and (b) $h_D^x = 12.8 \pi \simeq 
40.212$. In Fig.~\ref{mmaxvshx1} (a), we see large dips for $h^x$ equal to 
{\it all} integer multiples of $2\pi/T$. In Fig.~\ref{mmaxvshx1} (b), we see 
large dips for $h^x$ equal to {\it odd} integer multiples of $2\pi/T$, but 
the dips are much smaller for $h^x$ equal to {\it even} integer multiples of 
$2\pi/T$. 

We can understand these results using the FDPT. In
Fig.~\ref{mmaxvshx1} (a), we have $h_D^x = 40$; hence $\cos (h_D^x T/2) \ne 
\pm 1$, and the numerator of Eq.~\eqref{cmt10} can never vanish. We therefore
obtain large dips for $h^x$ equal to all integer multiples of $2\pi/T$ where
the denominator of Eq.~\eqref{cmt10} vanishes (case (ii)). However, in 
Fig.~\ref{mmaxvshx1} (b), $h_D^x = 12.8 \pi$ so that $\cos (h_D^x T/2) = 1$. 
Hence both the numerator and denominator of Eq.~\eqref{cmt10} vanish when 
$h^x$ is equal to even integer multiples of $2\pi/T$ (case (iii)). This 
explains why the dips in $m^x_{max}$
are much smaller for $h^x$ equal to even integer multiples of $2\pi/T$, but
they continue to be large for $h^x$ equal to odd integer multiples of $2\pi/T$. \\

{\it Form of the Floquet operator in different cases:} We now present 
expressions for the Floquet operator $U$ in Eq.~\eqref{u2} based on the exact 
results derived in App.~\ref{app:singlespin}. The purpose of this exercise
is to show that the form of $U$ is quite different in cases (i-iii).

Assuming that $h_D^x$ is positive and much larger than $|h^x|$ and $|h^z|$, 
we find, to zero-th order in $h^z / h^x_D$, that
\beq \cos \left( \frac{\ga}{2} \right) ~=~ \cos \left( \frac{h^x T}{2} \right),
~~~~{\rm and} ~~~~ {\hat k} ~=~ {\hat x}, \label{gak1} \eeq
provided that $e^{ih^x T} \ne 1$ (case (i)). Eq.~\eqref{gak1} implies that 
the Floquet operator corresponds to a rotation about the $\hat x$ axis by 
an angle $\ga$.

If $e^{ih^x T} = 1$, i.e., $\cos (h^x T/2) = \pm 1$, but $\cos (h^x T/2) \ne 
e^{ih_D^x T/2}$, the denominator of Eq.~\eqref{cmt10} vanishes but the 
numerator does not (case (ii), called the resonance condition). It turns out 
that we then have to expand up to second order in $h^z / h^x_D$. This gives
\bea {\hat k} &=& \cos \left( \frac{h_D^x T}{4} \right) ~{\hat z} ~-~ \sin
\left( \frac{h_D^x T}{4} \right) ~{\hat y} \non \\
&& {\rm if} ~~ \cos \left( \frac{h^x T}{2} \right) ~=~ 1, \non \\
&=& \sin \left( \frac{h_D^x T}{4} \right) ~{\hat z} ~+~ \cos \left( \frac{h_D^x
T}{4} \right) ~{\hat y} \non \\
&& {\rm if} ~~ \cos \left( \frac{h^x T}{2} \right) ~=~ - 1. \label{gak2} \eea
This implies that the Floquet operator corresponds
to a rotation about an axis lying in the $y-z$ plane. This implies that
the expectation value of $S^x$ will be zero in all the eigenstates of the
Floquet operator. 

Finally, if $e^{ih^x T} = 1$ and $\cos (h^x T/2) = e^{ih_D^x T/2}$, both the 
numerator and denominator of Eq.~\eqref{cmt10} vanish (case (iii)). We then 
discover that
\beq {\hat k} ~=~ \frac{h^x ~{\hat x} ~-~ h^z ~{\hat z}}{\sqrt{(h^z)^2 ~+~
(h^x)^2}}. \label{gak3} \eeq
Hence, the Floquet operator corresponds to a rotation about an axis lying in 
the $x-z$ plane.
 
{To summarize, assuming that $h^z/h_D^x$ is small, we obtain quite 
different results depending on which of the three cases (i-iii) arise. We see 
these differences both in the numerical results for $m^x_{max}$ shown in 
Fig.~\ref{mmaxvshx1} and in the forms of the Floquet operator in 
Eqs.~(\ref{gak1}-\ref{gak3}) which are obtained by an exact calculation.}


\begin{figure*}[ht]
\begin{center}
\includegraphics[width=0.32\linewidth]{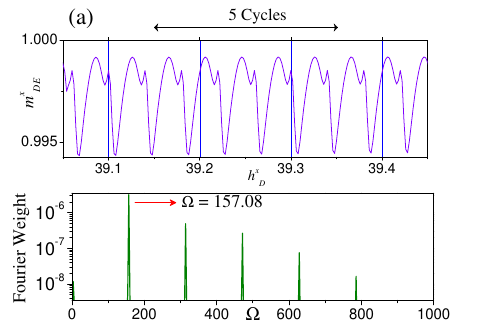}
\includegraphics[width=0.32\linewidth]{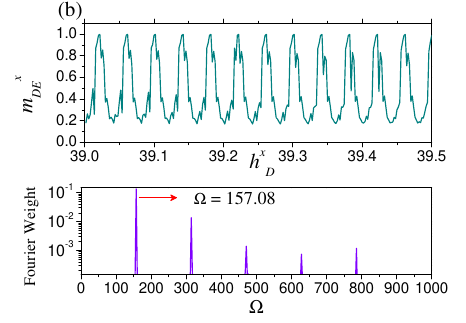}
\end{center}
\caption{Periodicity in drive strength, $h_{D}^{x}$, of the magnetization 
response (diagonal ensemble average $m^{x}_{DE}$, Eq.~\eqref{DEA_formal}). 
Top row shows periodicity for both off-resonance (left, $h^{x}_{0}=-0.2$)
and on-resonance (right, $h^{x}_{0}=-0.21$) drives. Other parameters, and 
initial low-temperature state, as in Fig.~\ref{EE} (b). In both cases the 
leading frequency of oscillations is $\Omega \approx 157.08 \approx 2\pi /\om$,
visible in the bottom panel, as predicted by Eq.~\eqref{mx2}. The other 
parameters are the same as in Fig.~\ref{Peak_Valley_Steps} (a).} 
\label{Oscillations} 
\end{figure*}
\subsection{FDPT for the Interacting Ising Chain}

Now we apply FDPT to our interacting Ising chain (Eq.~\eqref{ham3}) studied 
numerically above. We set $h^z \ll h_D^x,$ and treat 
$V$ as the perturbation. We use periodic boundary conditions.

The eigenstates $|n\ra$ of $H_0 (t)$ are diagonal in the basis of the 
operators $\si_n^x$. In particular, the state in which all spins $\si_n^x 
= +1$, will be denoted as $| 0 \ra,$ and we start by calculating the Floquet 
state 
$|m^{x}_{max}\ra$ (maximally polarized Floquet state) obtained by perturbing 
this state to first-order in $h^{z}/h_{D}^{x}.$ While calculating $m^x$ from 
the perturbation theory we use this Floquet state. 

The rationale for this is as follows. First,
if we start with a fully polarized state in the $+x$ direction 
(as is done, for example, in the experiments by Monroe~\cite{Monroe}), 
or, with the ground state of $H(0),$ with 
$h_{D}^{x} \gg h^{z},\kappa$, then the initial state is expected to have 
a strong overlap with this particular Floquet state. Hence at very long 
times, the expectation values of the observables in the wave function will 
be well approximated by the expectation value in this Floquet state. 

Secondly, in this setting, the insights from the single-spin problem studied 
above are most directly transferable; in particular, we again encounter the 
ideas of resonances and scars. With these in hand, we can then identify a 
number of features present in the data more generally, in particular for 
high-temperature states (which are of interest in the context of the 
NMR experiments by Rovny~\cite{Rovny}).
We find that the perturbation theory works best in the vicinity of the scars 
with their emergent integrability (see below), and present a limited 
exploration of the performance of FDPT away from these in App.~\ref{app:FDPT}. 

For the expansion of the Floquet state to leading order, the computation 
proceeds entirely along the lines of that presented for the single spin model. 
We denote the state in which all spins $\si_n^x = +1$ except for the site $m$ 
where $\si_m^x = -1$ as $|m \ra$.
In the limit in which $h_D^x$ is much larger than $J, ~\ka$ and $h_0^x$, we 
find that, to leading order in $h^{z}/h_{D}^{x}$, Eq.~\eqref{cmt5} takes the 
form
\bea c_m (0) &\simeq& \frac{h^z}{h_D^x} ~\frac{e^{iA T/2} ~[e^{ih_D^x T} ~-~ 
\cos (A T/2)]}{e^{iA T} ~-~ 1}, \non \\
A &=& 4 (J ~-~ \ka) ~+~ 2 h_0^x. \label{cmt6} \eea

The magnetization of this maximally polarized Floquet state is given as follows.
The expectation value of $\sum_{n=1}^L \si_n^x$ in each of the $m$ states
is $L-2$ and in the state $|0\ra$ is $L$. This gives
\beq m^x = 
1 ~-~ \frac{2}{L} ~\sum_{m=1}^L ~|c_m (0)|^2. \label{mx1} \eeq


\subsubsection{Resonances and stability of the scar}
\label{Reso_Scar}

The resonance condition, Eq.~\eqref{res1},\eqref{cmt6},
\beq e^{iA T} ~=~ 1 ~~{\rm where}~~ A ~=~ 4(J ~-~ \ka) ~+~ 2 h_{0}^{x} \ ,
\label{Res1} \eeq
signals the singularities of our expansion, where $c_m (0)$ diverges.
For our Hamiltonian this occurs for 
\beq h_0^x ~=~ -2J ~+~ 2\ka ~+~ \frac{p\om}{2}. \label{h0x} \eeq
\noindent
Here $p$ is an integer which corresponds to the number of photons absorbed or 
emitted in this transition. Thus, our first-order theory does not preclude 
multi-photon transitions.

This suggests considering all possible first-order resonances based on 
Eq.~\eqref{res1}, by considering the resonance condition more generally: 
evaluating the change $E_{m} - E_{n}$ due to the flip of only a single spin, 
$\si_{0}$, with $n$-th nearest-neighbor spins on the right/left denoted by 
$\si_{\pm n}$ yields the first-order resonance condition 
\beq h^{x}_{0}\si_{0} ~+~ J\si_{0}(\si_{-1} + \si_{1}) ~-~ \kappa\si_{0}(
\si_{-2} + \si_{2}) ~=~ \frac{p\om}{2}. \label{Gen_Res} \eeq 
Of course, individual resonances may be absent if there are no matrix 
elements between the states in question. 

This approach can be rather successful at identifying the locations of the 
numerically observed isolated resonances, as displayed in 
Fig.~\ref{Resonances}. There, the strength of the freezing is displayed
as a function of driving strength, for both slow and very slow drives, 
$\om=0.4, 0.04$, respectively.

The right panel of Fig.~\ref{Resonances} emphasizes the generality of this 
result: the considerations of the first-order resonances obtained
above yield the response even for the initially weakly-polarized ($m^x=0.05$) high-temperature initial state.

For a many-body Floquet system, a proliferation of Floquet resonances may lead 
to unbounded heating. Hence a stable non-thermal state (e.g., a scar) 
a priori requires the absence of 
resonances. Eq.~(\ref{Gen_Res}) shows this is 
straightforwardly possible to first order, since the resonances are isolated 
and can be well separated in parameter space. This stems from the fact that
the gap $E_{m} - E_{n}$ between two distinct (possibly degenerate) levels 
of $H_{0}^{x}$ (Eq.~\eqref{ham3}) does not necessarily vanish even in the 
thermodynamic limit (for example, if we take all the couplings in $H_{0}^{x}$ 
to be rational numbers). The absence of any signature of the higher order 
resonances in the exact numerical 
result at very low frequencies ($\omega = 0.04$) and large $h_{D}^{x} = 40$
in the neighborhoods of the scar points indicates that the first-order theory 
is sufficient there, and the FDPT-series is as least asymptotic in nature.\\

\XX{
Higher order resonances start
gaining importance as $h_{D}^{x}$ is reduced below a freezing cutoff ($h_{D}^{x}\approx 18$) as shown in Fig.~\ref{Peak_Valley_Steps}(b). The
choice of parameters rules out first order resonances in this case. The results are consistent with this when the drive amplitude is above  the cutoff -- we see no resonant dip in $m^{x}_{DE}.$ But as $h_{D}^{x}$ is tuned below the cutoff, rapid irregular fluctuations appear due to sharp resonant dips 
in $m^{x}_{DE}.$ The first order resonances being ruled out, these dips are due to higher order resonances. This 
implies the first order perturbation theory
is insufficient below the cutoff. With further lowering
of $h^{D}_{x}$ a sharp drop to the Floquet thermalized regime $m^{x}_{DE} \sim 0$ eventually appears below a threshold ($h_{D}^{x} \approx 5$).
}
\\

\begin{figure*}[htb]
\includegraphics[width=0.42\linewidth]{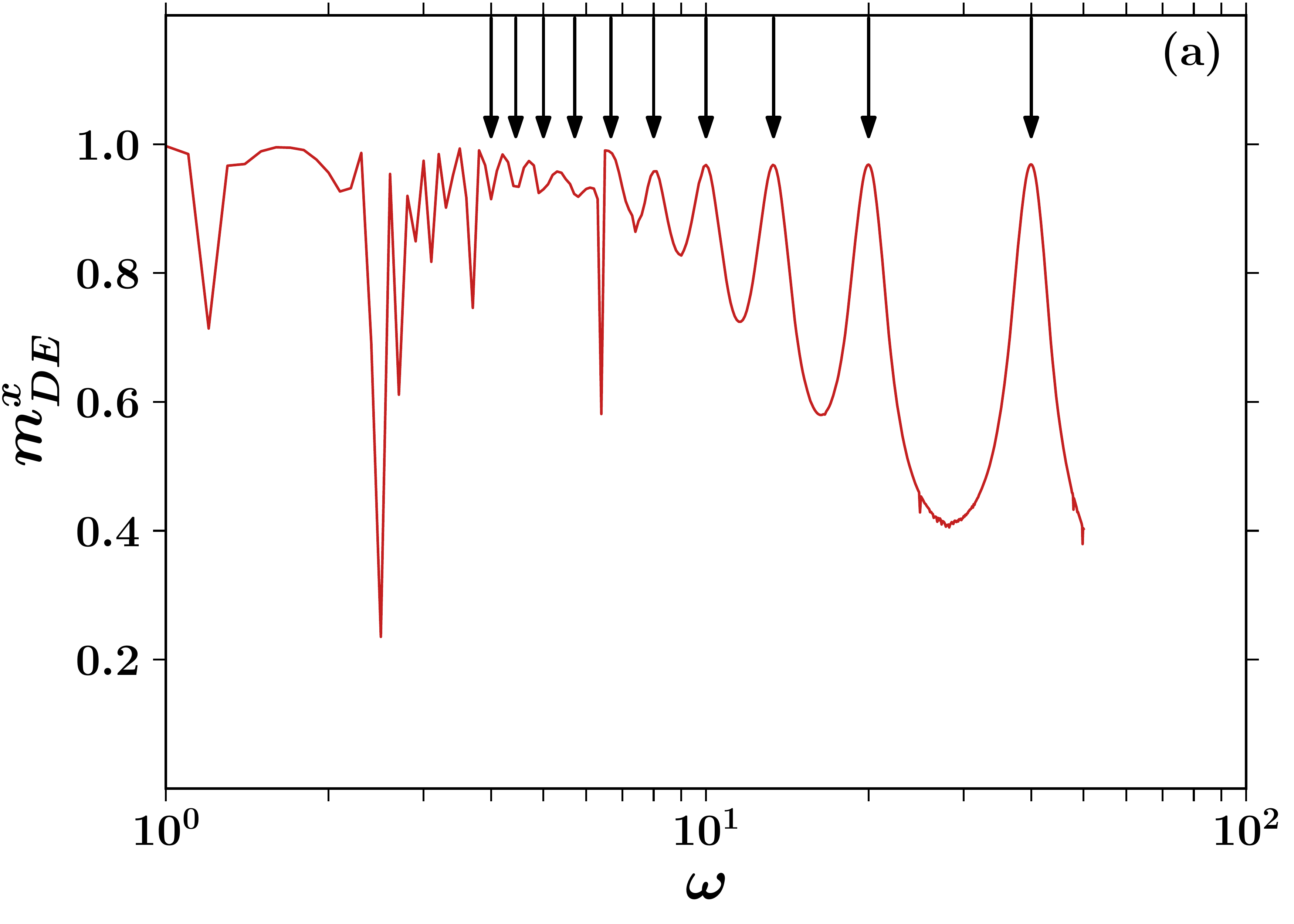}
\includegraphics[width=0.42\linewidth]{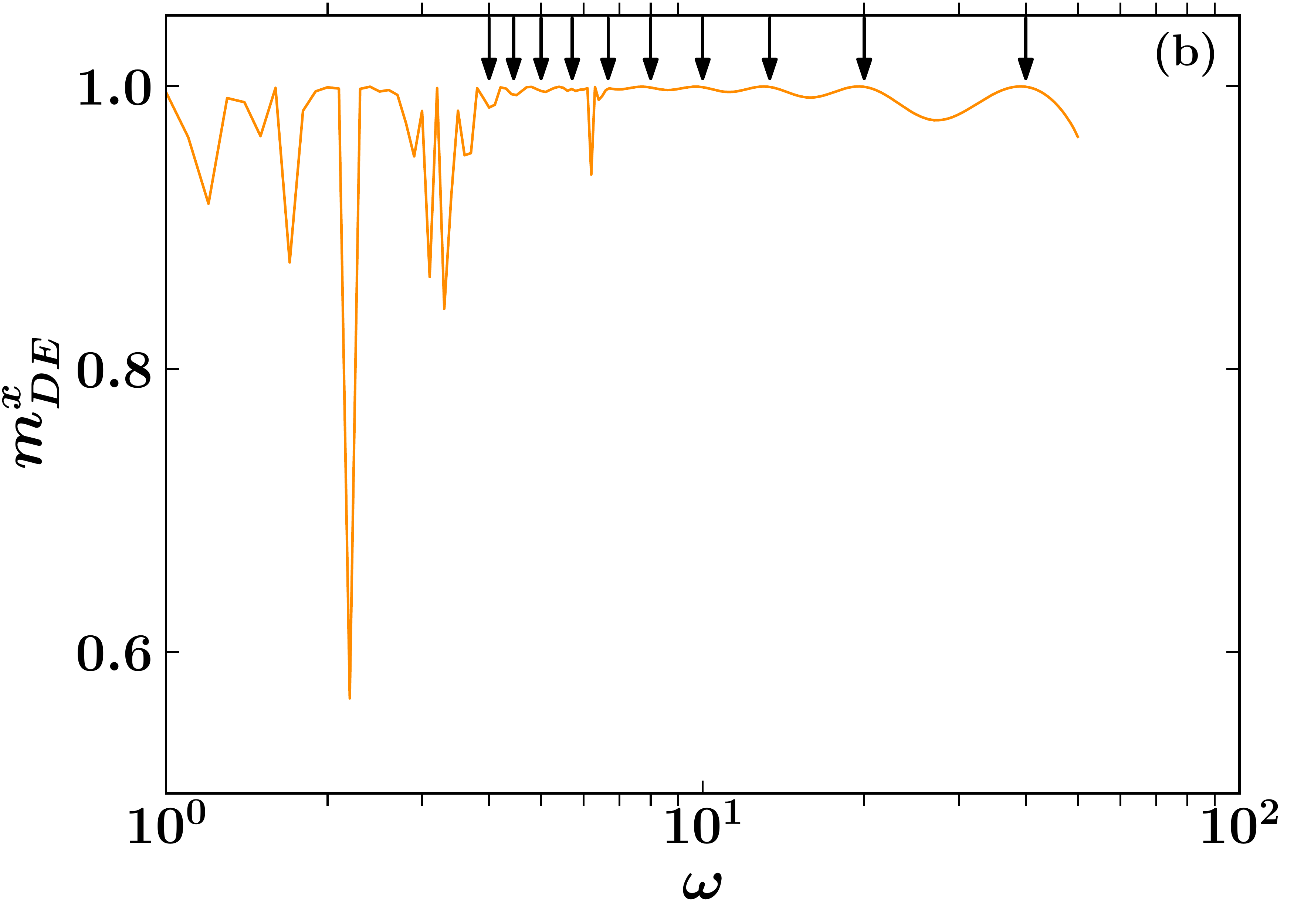}
\caption{$m^x_{DE}$, the infinite-time limit of the magnetization versus 
driving frequency, $\omega$, for Hamiltonians of the form in Eq.~(\ref{ham3}), 
but with different types of $H^{x}_{0}$, namely, three-spin interactions (Left 
panel: $H^{x}_{0} = H_{0}^{x(3Spin)}$, Eq.~\eqref{Gen:H0_3Spn}) and long-range 
interactions (Right panel: $H_{0}^{x} = H_{0}^{x(LR)}$, Eq.~\eqref{Gen:H0_LR}).
The strong freezing of $m^{x}_{DE}$ near the freezing points 
($\omega = h^{x}_{D}/k$) is observed in agreement with the prediction of the 
Magnus expansion up to the two leading orders in both cases. Interestingly, 
small corrections due to the higher order terms are also observed, namely, 
small deviations of the peak heights from unity for the three-spin case, and 
tiny shifts of the peak from the freezing condition for the long-range case. 
Apart from these, the higher order terms do not appear to change any of the 
key aspects of the phenomenon (strong emergent conservation of $m^x$ at all 
times and consequent lack of unbounded heating). Here the parameters are 
$J_{xxx} = 0.5, J = 1, \kappa = 0.7\pi/3, h^x_{0} = e/10, h^z = 1.2, 
h_{D}^{x} = 40, L = 20.$} \label{Gen:Fig:3Spn_LR}
\end{figure*}

\noindent {\it Scar from FDPT}: Considering the expression for the 
magnetization, obtained by substituting the expression for $c_{m}(0)$ 
(Eq.~\eqref{cmt6}) into the expression of $m^x$ (Eq.~\eqref{mx1}),
\bea 1 ~-~ m^x ~=~ 2 \left(\frac{h^z}{h_D^x} \right)^{2}\times 
~~~~~~~~~~~~~~~~~~~~~~~~~~~~~~~~~ && \non \\
\frac{1 ~+~ \cos^2 (AT/2) ~-~ 2 \cos (AT/2) \cos (h_D^x T)}{4 \sin^2 
(AT/2)}, ~~ &&
\label{mx2} \eea
we would like to make the following observations. 

First, Eq.~\eqref{mx2} indicates that $m^x$ should keep oscillating with 
$h_D^x$ with a period $\om$ (except when $\cos (AT/2)$ is close to zero), as 
is indeed observed in Fig.~\ref{Oscillations}. Notice, therefore, that the 
`high-field limit' is not entirely simple but is still endowed with a 
fine-structured periodicity. 

Secondly, when $\om = 2\pi/T$ is large, we can approximate $\cos (AT/2) 
\simeq 1 - (AT)^2/8$ and $\sin (AT/2) \simeq AT/2$ in Eq.~\eqref{mx2}: 
\beq 1 ~-~ m^x ~=~ 2 ~\left(\frac{h^z}{h_D^x} \right)^2 ~\frac{4 (1 ~-~ 
A^2 T^2/8)~ \sin^2 (h_D^x T/2)}{A^2 T^2}. \label{mx3} \eeq
This shows that freezing becomes weaker with increasing $\om$. An exception to 
this occurs when the numerator in Eq.~\eqref{mx3} vanishes, namely, when 
$\om = h_D^x/k$, where $k$ is an integer. At these points, we have 
$m^{x}/m^{x}(0) = 1$, i.e., perfect freezing. Those are precisely the `scar' 
points given by Eq.~\eqref{scar:def}, where the peaks of freezing are obtained 
numerically (Fig.~\ref{Peak_Valley_Steps}).

As encountered in the single spin model, there is an interesting interplay 
between the scars -- where $m^x$ is frozen -- and the resonances, 
where heating is hugely amplified. When the two coincide, this can destroy the 
inertness of the scar point. This is manifested as sharp dips in $m^{x}_{DE}$ 
in the numerical results discussed above, and for intermediate values of 
$h_{D}^{x}$ in the inset of Fig.~\ref{Peak_Valley_Steps}~(a). The FDPT 
predicts isolated resonances in parameter space and provides a guide for 
choosing the Hamiltonian parameters to avoid 
resonances and observe stable scars. Our choice of parameters for 
Fig.~\ref{Peak_Valley_Steps} is guided by the theory (Eq.~\eqref{Gen_Res}), 
and we indeed observe resonance-free strong freezing at the scar points. 

It would clearly be desirable to embark on a more detailed study, both with 
respect to the role of higher-order resonances (visible in the left panel of 
Fig.~\ref{Resonances}), and with regard to the statistics of the resonances 
as the system size increases.

\section{Robustness and Generality of the Scarring and Emergent Conservation}
\label{Sec:Gen}

\begin{figure*}[t]
\includegraphics[width=0.42\linewidth]{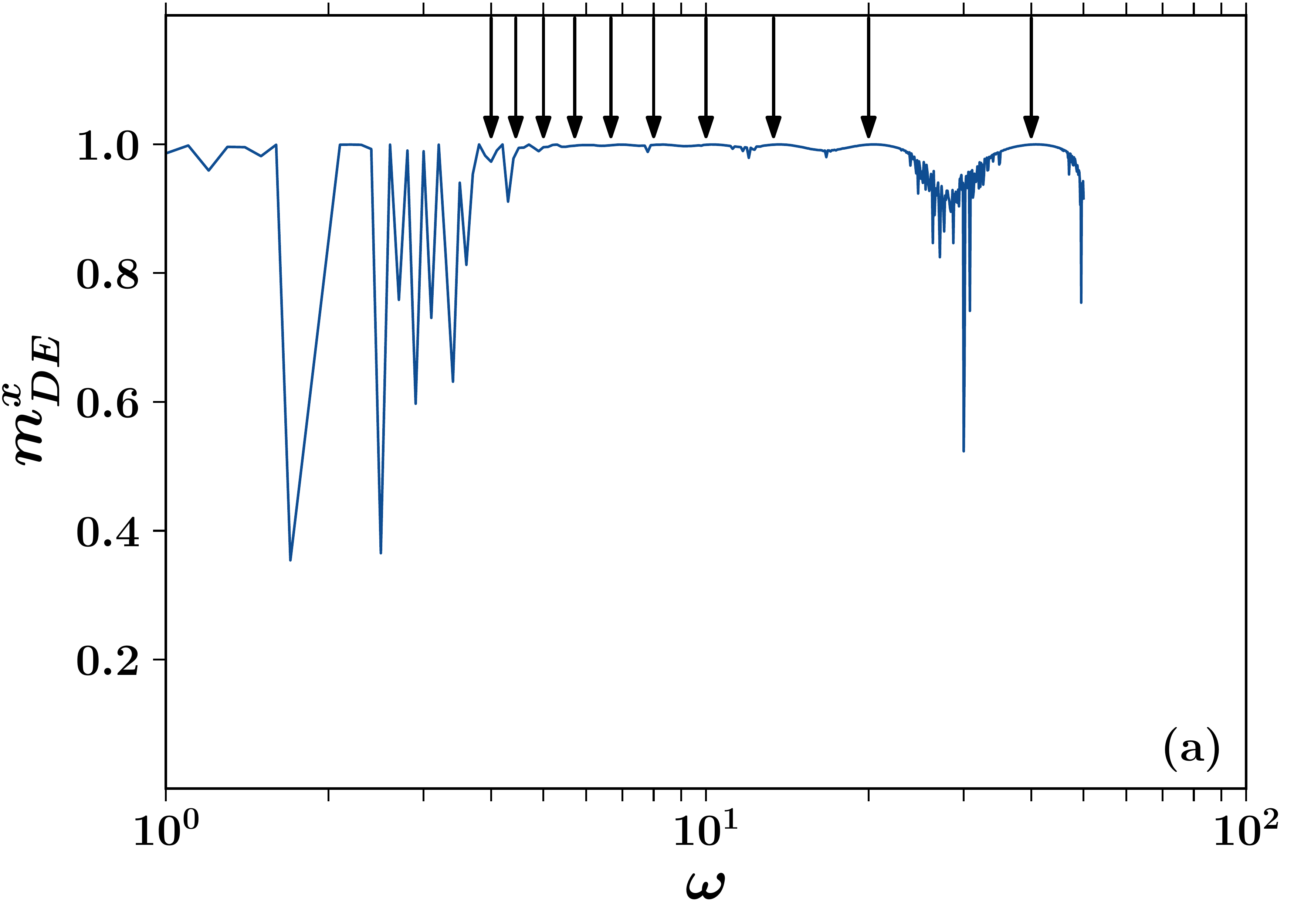}
\includegraphics[width=0.42\linewidth]{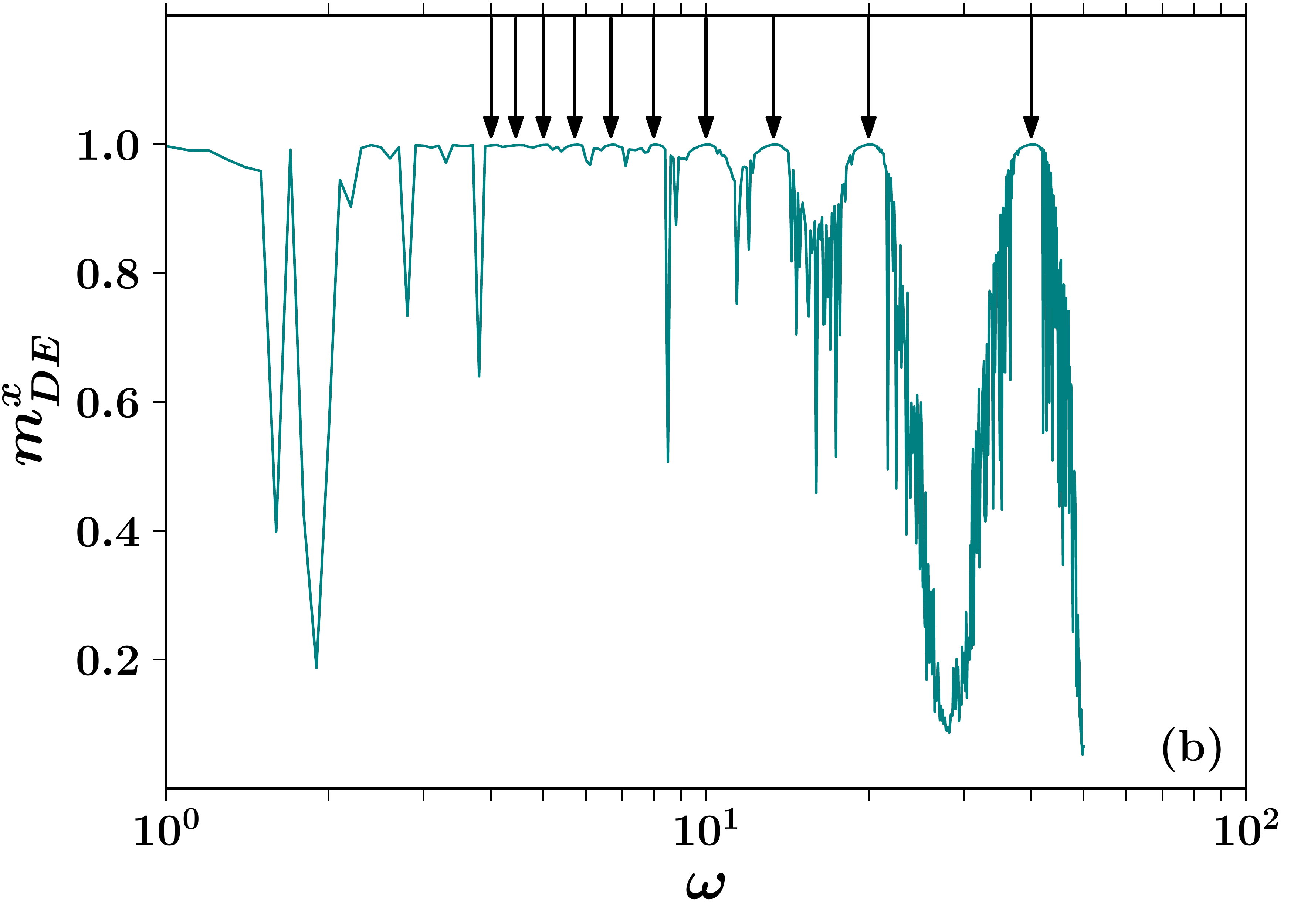}
\caption{$m^x_{DE}$, the infinite-time limit of the magnetization versus 
driving frequency, $\omega$, for Heisenberg interactions. Left panel: 
For the Hamiltonian in (Eq.~\eqref{ham3}), with $V$ replaced by $V^{HB}$ 
(Eq.~\eqref{V_Heisenberg}) and $H_{0}^{x}$ by $H_{0}^{x(HB)}$ 
(Eq.~\eqref{H0x_Heisenberg}), with translationally invariant isotropic 
interaction strengths ($J_{i}^{x} = J_{i}^{y} = J_{i}^{z} = 1.0$). Right 
panel: Same model as left panel, but with anisotropic interaction strengths 
$J_{i}^{x} = 1.0, J_{i}^{y} = 0.5, J_{i}^{z} = 0.6$. In both cases, the 
emergence conservation of $m^x$ and concomitant absence of thermalization are 
clearly visible. Slight shifts of the peaks from the predicted values are 
observed as in Fig.~\ref{Gen:Fig:3Spn_LR}. Data shown for $J = 1, \kappa = 
0.7\pi/3, h^x_{0} = e/10, h^z = 1.2, h_{D}^{x} = 40, L=20.$} \label{Gen:Fig:yz}
\end{figure*}

In this section we demonstrate the robustness and generality of the phenomenon
of emergent conservation and consequent absence of thermalization,
by comparing the diagonal ensemble average, $m^{x}_{DE}$, versus the
driving frequency $\omega$ for a range of qualitatively distinct models.
We also demonstrate the stability of the conservation law at the scar/freezing points upon
increasing the system size. In all cases, the drive strength is set to be
$h_{D}^{x} = 40,$ and the freezing peaks/scar points are thus expected to occur for $\omega = 40/k,$
where $k$ is an integer. This condition was derived for all Ising interactions in Sec.~\ref{sec:Magnus},
and will be derived for general two-body Heisenberg interactions in Sec.~\ref{subsec:HB}.

\subsection{Additional forms of Ising Interactions}
\label{subsec:Ising}

First, we confirm, as predicted by the moving-frame Magnus expansion in 
Sec.~\ref{sec:Magnus}, the robustness of the phenomenon under diverse 
variations of the form of $H_{0}^{x}$ in the total Hamiltonian partitioned 
in the form of Eq.~\eqref{ham3}. We recall that $H_{0}^{x}$ is the portion
of the static part of the Hamiltonian that commutes with $H_{D}$, and the 
nature of the whole static part can be tuned over a wide variety of many-body 
Hamiltonians depending on the form of $H_{0}^{x},$ ranging from 
non-interacting to interacting, integrable to non-integrable, low to high 
dimensional. We consider two forms for $H_{0}^{x}$. First we add a three-body 
interaction, of strength $J_{xxx}$,
\bea H_{0}^{x(3Spin)} &=& -J\sum_{i}\sigma_{i}^{x}\sigma_{i+1}^{x} 
+\kappa\sum_{i}\sigma_{i}^{x}\sigma_{i+2}^{x} \non \\ 
&~& + J_{xxx}\sum_{i}\sigma_{i}^{x}\sigma_{i+1}^{x}\sigma_{i+2}^{x}-h_{0}^{x}\sum_{i}^{L}\sigma_{i}^{x}. \label{Gen:H0_3Spn} \eea
\noindent 
Secondly, we consider long-range interactions as follows. Spins are placed 
equidistantly on a circle, and the distance $r_{ij}$ between the $i$-th and 
the $j$-th spin is measured along the chord connecting them, such that
\beq H_{0}^{x(LR)} = -J\sum_{ij}\frac{\sigma_{i}^{x}\sigma_{j}^{x}}{r_{ij}} -
h_{0}^{x}\sum_{i}^{L}\sigma_{i}^{x}. \label{Gen:H0_LR} \eeq
\noindent The increased effective coordination number is intended to mimic the 
phenomenology of higher-dimensional models.
The results are given in Fig.~\ref{Gen:Fig:3Spn_LR}. \\

\subsection{General Heisenberg Interactions}
\label{subsec:HB}

We consider the case where the static part of the Hamiltonian $H(t)$ 
(Eq.~\eqref{ham3}) consists not only of a simple transverse field but also 
includes a general Heisenberg interactions with arbitrary position dependence. 
The Heisenberg terms involving $\sigma_{i}^{y,z}$ are included in the $V$ term,
and those involving $\sigma_{i}^{x}$ are included in the $H^{0}_{x}$ term as 
follows. The total Hamiltonian $H(t) = H^{HB}(t)$ in this case has the same 
form in Eq.~\eqref{ham3}, but with with $V$ replaced by
\beq V^{HB} = -\sum_{i,j}\Jyij\si_i^y\si_{j}^y -\sum_{i,j}\Jzij\si_i^z
\si_{j}^z -h^{z}\sum_{i}\sigma^{z}, \label{V_Heisenberg} \eeq
\noindent and $H_{0}^{x}$ replaced by
\beq H^{x(HB)}_{0} = -\sum_{i,j}\Jxij\si_i^x \si_{j}^{x} + \ka \sum \si_i^x 
\si_{i+2}^x - h_0^x\sum \si_{i}^{x}. \label{H0x_Heisenberg} \eeq
The total static Hamiltonian $V + H^{x}_{0}$ can thus have a general 
Heisenberg term with arbitrary interaction graph (coordination number, 
spatial dimensionality and position dependence).

For the changed form of $V$, the moving frame Magnus expansion requires some 
additional lengthy steps (see Appendix~\ref{app:Heisenberg}), but 
eventually leads to the same conclusion as derived in Sec.~\ref{sec:Magnus}, 
namely, $m^x$ is exactly conserved in the first two orders of the expansion. \\

Interestingly, the first term (zeroth order in $1/\omega$) 
exhibits an attractive route to the emergent conserved quantity: all the terms in addition to $H_{0}^{x}$ 
do not in fact vanish, but their sum explicitly exhibits an ${\rm U(1)}$ symmetry present neither in $H(t),$
nor in $H^{mov}(t)$. This assures conservation of $m^x$ in the first order.
In the next order (first order in $1/\omega$), all the terms except $H^{x}_{0}$ vanish. In the
following we summarize the results, relegating the detailed calculation to the Appendix. \\

For the total Hamiltonian $H^{HB}(t),$ employing the unitary transformation induced 
by $W(t)$ (Eq.~\eqref{Urot}), we switch to the moving frame, 
in which our total Hamiltonian reads
\begin{widetext}
\bea H^{mov}_{_{HB}}(t) &=& H_0^x ~-~ \sum_{i,j} \Jyij \si_i^y \si_{j}^y
\left[\mathbb{I}\cos^2 (2 \theta) - \si_i^x \si_{j}^x \sin^2 (2\theta) 
+ \frac{i}{2} \sin (4 \theta) (\si_i^x +\si^x_{j}) \right] \non\\
&&- ~\sum_{i,j} \Jzij \si_i^z \si_{j}^z \left[ \mathbb{I}\cos^2 (2 \theta)
- \si_i^x \si_{j}^x \sin^2 (2\theta) + \frac{i}{2} \sin (4 \theta) 
(\si_i^x + \si^x_{j}) \right] \non\\
&&- ~h^z \cos (2 \theta) \sum_i \si_i^z ~+ ~h^z \sin (2 \theta) \sum_i \si_i^y.
\label{Gen:H_rot_C1C2} \eea
\end{widetext}
\noindent
In the following, we state the results of the Magnus expansion of 
$H^{mov}_{_{HB}}(t)$. \\

\noindent
The first term (zeroth order in $1/\omega$) is the average Hamiltonian, given by
\bea && H_{eff}^{(0)} ~=~ \frac{1}{T}\int_{0}^{T}dt~H^{mov}_{_{HB}}(t) \non \\
&& = H_0^{x(HB)} ~-~ \frac{1}{2}\sum_{i,j}(\Jyij + \Jzij)\left[ \si_i^y 
\si_{j}^y + \si_i^z \si_{j}^z \right], \label{Gen:Heff_0} \eea
\noindent under the freezing condition $h^{x}_{D} T = 2\pi k$ (or, $h^{x}_{D} 
= k\omega$). This term, though non-trivial and non-zero, is visibly 
${\rm U(1)}$ symmetric and commutes with $m^x.$
\\

\noindent The second term ($1$st order in $1/\omega$), reads 
\beq H_{eff}^{(1)}=\frac{1}{2!(i)T}\int_{0}^{T}dt_{1}\int_{0}^{t_{1}}
dt_{2}\left[H^{mov}_{_{HB}}(t_{1}),H^{mov}_{_{HB}}(t_{2})\right]. 
\label{Gen:Heff_1_def} \eeq
\noindent
Using Eq.~\eqref{Gen:H_rot_C1C2}, calculating all the commutators, and 
performing the integrals (see App.~\ref{app:Heisenberg}), we finally get, 
under the freezing condition $h^{x}_{D} T = 2\pi k,$
\beq H_{eff}^{(1)} = 0. \eeq
\noindent This finally yields
\bea H_{eff} = H_0^{x(HB)} - \frac{1}{2}\sum_{i,j}(\Jyij + \Jzij) \left[ 
\si_i^y \si_{j}^y + \si_i^z \si_{j}^z \right], \label{Gen:Heff_HB} \eea
up to $\mathcal{O}(1/\omega^2)$. This implies strong conservation of $m^x$ for 
large $h_{D}^{x}$ at the freezing points. We numerically check two well-known 
special cases, namely the translationally invariant, isotropic and anisotropic 
Heisenberg chains (in the presence of all the other interactions considered
earlier). The clean, 
translationally invariant, non-integrable chains with short-range interactions, 
as employed here for the demonstration, are probably the easiest to heat up 
(hence hardest to freeze), enjoying no protection from localization of any 
sort or additional stability that could occur due to high coordination number.
The results are summarized in Figs.~\ref{Gen:Fig:yz} (a-b).

\subsection{$L$-Dependence of the freezing}
\label{subsec:Ldep}

\begin{figure}[htb]
\includegraphics[width=0.94\linewidth]{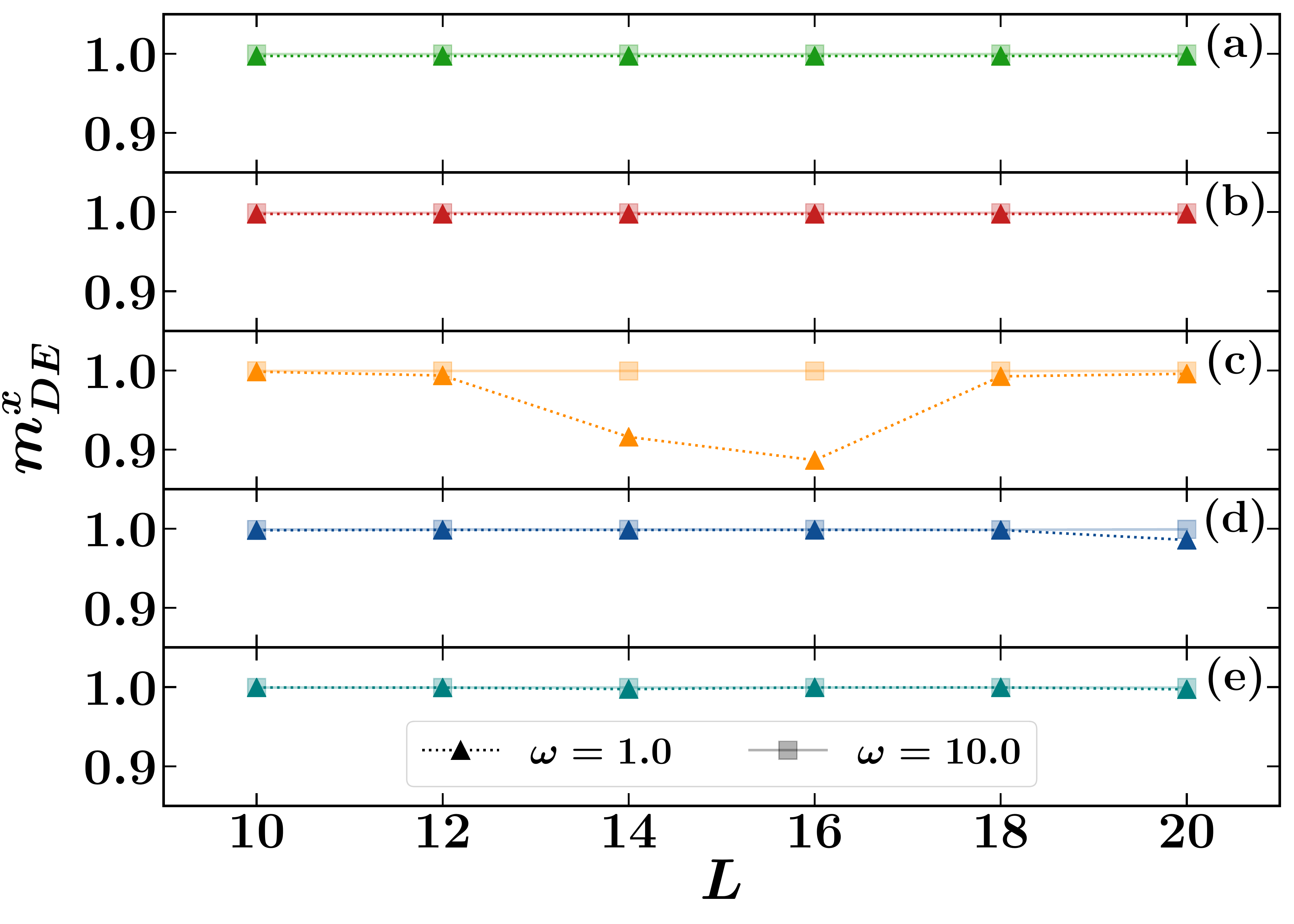}
\caption{The $L$-dependence of $m^{x}_{DE}$ at the freezing peaks 
corresponding to $h_{D}^{x}=40$ and for $\omega=1.0$ and $10.0$ are plotted 
against the system size $L$ for different variants of the drive Hamiltonian 
$H(t)$ (Eq.~\eqref{ham3}). Panels (a-c) show different variants of Ising 
interactions: (a): $H_{0}^{x}$ contains the nearest- and next-nearest-neighbor 
interactions and on-site fields (Figs.~\ref{Peak_Valley_Steps} (a), 
(b): $H_{0}^{x} = H_{0}^{x(3Spin)}$ (three-spin interactions; 
Fig.~\ref{Gen:Fig:3Spn_LR} (a)), (c): $H_{0}^{x} = H_{0}^{x(LR)}$ 
(additional long-range interactions; Fig.~\ref{Gen:Fig:3Spn_LR} (b). Panel 
(d-e) shows Heisenberg interactions: $H(t) = H^{HB}(t),$ results are 
shown for translationally invariant clean chains, (d): for the isotropic 
case (Fig.~\ref{Gen:Fig:yz} (a)), and (e) for the anisotropic case 
(Fig.~\ref{Gen:Fig:yz} (b)).} \label{Gen:Fig:L-dep} \end{figure}

The $L$-dependence of $m^{x}_{DE}$ at the scar points throws light on the 
stability of the emergent conservation with increasing system size. The 
conservation shows no perceptible degradation with increasing $L$, in agreement 
with the above analytical result for $\omega=10$. For $\omega=1,$ there is a 
non-monotonic behavior, but no systematic decline. These kinds of 
irregularities (see, e.g., Figs.~\ref{Gen:Fig:L-dep} (c-d)) often occur as 
"pathological" finite-size effects close to integrability~\cite{AD-RM}.

	With present-day numerical methods unable to access significantly larger system sizes,
our data can of course not rule out thermalization appearing beyond an -- as yet unknown -- 
much larger ``prethermal length scale". However, we emphasize that the results in Fig.~\ref{Gen:Fig:L-dep}
exhibit not even a discernible systematic {\it tendency} towards unfreezing as a precursor to thermalization 
as the system size is increased.

\noindent
{\it Resonances:} Finally, we note that our treatment of the resonances discussed in Sect.~\ref{Reso_Scar}
will essentially carry over to the expanded settings discussed in this section. In particular, the isolated
nature of first-order resonances discussed there, which underpins the stability of the conservation law,
remains intact, as follows. For the various forms of $H^{x}_{0}$, 
Eq.~\eqref{Gen_Res} will still equate a finite change in the eigenvalue
of $H^{x}_{0}$ due to a single spin flip with an integer multiple of $\omega/2.$ Since the single 
spin-flip energies remain finite as $L\to\infty$, 
the resonances are isolated. For the case of Heisenberg interactions,
in our formalism, the Heisenberg terms will be absorbed into the perturbation, and not
otherwise affect the resonance condition.

These numerical and analytical observations all point towards the
emergence of a stable conservation law -- not present in the undriven case --
and the absence of ergodic heating starting from any generic initial state 
across a wide range of quantum chaotic systems under a strong periodic drive. 

\section{Conclusions and Outlook}

In conclusion, we have demonstrated that generic interacting Floquet systems 
subjected to a strong periodic drive can exhibit scar points, i.e., points in 
the drive parameter space at which the system becomes non-ergodic due to the
emergence of constraints in the form of a quasi-conservation law
\newold{not present in the undriven system. This manifests itself in the absence of ergodicity
and unbounded heating starting from an arbitrary initial state at and around these points.}{} 
This is captured by our strong-field Magnus 
expansion in a time-dependent frame. For low drive frequencies, 
we formulate a novel perturbation theory (Floquet-Dyson perturbation theory) 
which works, even at first order, very accurately at or near integrability 
of the scar points. In particular, the resonances predicted by the 
theory accurately coincide with the sharp dips in the quasi-conserved quantity.
At the resonances, the system absorbs energy without bound from the drive, 
and hence the scars `compete' with the resonances. The resonances predicted by 
the theory appear to be isolated in parameter space, and thus, the theory 
provides a guideline for choosing parameters for observing 
resonance-free stable scars, as we demonstrate here. These results hold in 
particular for Ising systems in any dimension and with any form of the Ising 
interactions, as well as in the presence of additional pairwise 
Heisenberg interactions forming an arbitrary interaction graph. We also 
demonstrate the robustness of the phenomenon in the presence of anisotropic 
Heisenberg ($XYZ$) interactions. \\

\cyan{
The exact mechanism of this many-body phenomenon is actually still unknown, and 
the intuitions we have gathered are based on renormalization of the couplings, 
which are most effectively revealed under the non-perturbative, 
time-dependent frame transformation. For certain values of parameters, 
these renormalization factors vanish owing to destructive many-body 
quantum interference. These features are not captured by ordinary (lab-frame) 
Magnus expansion because it misses the effective re-summation necessary for 
these factors to manifest, done by the frame transformation.} \\

\XX{
The emergence and stability of a conservation law in an
interacting, quantum chaotic many-body system due to strong periodic drive 
is an unexpected and intriguing phenomenon, which warrants extensive investigations. One important
question is to study the nature of the state through continuous (non-stroboscopic) 
time - the so-called micro-motions. A powerful technique to study this is  
the so-called van Vleck expansion (also known as the "High-frequency expansion"  
see, e.g.,~\cite{Anatoli_Rev,Andre_Anisimovas_Rev,Bukov_Thesis} and references therein) 
of the Floquet Hamiltonian. This is also a potential alternative to our approach to
study the stroboscopic problem.}\\ 

Our work also touches on various Floquet experiments. In the original 
experimental work on Floquet many-body localization~\cite{Bordia_Knap_Bloch}, 
the interest of a large drive was already noted. In the context of the studies
of Floquet time crystals, the two kinds of states studied above have also 
played a central role: the trapped ion experiment~\cite{Monroe} used a fully 
polarized starting state, while the NMR experiment~\cite{Rovny} employed a 
high temperature state.

Our work points towards the important role in non-equilibrium settings played 
by the generation of emergent conservation laws and constraints, in contrast 
to only focusing on those 
existing in the static (undriven) system, and their demise under an external 
drive. Our work also opens a door for stable Floquet engineering in 
interacting systems, and indicates a recipe for tailoring interesting states 
and structured Hilbert spaces by choosing suitable drive Hamiltonians.

\vspace{.6cm}

\centerline{\bf Acknowledgments}

\vspace{.3cm} AD thanks Subinay Dasgupta and Sirshendu Bhattacharyya for 
collaborating on a non-interacting version of the phenomenon studied 
here~\cite{SB_AD_SDG}. \newold{We acknowledge useful discussions with Marin 
Bukov. The Quspin python package~\cite{Quspin_D1,Quspin_D2} was used in this 
work. This research was supported in part by the International Centre for 
Theoretical Sciences (ICTS) during a visit for the program - Thermalization, 
Many body localization and Hydrodynamics (Code: ICTS/hydrodynamics2019/11).}{}
AD and AH acknowledge the partner group program ``Spin liquids: correlations, 
dynamics and disorder" between IACS and MPI-PKS, and the visitors program of
MPI-PKS for supporting visits to PKS during the collaboration. This research 
was in part \newold{supported by the Deutsche Forschungsgemeinschaft under 
the cluster of excellence EXC2147 ct.qmat (project-id 39085490) and}{}
developed with funding from the Defense Advanced Research Projects 
Agency (DARPA) via the DRINQS program. The views, opinions and/or findings 
expressed are those of the authors and should not be interpreted as 
representing the official views or policies of the Department of Defense or 
the U.S. Government. RM is grateful to Vedika Khemani, David Luitz and Shivaji 
Sondhi for collaboration on related work \cite{LMSK}. 
DS thanks DST, India for Project No. SR/S2/JCB-44/2010 for financial support.

\appendix

%

\section{Strong-field Floquet expansion}
\label{SM_Magnus}

\subsection{The Ising Case}
\label{app:Ising}

Here, we provide the details of the derivation of the effective Hamiltonian
in Eqs.~(\ref{HF_0} - \ref{HF2_S1_S2_S3}). Carrying out the Pauli algebra gives
\bea H_{mov} &=& H_{0}^{x} -h^{z}\sum_{i}\left[ \cos{(2\theta)}\si_{i}^{z} 
+ \sin{(2\theta)}\si_{i}^{y} \right], ~ {\rm where} \non \\
\theta(t) &=& - ~h_{D}^{x}\int_{0}^{t} dt' ~\Sgn (\sin{\om t'}).
\label{H_mov_CosSin} \eea
We note that the frame change does not affect $m^{x},$ since it commutes with 
$W(t).$ \\

Next, we do the Magnus expansion of $H_{mov}.$ The initial orders are given by 
\bea
H_{eff} &=& \sum_{n=0}^{\infty} H_{F}^{0}, ~{\rm where} \non \\
H_{F}^{(0)} &=& \frac{1}{T}\int_{0}^{T}H_{mov}(t) dt, \non \\
H_{F}^{(1)} &=& \frac{1}{2!iT}\int_{0}^{T}dt_{1}\int_{0}^{t_{1}}dt_{2}
[H_{mov}(t_{1}), H_{mov}(t_{2})], \non \\
\label{Magnus_Gen} \eea
\noindent etc. We first consider the term $H_{F}^{(0)}$. 
It is easy to check that $H^{x}_{0}$ remains unaffected by the rotation, 
and the integrals in the first term vanish, giving
\beq H_{F}^{(0)} = H^{x}_{0}. \label{H_eff0_Ising} \eeq

Next we consider the second-order term 
\beq H^{(1)}_{F} ~=~ \frac{1}{2 ! (i) T}\int_{0}^{T}dt_{1}\int_{0}^{t_{1}}
dt_2 [H(t_1),H(t_2)]. \label{H_F1a} \eeq
Arranging the terms in the commutator, we get
\bea [H(t_{1}), H(t_{2})] &=& K_{1} + K_{2} + K_{3}, ~~ {\rm where} \non \\
K_{1} = &-& h^{z}~\{\cos{(\theta(t_{2}))} - \cos{(\theta(t_{1}))} \}~
[H_{0}^{x},\St_{z}], \non \\
K_{2} = &-& h^{z}~\{\sin{(\theta(t_{2}))} - \sin{(\theta(t_{1}))} \}~
[H_{0}^{x},\St_{y}], \non \\
K_{3} = && (h^{z})^{2}\sin{[\theta(t_{2}) - \theta(t_{1})]}[\St_{z},\St_{y}],
\label{I1_I2_I3} \eea
where ${\mathcal S}_{x/y/z} = \sum_{i}^{L}\si^{x/y/z}_{i}.$

Next we note that the integral in Eq.~\eqref{H_F1a} can be broken up in the 
following way,
\bea I[f(\theta(t_{1}),\theta(t_{2}))] &=& \int_{0}^{T}dt_{1}\int_{0}^{t_{1}} 
dt_2 [f(\theta(t_{1}),\theta(t_{2}))] \non \\ 
= I_{1}[f(\theta(t_{1}),\theta(t_{2}))] &+& I_{2}[f(\theta(t_{1}),
\theta(t_{2}))], \non \\
+ I_{3}[f(\theta(t_{1}),\theta(t_{2}))], && {\rm where} \non \\
I_{1}[f(\theta(t_{1}),\theta(t_{2}))] &=& \int_{0}^{T/2}dt_{1}\int_{0}^{t_{1}} 
dt_2 [f(\theta(t_{1}),\theta(t_{2}))], \non \\
I_{2}[f(\theta(t_{1}),\theta(t_{2}))] &=& \int_{T/2}^{T}dt_{1}\int_{0}^{T/2} 
dt_2 [f(\theta(t_{1}),\theta(t_{2}))], \non \\
I_{3}[f(\theta(t_{1}),\theta(t_{2}))] &=& \int_{T/2}^{T}dt_{1}
\int_{T/2}^{t_{1}} dt_2 [f(\theta(t_{1}),\theta(t_{2}))]. \non \\
\label{Integrals} \eea
Finally, we note that 
\bea {\rm For} ~ I_{1}, && \theta(t_{1}) = - h_{D}^{x}t_{1}, ~~\theta(t_{2}) 
= - h_{D}^{x}t_{2}, \non \\
{\rm For} ~ I_{2}, && \theta(t_{1}) = - h_{D}^{x}(T - t_{1}), ~~\theta(t_{2})
= - h_{D}^{x}t_{2}, \non \\
{\rm For} ~ I_{3}, && \theta(t_{1})=- h_{D}^{x}(T - t_{1}),  \theta(t_{2})= 
- h_{D}^{x}(T - t_{2}). \label{Theta_limits} \eea

Using Eqs.~\eqref{H_F1a}, \eqref{I1_I2_I3}, \eqref{Integrals} and 
\eqref{Theta_limits}) and evaluating the integrals, we obtain
Eqs.~(\ref{HF_0}) - (\ref{HF2_S1_S2_S3}) [see ~\ref{app:Integrals} for 
further details.]

%

\subsection{The Heisenberg Case}
\label{app:Heisenberg}

In the Heisenberg case, the total Hamiltonian $H(t) = H^{HB}(t)$ has the same 
form as that in Eq.~\eqref{ham3}, except with $V$ replaced by
\beq V^{HB} = -\sum_{i,j}\Jyij\si_i^y\si_{j}^y -\sum_{i,j}\Jzij\si_i^z\si_{j}^z
-h^{z}\sum_{i}\sigma^{z}, \label{V_HB} \eeq
\noindent
and $H_{0}^{x}$ replaced by
\beq H^{x(HB)}_{0} = -\Jxij\sum \si_i^x \si_{j}^{x} 
+ \ka \sum \si_i^x \si_{i+2}^x - h_0^x\sum \si_{i}^{x}. \label{H0x_HB} \eeq

Now, following Sec.~\ref{sec:Magnus}, we switch to the moving frame by 
acting on the total Hamiltonian $H^{HB}(t)$ with the unitary transformation 
given by
\bea \label{eq:e5}
V(t)&=& \exp\left[ih_D^x \sum_j \si _j^x \int_{t_0}^t \Sgn (\sin( \om
t'))dt'\right]\non\\
&=& \prod_j \exp\left[ih_D^x \si _j^x \int_{t_0}^t \Sgn (\sin( \om
t'))dt'\right], \eea
\noindent
where
\bea \label{eq:e6}
\theta(t)&=& h_D^x \int_{t_0}^t \Sgn (\sin( \om t'))dt', \eea
\noindent This gives our moving-frame Hamiltonian
\bea H^{mov}_{_{HB}}(t)&=& \prod_i \exp[-i \si_i^x \theta(t)]H_0
\exp[i \si_i^x \theta(t)]\non \\ \non
\eea
\begin{widetext}
\bea &=& H_{0}^{x(HB)} - \prod_i \exp\left[-i\si_i^x \theta(t)\right]
\left[\sum_{k,l} J^y_{k,l} \si_k^y \si_{l}^y\right] \prod_j \exp\left[
i \si_j^x \theta(t)\right] - \prod_i \exp\left[-i\si_i^x \theta(t)\right]
\left[\sum_{k,l} J^z_{k,l} \si_k^z \si_{l}^z\right] \prod_j \exp\left[
i \si_j^x \theta(t)\right]\non\\
&-& h_z \prod_i \exp\left[-i\si_i^x \theta(t)\right] \left[\sum_{k,l} \si_k^z 
\right] \prod_j \exp\left[i \si_j^x \theta(t)\right], \label{eq:e7} \eea
where
\bea H_{0}^{x(HB)} &=& -\Jxij\sum \si_i^x \si_{j}^x + \ka \sum \si_i^x 
\si_{i+2}^x - h_0^x\sum \si_i^x. \label{eq:Hx0} \eea
\noindent This gives 
\bea \label{eq:Hrot1}
H^{mov}_{_{HB}}(t) &=& H_{0}^{x(HB)} ~-~ \sum_{k,l} J^y_{k,l} e^{-i \si_k^x 
\theta(t)} e^{-i \si_{l}^x \theta(t)} \left( \si_k^y \si_{l}^y \right)
e^{i \si_k^x \theta(t)}e^{i \si_{l}^x \theta(t)} \non\\
&&-~ \sum_{k,l} J^z_{k,l} e^{-i \si_k^x \theta(t)}e^{-i \si_{l}^x
\theta(t)} \left( \si_k^z \si_{l}^z \right) e^{i \si_k^x \theta(t)}
e^{i \si_{l}^x \theta(t)} ~-~ h^z \sum_{k,l} e^{-i \si_k^x \theta(t)}\si_k^z
e^{i \si_k^x \theta(t)}. \eea

The $\Jyij$ term can be simplified to
\bea \label{eq:e8}
&=& - ~\sum_{i,j} \Jyij \si_i^y \si_{j}^y e^{2i \si_i^x \theta(t)}
e^{2i \si_{j}^x \theta(t)} \non \\
&=& -~ \sum_{i,j} \Jyij \si_i^y \si_{j}^y \left[ \mathbb{I}\cos^2 (2 \theta)
- \si_i^x \si_{j}^x \sin^2 (2\theta) 
+  i \sin (2 \theta) \cos (2 \theta) 
(\si_i^x +\si^x_{j}) \right]. \eea
Similarly, the $\Jzij$ term becomes
\bea \label{eq:e9}
&=& - ~\sum_{i,j} \Jzij \si_i^z \si_{j}^z \left[ \mathbb{I}\cos^2 (2 \theta)
- \si_i^x \si_{j}^x \sin^2 (2\theta) +  i \sin (2 \theta) \cos (2 \theta) 
(\si_i^x +\si^x_{j}) \right]. \eea
The $h_z$ term is similar to our previous case, namely,
\bea \label{eq:e10}
&=& -h^z \cos (2 \theta) \sum_i \si_i^z + h^z \sin (2 \theta) 
\sum_i \si_i^y. \eea
Hence,

\bea H^{mov}_{_{HB}}(t)&=& H_{0}^{x(HB)} ~-~ \sum_{i,j} \Jyij \si_i^y \si_{j}^y
\left[\mathbb{I}\cos^2 (2 \theta) - \si_i^x \si_{j}^x \sin^2 (2\theta) 
+ \frac{i}{2} \sin (4 \theta) (\si_i^x +\si^x_{j}) \right] \non\\
&&- ~\sum_{i,j} \Jzij \si_i^z \si_{j}^z \left[ \mathbb{I}\cos^2 (2 \theta)
- \si_i^x \si_{j}^x \sin^2 (2\theta) + \frac{i}{2} \sin (4 \theta) (\si_i^x +
\si^x_{j}) \right] \non\\
&&- ~h^z \cos (2 \theta) \sum_i \si_i^z ~+~ h^z \sin (2 \theta) \sum_i \si_i^y.
\label{eq:Hrot} \eea
\end{widetext}
Next we do the Magnus expansion on Eq.~\eqref{eq:Hrot}. The zeroth order term 
is
\bea \label{eq:e12}
H_{\mathrm{eff}}^{(0)}&=& \frac{1}{T} \int_0^T dt H^{mov}_{_{HB}}(t). \eea
Now, with the definition of $\theta(t)$ given in Eq.~\eqref{eq:e6}, we get 
\begin{subequations}
\label{eq:e13}
\bea 
\int_0^T \cos^2 (2\theta) dt &=& \frac{T}{2} + \frac{\sin 2 h_D^xT}{4h_x^D},
\\
\int_0^T \sin^2 (2\theta) dt &=& \frac{T}{2} - \frac{\sin 2 h_D^xT}{4h_x^D}, 
\\
\int_0^T \sin (4\theta) dt &=& \frac{1}{4h_x^D} (1 - \cos 2 h_D^xT),
\\
\int_0^T \sin (2\theta) dt &=& \frac{1}{h_x^D} (1 - \cos  h_D^xT), 
\\
\int_0^T \cos (2\theta) dt &=& \frac{1}{\hd} \sin (\hd T). 
\eea
\end{subequations}
Applying the freezing condition, 
\bea \label{eq:14}
h_D^x T&=& 2\pi n, \eea
we get
\begin{subequations} \label{eq:e14}
\bea \int_0^T \cos^2 (2\theta) dt&=& \frac{T}{2}, \\
\int_0^T \sin^2 (2\theta) dt&=& \frac{T}{2}, \\
\int_0^T \sin (4\theta) dt&=& 0, \\
\int_0^T \sin (2\theta) dt&=& 0, \\
\int_0^T \cos (2\theta) dt&=& 0. \eea
\end{subequations}
Putting everything in Eq.~\eqref{eq:e12}, we obtain
\bea \label{eq:e15}
H^{(0)}_{eff}&=& H_{0}^{x(HB)} ~-~ \sum_{i,j} \Jyij \si_i^y \si_{j}^y\frac{1}{T}
\left[ \frac{T}{2} - \si_i^x \si_{j}^x \frac{T}{2}\right]\non\\
&&- ~\sum_{i,j} \Jzij \si_i^z \si_{j}^z\frac{1}{T}
\left[ \frac{T}{2} - \si_i^x \si_{j}^x \frac{T}{2}\right] \\
= H_{0}^{x(HB)} &-& \frac{1}{2}\sum_{i,j} \left[ \si_i^y \si_{j}^y +
\si_i^z \si_{j}^z \right](\Jyij + \Jzij). \eea
We already have,
\bea \label{eq:e16}
[H_{0}^{x(HB)},m^x]= 0,\non \eea
and one can easily show that 
\bea \label{eq:e17}
\left[ \sum_{i,j}( \si_i^y \si_{j}^y +\si_i^z \si_{j}^z), m^x \right]=0. \eea
The first-order term of the Magnus expansion is
\beq \label{eq:e18}
H_{\mathrm{eff}}^{(1)}=\frac{1}{2!Ti}\int_0^{T}dt_1\int_0^{t_1}dt_2\left[
H^{mov}_{_{HB}}(t_1),H^{mov}_{_{HB}}(t_2)\right]. \eeq
Rearranging all the terms in Eq.~\eqref{eq:Hrot}, we can write
\bea H^{mov}_{_{HB}}(t)&=& H_{0}^{x(HB)} + A \cos^2 (2\theta) + 
B \sin^2 (2\theta) \non\\ 
&&+ C \sin (4\theta) + D \cos (2\theta) + E \sin (2\theta). \non \\
\eea
Hence
\begin{widetext}
\bea \left[H^{mov}_{_{HB}}(t_1),H^{mov}_{_{HB}}(t_2)\right]&=& [H_{0}^{x(HB)},
A]~I_1 + [H_{0}^{x(HB)},B]~I_2 \non \\
+ [H_{0}^{x(HB)},C]~I_3 + [H_{0}^{x(HB)},D] ~I_4 &+& [H_{0}^{x(HB)},E] ~I_5 + 
[A,B] ~I_6 + [A,C] ~I_7 + [A,D] ~I_8 \non \\
+ [A,E] ~I_9 + [B,C] ~I_{10} + [B,D] ~I_{11} &+& [B,E] ~I_{12} + [C,D] ~I_{13} 
+ [C,E] ~I_{14} + [D,E] ~I_{15}, \eea
\noindent
where
\bea A = -\sum_{i,j}[J^{y}_{ij}\sigma_{i}^{y}\sigma_{j}^{y} + J^{z}_{ij} 
\sigma_{i}^{z}\sigma_{j}^{z}], &~&
B = \sum_{i,j}[(J^{y}_{ij}\sigma_{i}^{y}\sigma_{j}^{y} + J^{z}_{ij} 
\sigma_{i}^{z}\sigma_{j}^{z})\sigma_{i}^{x}\sigma_{j}^{x}], ~C = -\sum_{i,j} 
\frac{i}{2} [(J^{y}_{ij}\sigma_{i}^{y}\sigma_{j}^{y} + J^{z}_{ij}\sigma_{i}^{z}
\sigma_{j}^{z})(\sigma_{i}^{x} + \sigma_{j}^{x})], \non 
\eea
\bea D = -h^{z}\sum_{i}\sigma_{i}^{z}, ~ &{\rm and}& ~ E = h^{z}\sum_{i}
\sigma_{i}^{y}, \eea
\end{widetext}
\noindent and
\bea
I_1 &=& \cos^2 (2\ta(t_2)) - \cos^2 (2\ta(t_1)), \non \eea
\bea I_2 &=& \sin^2 (2\ta(t_2)) - \sin^2 (2\ta(t_1)), \non \eea
\bea I_3 &=& \sin (4\ta(t_2)) - \sin (4\ta(t_1)), \non \eea
\bea I_4 &=& \cos (2\theta (t_2)) - \cos (2\theta(t_1)), \non \eea
\bea I_5 &=& \sin (2\theta (t_2)) - \sin (2\theta(t_1)), \non \eea
\bea I_6 &=& \cos^{2} (2\theta(t_1)) \sin^{2} (2\ta(t_2)) - \sin^{2} (2
\theta(t_1)) \cos^{2} (2\ta(t_2)), \non \eea
\bea I_7 &=& \cos^{2} (2\theta(t_1)) \sin (4\ta(t_2)) - \sin (4\theta(t_1))
\cos^{2} (2\ta(t_2)), \non \eea
\bea I_8 &=& \cos^{2} (2\theta(t_1)) \cos (2\ta(t_2)) - \cos (2\theta(t_1))
\cos^{2} (2\ta(t_2)), \non \eea
\bea I_9 &=& \cos^{2} (2\theta(t_1)) \sin (2\ta(t_2)) - \sin (2\theta(t_1))
\cos^{2} (2\ta(t_2)), \non \eea
\bea I_{10} &=& \sin^{2} (2\theta(t_1)) \sin (4\ta(t_2)) - \sin (4\theta(t_1))
\sin^{2} (2\ta(t_2)), \non \eea
\bea I_{11} &=& \sin^{2} (2\theta(t_1)) \cos (2\ta(t_2)) - \cos (2\theta(t_1))
\sin^{2} (2\ta(t_2)), \non \eea
\bea I_{12} &=& \sin^{2} (2\theta(t_1)) \sin (2\ta(t_2)) - \sin (2\theta(t_1))
\sin^{2} (2\ta(t_2)), \non \eea
\bea I_{13} &=& \sin (4\theta(t_1))\cos (2\ta(t_2)) - \cos (2\theta(t_1))
\sin{4\ta(t_2)}, \non \eea
\bea I_{14} &=& \sin (4\theta(t_1))\sin (2\ta(t_2)) - \sin (2\theta(t_1))
\sin (4\ta(t_2)), \non \eea
\bea I_{15} &=& \sin [2(\theta(t_1) - \theta(t_2))]. \label{eq:integrants} \eea
Now, one can show that all the $15$ integrals of $I_n$ in
Eq.~\eqref{eq:integrants} vanish for the freezing condition. Hence, 
\bea H_{eff} = H_{0}^{x(HB)} - \frac{1}{2}\sum_{i,j} \left[ \si_i^y \si_{j}^y 
+ \si_i^z \si_{j}^z \right](\Jyij + \Jzij) \non \eea
up to the first two orders.

\subsection{Explicit Calculation of the Integrals in the Moving Frame Magnus Expansion: 
The Ising Case (self contained)}
\label{app:Integrals}

The Hamiltonian can be written as:
\bea
\label{eq:supp1}
H(t)&=& H_0+ r(t)H_D.
\eea
We go to the rotating frame using the transformation:
\bea
\label{eq:supp2}
H^{\mathrm{mov}}(t)&=& W^\da(t)H_0 W(t),
\eea
where the rotation operator is 
\bea
\label{eq:supp21}
W(t)=\exp\left[ -i\int^t_{t_0}
r(t')dt' H_D\right]. 
\eea
The first case is an Ising model with next-nearest neighbour terms:
\bea
H_0&=& 
-\sum_{i} J \si_i^x \si_{i+1}^x  
+ \ka \sum \si_i^x \si_{i+2}^x \non \\
&-& h_0^x\sum \si_i^x  - h^z\sum \si_i^z, 
\label{eq:e31}\\
&=& H^x_0 + V\\
H_D&=& - \hd \sum \si_i^x,\label{eq:supp32}
\eea
and 
\bea
\label{eq:supp4}
r(t)&=& \Sgn(\sin(\om t)).
\eea
From eq. \ref{eq:supp21}, eq. \ref{eq:supp32}, and eq. \ref{eq:supp4},
\bea
\label{eq:supp5}
W(t)&=& \exp\left[ih_D^x \sum_j \si _j^x \int_{t_0}^t \Sgn(\sin( \om
t'))dt'\right]\non\\
&=& \prod_j \exp\left[ih_D^x \si _j^x \int_{t_0}^t \Sgn(\sin( \om
t'))dt'\right].
\eea
Defining, 
\bea
\label{eq:supp6}
\theta(t)&=& - h_D^x \int_{t_0}^t \Sgn(\sin( \om t'))dt',
\eea
putting all these together,we get,
\bea
H^{\mathrm{mov}}(t)&=& \prod_i
\exp[-i \si_i^x \theta(t)]H_0
\exp[i \si_i^x \theta(t)]\non\\\non
\eea
\begin{widetext}
\bea
&=& H_0^x - h_z \prod_i \exp\left[-i\si_i^x \theta(t)\right]
\left[\sum_{k} \si_k^z \right]
\prod_j \exp\left[i \si_j^x \theta(t)\right]\\
&=&  H_0^x
- h^z \sum_{k} e^{-i \si_k^x \theta(t)}\si_k^z
e^{i \si_k^x \theta(t)}
\label{eq:supp7}
\eea
\bea
\therefore H^{\mathrm{mov}}(t)&=& 
H_0^x - h^z \cos 2 \theta \sum_i \si_i^z - h^z \sin 2 \theta \sum_i
\si_i^y.
\label{eq:Hrotsupp}
\eea
\end{widetext}

Now we can do Magnus expansion on eq. \ref{eq:Hrotsupp} . The zeroth order term is:
\bea
\label{eq:Heff0supp}
H_{\mathrm{eff}}^{(0)}&=&  \frac{1}{T} \int_0^T  H^{\mathrm{mov}}(t)~dt\non \\
&=& \frac{1}{T} \int_0^T H_{0}^{x}~dt - \frac{h^z}{T} \sum_i \si_i^z \int_0^T \cos{2 \theta}~dt \non \\
&-& \frac{h^z}{T} \sum_i \si_i^y \int_0^T  \sin{2 \theta}~dt
\eea

Now, with the definition of $\theta(t)$ as given in eq. \ref{eq:supp6} 
(note: $\ta(t) = - \hd t $ for $0<t \le \frac{T}{2}$; and $\ta(t) = -(\hd T - \hd t)$ for $\frac{T}{2} \le t \le T$) the integral simplifies to:
\bea
\label{eq:cosfirst}
\int_0^T \cos 2 \theta~dt &=& \int_0^{\frac{T}{2}} \cos{2 \theta}~dt + \int_{\frac{T}{2}}^T \cos{2 \theta}~dt \non\\
&=& \int_0^{\frac{T}{2}} \cos{(2 \hd t)}~dt + \int_{\frac{T}{2}}^T \cos{2 (\hd T - \hd t)}~dt \non \\
&=& \frac{1}{\hd} \sin{(\hd T)}
\eea

Similarly, 
\bea
\label{eq:sinfirst}
\int_0^T \sin 2 \theta~dt &=& \frac{1}{\hd} (\cos{(\hd T)} - 1)
\eea

Putting eq. \ref{eq:cosfirst} and eq. \ref{eq:sinfirst} into eq. \ref{eq:Heff0supp}, we get
\bea
\label{eq:Heff0final}
H_{\mathrm{eff}}^{(0)}&=& H^x_0 - \frac{h^z}{\hd T} \sum_i \si_i^z \sin{(\hd T)} \non \\
&+& \frac{h^z}{\hd T} \sum_i \si_i^y (1 - \cos{(\hd T)}) 
\eea
 
Now putting the freezing condition: $\hd T = 2n \pi$ in eq. \ref{eq:Heff0final}, i.e. $\sin \hd T= 0 $ and $\cos \hd T= 1$,  one gets,
\bea
\label{eq:HF0freez}
H_{\mathrm{eff}}^{(0)}|_{freezing}&=& H^x_0
\eea


Next we evaluate the $1^{st}$ order term.
\begin{widetext}
\bea
\label{eq:Heff1supp}
H_{\mathrm{eff}}^{(1)}&=&  \frac{1}{2! Ti} \int_0^T dt_1 \int_0^{t_1} dt_2 \left[H^{\mathrm{mov}}(t_1),H^{\mathrm{mov}}(t_2)\right]
\eea
\end{widetext}
Calling $\ta(t_1) = \ta_1$; $\ta(t_2) = \ta_2$ and $\sum_i \si_i^z = S^z$ ; $\sum_i \si_i^y = S^y$ and using the form of $H^{\mathrm{mov}}$
from eq. \ref{eq:Hrotsupp}, the commutator in eq. \ref{eq:Heff1supp} simplifies to:

\begin{widetext}
\bea
\left[H^{\mathrm{mov}}(t_1),H^{\mathrm{mov}}(t_2)\right] \non
\eea
\bea
\label{eq:Heff1new}
= \left[S^z,H^x_0 \right] h^z (\cos 2\ta_2 - \cos 2 \ta_1) +
\left[S^y,H^x_0 \right] h^z (\sin 2 \ta_2 - \sin 2 \ta_1)  + \left[S^y,S^z \right] (h^z)^2 \sin (2\ta_1 - 2\ta_2).
\eea
Now, for example, the integral corresponding to the first term is:

\bea
I_1 = \int_0^T \int_0^{t_1} dt_1 dt_2 (\cos 2\ta_2 - \cos 2 \ta_1) 
\eea
\bea
&=& \int_0^{\frac{T}{2}} dt_1 \int_0^{t_1} dt_2 \cos 2\ta(t_2)  + \int_{\frac{T}{2}}^T dt_1 \int_0^{\frac{T}{2}} dt_2 \cos 2\ta(t_2)
 + \int_{\frac{T}{2}}^T dt_1 \int_{\frac{T}{2}}^{t_1} dt_2 \cos 2\ta(t_2) - \int_0^T dt_1 \cos 2\ta(t_1) t_1 \non
\eea
\bea
&=& \int_0^{\frac{T}{2}} dt_1 \int_0^{t_1} dt_2 \cos (2\hd t_2) + \int_{\frac{T}{2}}^T dt_1 \int_0^{\frac{T}{2}} dt_2 \cos (2\hd t_2)
	+ \int_{\frac{T}{2}}^T dt_1 \int_{\frac{T}{2}}^{t_1} dt_2 \cos (2\hd T - 2\hd t_2)\non\\
	&-& \int_0^{\frac{T}{2}}dt_1 t_1 \cos(2\hd t_1) - \int_{\frac{T}{2}}^{T}dt_1 t_1 \cos (2\hd T - 2\hd t_1)
\eea

\bea
\label{eq:I1}
I_1 &=& I_1^A + I_1^B + I_1^C - I_1^D - I_1^E
\eea
Now,

\bea
\label{eq:I1A}
I_1^A = \int_0^{\frac{T}{2}} dt_1 \int_0^{t_1} dt_2 \cos (2\hd t_2) = \frac{1}{2\hd} \int_0^{\frac{T}{2}} dt_1 \sin 2\hd t_1
= \frac{1}{(2\hd)^2} \left[1 - \cos \hd T \right]
\eea

\bea
\label{eq:I1B}
I_1^B = \int_{\frac{T}{2}}^T dt_1 \int_0^{\frac{T}{2}} dt_2 \cos (2\hd t_2) = \frac{1}{2\hd} \int_{\frac{T}{2}}^T dt_1 \sin \hd T
= \frac{T}{4\hd} \sin \hd T
\eea

\bea
\label{eq:I1C}
I_1^C = \int_{\frac{T}{2}}^T dt_1 \int_{\frac{T}{2}}^{t_1} dt_2 \cos (2\hd T - 2\hd t_2) 
= \frac{1}{2\hd} \int_{\frac{T}{2}}^T dt_1 \left[ \sin \hd T - \sin(2\hd T - 2\hd t_1) \right] \non\\
= \frac{T}{4\hd} \sin \hd T - \frac{1}{(2\hd)^2} + \frac{1}{(2\hd)^2} \cos \hd T
\eea

\bea
\label{eq:I1D}
I_1^D = \int_0^{\frac{T}{2}}dt_1 t_1 \cos(2\hd t_1) = \frac{1}{2\hd} \left(\frac{T}{2} \sin \hd T - \int_0^{\frac{T}{2}}dt_1 \sin 2\hd t_1 \right) \non\\
= \frac{T}{4 \hd} \sin \hd T + \frac{1}{(2\hd)^2} \cos \hd T - \frac{1}{(2\hd)^2}
\eea

\bea
\label{eq:I1E}
I_1^E = \int_{\frac{T}{2}}^{T}dt_1 t_1 \cos (2\hd T - 2\hd t_1) 
= \frac{T}{4 \hd} \sin \hd T + \frac{1}{2\hd} \int_{\frac{T}{2}}^{T}dt_1 \sin (2\hd T - 2\hd t_1) \non\\
= \frac{T}{4 \hd} \sin \hd T + \frac{1}{(2\hd)^2} (1 - \cos \hd T)
\eea
\end{widetext}
Now, putting eqs. \ref{eq:I1A},\ref{eq:I1B},\ref{eq:I1C},\ref{eq:I1D},\ref{eq:I1E} into eq. \ref{eq:I1}, yields:
\bea
I_1 = 0
\eea

Carrying out the integrals corresponding to the other two commutators in \ref{eq:Heff1new}, one can similarly get:

\bea
I_2 = \int_0^T \int_0^{t_1} dt_1 dt_2 (\sin 2 \ta_1 - \sin 2 \ta_2) = 0 \\
I_3 = \int_0^T \int_0^{t_1} dt_1 dt_2 \sin (2\ta_1 - 2\ta_2) = 0
\eea

\section{Floquet-Dyson Perturbation Theory}
\label{app:FDPT}

We start from Eq.~\eqref{sch1}, which implies that
\bea && i ~\sum_m \dot{c}_m (t) ~e^{-i \int_0^t dt' E_m (t')} ~|m \ra \non \\ 
&=& V ~\sum_m ~c_m (t) ~e^{-i \int_0^t dt' E_m (t')} ~|m \ra, \label{sch2} \eea 
where the dot over $c_m$ denotes $d/dt$. Taking the inner product of 
Eq.~\eqref{sch2} with $\la n |$ and using Eq.~\eqref{nvn}, we find, to first
order in $V$, that
\beq \dot{c}_n ~=~ 0. \eeq
We can therefore choose 
\beq c_n (t) ~=~ 1 \label{cnt1} \eeq
for all $t$. We thus have
\bea |\psi_n (t)\ra &=& e^{-i \int_0^t dt' E_n (t')} ~|n \ra \non \\
&& +~ \sum_{m \ne n} ~c_m (t) ~e^{-i \int_0^t dt' E_m (t')} ~|m \ra. 
\label{psi1} \eea
Hence Eq.~\eqref{floeig1} implies that the Floquet eigenvalue is still given 
by $\mu_{n}^{(0)} = \int_{0}^{T} dt E_{n}(t)$ up to first order in $V$. 

Next, taking the inner product of Eq.~\eqref{sch2} with $\la m |$, where 
$m \ne n$, we find, to first order in $V$, that
\beq \dot{c}_m ~=~ - i ~\la m | V | n \ra ~e^{i \int_0^t dt' [E_m (t') - E_n 
(t')]}, \label{cmdot} \eeq
so that
\bea c_m (T) &=& c_m (0) - i ~\la m | V | n \ra \non \\
&& \times ~\int_0^T dt ~e^{i \int_0^t dt' [E_m (t') - E_n (t')]}. 
\label{cmt1} \eea

We now impose the condition on $|\y_{n}(T)\ra$ of Eq.~\eqref{nine} such that 
$|\y_{n}(0)\ra$ turns out to be a Floquet state, i.e., 
from Eq.~\eqref{psi1} we must have
\beq \psi_n (T) ~=~ e^{-i \int_0^T dt E_n (t)} ~\psi_n (0), \eeq
namely, we must have
\beq c_m (T) ~=~ e^{i \int_0^T dt [E_m (t) - E_n (t)]} ~c_m (0) \eeq
for all $m \ne n.$ Clearly, $|\y_{n}(0)\ra$ satisfying this condition can be 
identified as the Floquet state $|\mu_{n}\ra$.

\subsection{Single spin model}

\subsubsection{Model}
\label{app:singlespin}

We consider a single spin-$S$ object which evolves according to the
time-dependent Hamiltonian
\beq H (t) ~=~ -~ h^x S^x ~-~ h^z S^z ~-~ h_D^x ~\Sgn (\sin (\om t)) ~S^x.
\label{ham4a} \eeq
Since $\sin (\om t)$ is positive for $0 < t < T/2$ and negative for 
$T/2 < t < T$, where $T=2\pi/\om$, the Floquet operator is given by
\bea U &=& e^{(iT/2)~[(h^x - h_D^x) S^x ~+~ h^z S^z]} \non \\
&& \times ~e^{(iT/2)~[(h^x + h_D^x) S^x ~+~ h^z S^z]}. \label{u1a} \eea
The group properties of matrices of the form $e^{i{\vec a} \cdot {\vec S}}$ 
imply that $U$ in Eq.~\eqref{u1a} must be of the same form and can therefore
be written as
\bea U &=& e^{i\ga {\hat k} \cdot {\vec S}}, \non \\
{\rm where} ~~~ {\hat k} &=& (\cos \ta, \sin \ta \cos \phi, \sin \ta \sin \phi)
\label{u2a} \eea
is a unit vector. We will work in the basis in which $S^x$ is diagonal; hence 
we choose the polar angles in such a way that the $x$-component of ${\hat k}$ 
is equal to $\cos \ta$. The eigenstates of $U$ in Eq.~\eqref{u2a} are the same 
as the eigenstates of the matrix $M = {\hat k} \cdot {\vec S}$. 
It is then clear that the expectation values of $S^x$ in the different
eigenstates take the values $\cos \ta$ times $S, S-1, \cdots, -S$.
The maximum expectation value is given by $s_{max} = S \cos \ta$.

An important point to note is that if the parameters $h^x, h^z, h_D^x$
and $T$ are fixed and only the spin $S$ is varied, the values of $\ga$ and
$\hat k$ in Eqs.~\eqref{u2a} do not change. This means that if
we can calculate these quantities for one particular value of $S$, the results
will hold for all $S$. In particular, $m^x_{max} \equiv s_{max}/S = \cos \ta$
will not depend on $S$. We have confirmed this numerically for a variety of
parameter values.

\subsubsection{Results from FDPT}

Next, we apply the perturbation theory developed in Sec.~\ref{sec:pt}. 
Writing the Hamiltonian as $H = H_0 (t) + V$, where
\bea H_0 (t) &=& -~ h^x S^x ~-~ h_D^x ~\Sgn (sin (\om t)) ~S^x, \non \\
V &=& -~ h^z S^z, \eea
we can do perturbation theory to study how the state given by $|0 \ra \equiv
|S^x = S\ra$ mixes with the state $|1 \ra \equiv | S^x = S-1 \ra$. Following
the steps leading up to Eq.~\eqref{cmt2}, and using the fact that $\la 0 | 
S^z | 1 \ra = \sqrt{S/2}$, we find that
\beq c_1 (0) ~=~ \frac{\sqrt{2S} ~h^z}{h_D^x} ~\frac{e^{ih^x T/2} ~
[e^{ih_D^x T/2} - \cos (h^x T/2)]}{e^{ih^x T} ~-~ 1}, \label{cmt10a} \eeq

\subsubsection{Exact results}

It is instructive to look at the form of the Floquet operator $U$ in different 
cases. We first derive an exact expression for $U$ using the identity that if
\beq e^{i\al {\hat m} \cdot {\vec S}} ~e^{i\chi {\hat n} \cdot {\vec S}} ~=~
e^{i\ga {\hat k} \cdot {\vec S}}, \label{mnk1a} \eeq
then
\bea
\cos \left( \frac{\ga}{2} \right) &=& \cos \left( \frac{\al}{2} \right) \cos 
\left( \frac{\chi}{2} \right)\non \\
&& -~ {\hat m} \cdot {\hat n} ~\sin \left( \frac{\al}{2} \right) \sin \left( 
\frac{\chi}{2} \right), \non \\
{\hat k} &=& \frac{1}{\sin \left( \ga/2 \right)} ~\Big[ ~{\hat m} ~\sin
\left(\frac{\al}{2} \right) \cos \left( \frac{\chi}{2} \right) \non \\
&& ~~+~ {\hat n}~\sin \left( \frac{\chi}{2} \right) \cos \left( 
\frac{\al}{2} \right) \non \\
&& ~~- ~{\hat m} \times {\hat n} ~\sin \left( \frac{\al}{2} \right)
\sin \left( \frac{\chi}{2} \right) \Big]. \label{mnk2a} \eea
We can derive Eq.~\eqref{mnk2a} from Eq.~\eqref{mnk1a} for the case $S=1/2$ 
when ${\vec S} = {\vec \si}/2$. Eq.~\eqref{mnk2a} then follows for any value 
of $S$ due to the group properties of the matrices given in Eq.~\eqref{mnk1a}.

We will now use Eqs.~(\ref{mnk1a}-\ref{mnk2a}) along with Eq.~\eqref{u2a}
which can be written in the form
\bea \al &=& \frac{T}{2} ~\sqrt{(h_D^x ~-~ h^x)^2 ~+~ (h^z)^2}, \non \\
{\hat m} &=& -~ \frac{(h_D^x ~-~ h^x) ~{\hat x} ~-~ h^z ~{\hat z}}{\sqrt{(h_D^x
~-~ h^x)^2 ~+~ (h^z)^2}}, \non \\
\chi &=& \frac{T}{2} ~\sqrt{(h_D^x ~+~ h^x)^2 ~+~ (h^z)^2}, \non \\
{\hat n} &=& \frac{(h_D^x ~+~ h^x) ~{\hat x} ~+~ h^z ~{\hat z}}{\sqrt{(h_D^x
~-~ h^x)^2 ~+~ (h^z)^2}}, \label{albe1a} \eea
where we have assumed that $h_D^x$ is positive and much larger than $|h^x|$
and $|h^z|$. 

If $e^{ih^x T} \ne 1$, we can write the expressions in Eqs.~\eqref{albe1a} 
to zero-th order in the small parameter $h^z/h_D^x$ to obtain
\bea \al &=& \frac{T}{2} ~(h_D^x ~-~ h^x), ~~~~~~{\hat m} ~=~ -~{\hat x},
\non \\
\chi &=& \frac{T}{2} ~(h_D^x ~+~ h^x), ~~~~~~{\hat n} ~=~{\hat x}. 
\label{albe2a} \eea
Eqs.~(\ref{mnk1a}-\ref{mnk2a}) then imply that
\beq \cos \left( \frac{\ga}{2} \right) ~=~ \cos \left( \frac{h^x T}{2} \right),
~~~~{\rm and} ~~~~ {\hat k} ~=~ {\hat x}. \label{gak1a} \eeq
We thus find that the Floquet operator for the time period $T$ corresponds to
a rotation about the $\hat x$ axis. 

If $e^{ih^x T} = 1$, i.e., $\cos (h^x T/2) = \pm 1$, the denominator of 
Eq.~\eqref{cmt10a} vanishes. If 
$e^{ih_D^x T/2} \ne \cos (h^x T/2)$, we have to expand the expressions in 
Eqs.~\eqref{albe1a} up to second order in $h^z/h_D^x$ to find that
\bea {\hat k} &=& \cos \left( \frac{h_D^x T}{4} \right) ~{\hat z} ~-~ \sin
\left( \frac{h_D^x T}{4} \right) ~{\hat y} \non \\
&& {\rm if} ~~ \cos \left( \frac{h^x T}{2} \right) ~=~ 1, \non \\
&=& \sin \left( \frac{h_D^x T}{4} \right) ~{\hat z} ~+~ \cos \left( \frac{h_D^x
T}{4} \right) ~{\hat y} \non \\
&& {\rm if} ~~ \cos \left( \frac{h^x T}{2} \right) ~=~ - 1. \label{albe3a} \eea
Hence the Floquet operator corresponds
to a rotation about an axis lying in the $y-z$ plane. This implies that
the expectation value of $S^x$ will be zero in all the eigenstates of the
Floquet operator. 

If $e^{ih^x T} = 1$ and $e^{ih_D^x T/2} = \cos (h^x T/2) = \pm 1$, both the 
numerator and denominator of Eq.~\eqref{cmt10a} vanish. We then discover that
\beq {\hat k} ~=~ \frac{h^x ~{\hat x} ~-~ h^z ~{\hat z}}{\sqrt{(h^z)^2 ~+~
(h^x)^2}}. \label{gak3a} \eeq
In this case, the Floquet operator corresponds to a rotation about an
axis lying in the $x-z$ plane.

\subsection{FDPT for the Ising chain}

\beq E_m (t) - E_0 (t) = 4(J - \ka) + 2 h_0^x + 2 h_D^x ~\Sgn 
(\sin (\om t)). \label{em0} \eeq
We now use the notations and results from Sec.~\ref{sec:pt} to 
construct the Floquet state $|\psi(0)\ra$ obtained by perturbing the 
unperturbed (Floquet) eigenstate $| 0 \ra$ to first order in $V$ given by
\bea \psi (0) &=& c_{0}|0\ra ~+~ \sum_{m\ne 0}^L ~c_m (0) |m\ra \non \\
&=& c_{0}|0\ra ~+~ \sqrt{L} ~c_m (0) |L-2 \ra, \label{psi4} \eea
where 
\beq |L-2 \ra ~\equiv~ \frac{1}{\sqrt{L}} ~\sum_{m=1}^L ~|m \ra \label{lm2} \eeq
is a translation invariant and normalized state in which $\sum_m \si_m^x = L-2$.
Taking $c_0 (t) = 1$ for all $t$ and using 
$\la m | V | 0 \ra = - h^z$ in Eq.~\eqref{cmt2}, we get 
\beq c_m (0) ~=~ i h^z ~\frac{\int_0^T dt ~e^{i \int_0^t dt' [E_m (t') - 
E_0 (t')]}}{e^{i \int_0^T dt [E_m (t) - E_0 (t)]} ~-~ 1}, \label{cmt5} \eeq
where $E_m (t) - E_0 (t)$ is given in Eq.~\eqref{em0}.

\subsection{Failure of FDPT and Emergent Integrability at the Scars}

\begin{figure}
\begin{center}
\includegraphics[width=0.65\linewidth]{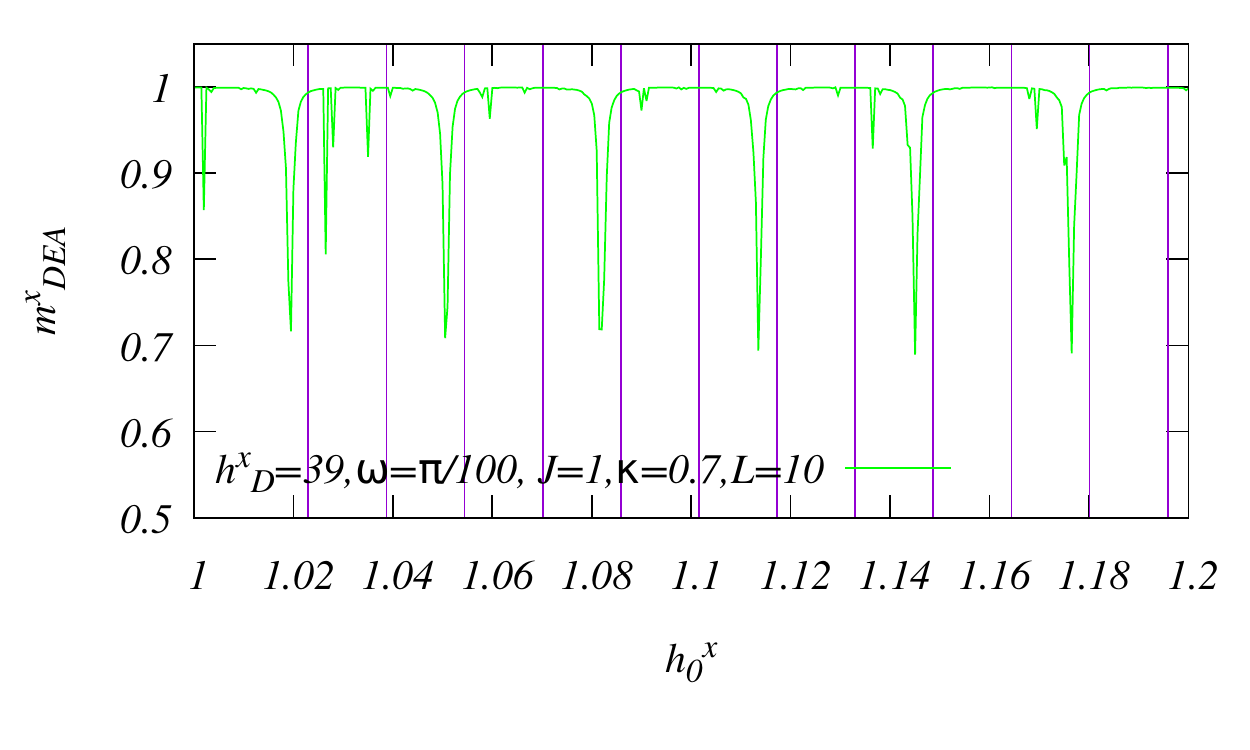}
\end{center}
\caption{The result shows mismatches between the numerical dips in $m^{x}_{DE}$
and the first-order FDPT prediction (vertical lines obtained from 
Eq.~\eqref{Gen_Res}). The parameters are chosen such that the condition for 
a scar is not satisfied, i.e., $h_{D}^{x} \ne n\omega$.} \label{Mismatch} 
\end{figure}

The FDPT always works well in integrable systems (e.g., the single large spin 
case discussed here, and also other studied examples not reported here).
However, FDPT seems to lose accuracy away from integrability, and hence from 
the scar points. This is an interesting indirect indication of the fact that 
integrability emerges at the scar points. In contrast to the very accurate 
prediction of resonances in the Fig.~\ref{Resonances} (main text), 
Fig.~\ref{Mismatch} shows a substantial mismatch between the FDPT predictions 
and the true numerical resonances (dips) away from the scars.

\section{Freeze an Arbitrary Bit-String by Tailoring the Emergent Conservation Law}
\label{app:varfield}

\begin{figure}[h!]
\begin{center}
\includegraphics[width=0.65\linewidth]{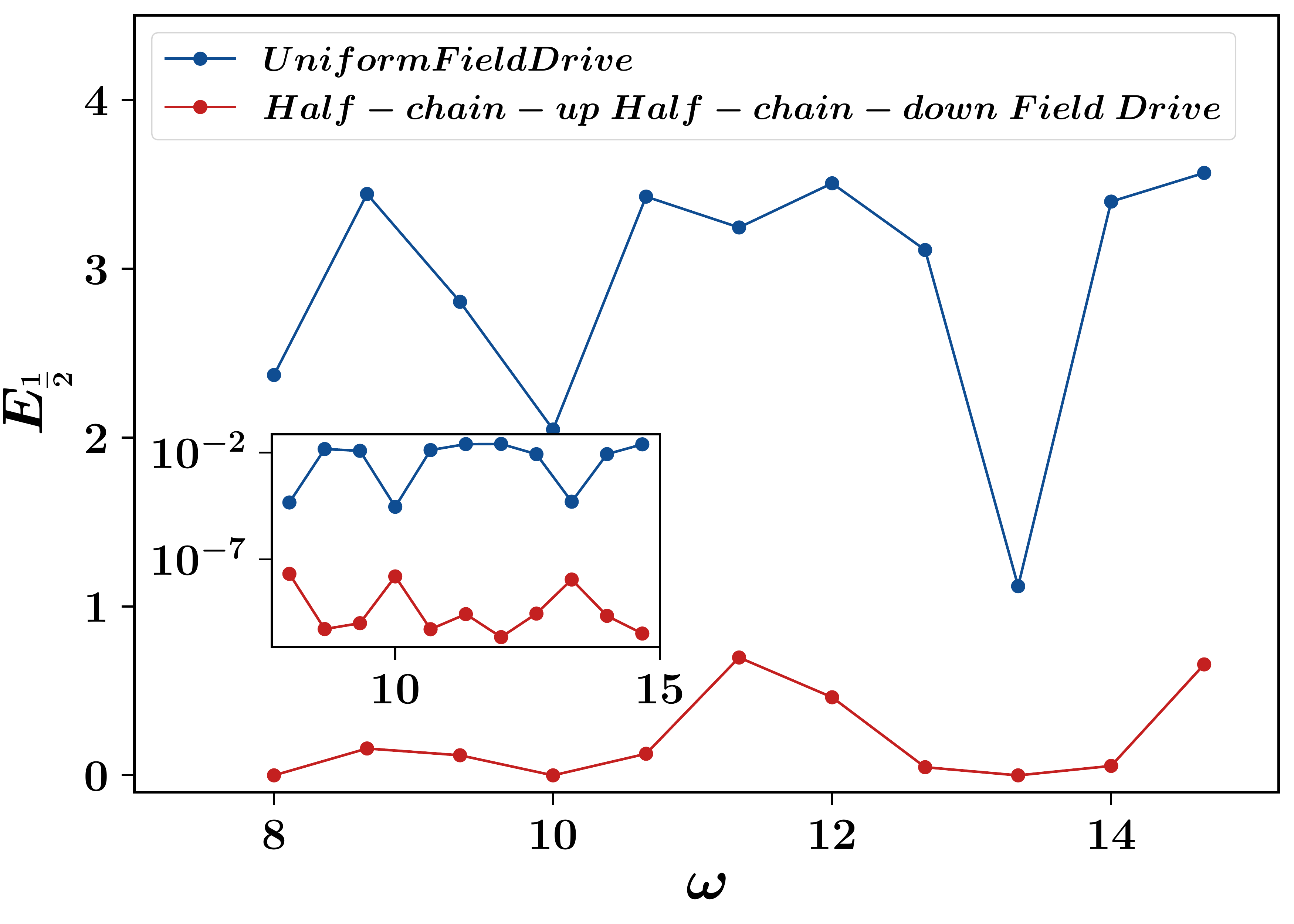}
\end{center}
\caption{Freezing the entanglement growth of the $L/2$-domain-pair $x$-basis 
state under half-up half-down field drive (Eq.~\eqref{HD_Domain}). Fate of a 
fully polarized state under the same drive is also shown for comparison.
The main frame shows $E_{\frac{1}{2}},$ while the inset shows $m^x,$ averaged 
over $10^{4}$ cycles, after driving for $10^{10}$ cycles. The results are for 
$J=1, ~\kappa = 0.7, ~h^{x}_{0}= e/ 10, ~h_{D}^{x}=40, ~L=14$.} 
\label{Pattern_Freezing} \end{figure}

From Fig.~\ref{EE} we note that the fully-polarized state is quite 
special -- at the scar points not only does its magnetization remain strongly 
frozen close to unity, its entanglement entropy also does not grow. This is 
in stark contrast with other $x$-basis states for which, though $m^x$ remains 
conserved, the entanglement entropy experiences substantial growth. This 
can be understood from the step-like structure (Fig.~\ref{Peak_Valley_Steps}) 
appearing at the scar points. We expect this phenomenology to be present for 
other strong drives which divide up the Hilbert space into sectors which are at 
most weakly mixed as long as these sectors are separated by finite gaps.

We illustrate this by arresting the entanglement dynamics of the 
$L/2$-domain-pair 
state, which sees substantial growth of $E_{\frac{1}{2}}$ under the drive with 
uniform longitudinal field (Fig.~\ref{EE}, middle column).
Instead of a uniform field, we choose the following drive Hamiltonian
\beq H_{D} = - h_{D}^{x}\sum_{i=1}^{L/2}\sigma_{i}^{x} + h_{D}^{x}
\sum_{i=L/2 + 1}^{L}\sigma_{i}^{x}, \label{HD_Domain} \eeq
\noindent
keeping the rest of the set-up the same as given by Eq.~(\ref{ham3}). 
For $H_{D}$ of the above form, $L/2$-domain-pair state is in an 
eigensector of its own. As 
expected, the entanglement growth is strongly suppressed for the 
$L/2$-domain-pair state, especially, at $\omega = 8, 10$ and $13.33 \cdots$ 
which are the scar points corresponding to the applied drive amplitude 
$h_{D}^{x}=40,$ while a substantial growth of entanglement is observed for the 
fully polarized initial state. This is in stark contrast to the
results for the uniform drive (left and middle columns of Fig.~\ref{EE}).

\bibliography{HSMD_Final_PRX_V1} 

\begin{thebibliography}{61}
\expandafter\ifx\csname natexlab\endcsname\relax\def\natexlab#1{#1}\fi
\expandafter\ifx\csname bibnamefont\endcsname\relax
  \def\bibnamefont#1{#1}\fi
\expandafter\ifx\csname bibfnamefont\endcsname\relax
  \def\bibfnamefont#1{#1}\fi
\expandafter\ifx\csname citenamefont\endcsname\relax
  \def\citenamefont#1{#1}\fi
\expandafter\ifx\csname url\endcsname\relax
  \def\url#1{\texttt{#1}}\fi
\expandafter\ifx\csname urlprefix\endcsname\relax\def\urlprefix{URL }\fi
\providecommand{\bibinfo}[2]{#2}
\providecommand{\eprint}[2][]{\url{#2}}

\bibitem[{\citenamefont{Srednicki}(1994{\natexlab{a}})}]{Srednicki}
\bibinfo{author}{\bibfnamefont{M.}~\bibnamefont{Srednicki}},
  \bibinfo{journal}{Phys. Rev. E} \textbf{\bibinfo{volume}{50}},
  \bibinfo{pages}{888} (\bibinfo{year}{1994}{\natexlab{a}}),
  \urlprefix\url{https://link.aps.org/doi/10.1103/PhysRevE.50.888}.

\bibitem[{\citenamefont{Rigol et~al.}(2016)\citenamefont{Rigol, Dunjko, and
  Olshanii}}]{Rigol_Nature}
\bibinfo{author}{\bibfnamefont{M.}~\bibnamefont{Rigol}},
  \bibinfo{author}{\bibfnamefont{V.}~\bibnamefont{Dunjko}}, \bibnamefont{and}
  \bibinfo{author}{\bibfnamefont{M.}~\bibnamefont{Olshanii}},
  \bibinfo{journal}{Nature} \textbf{\bibinfo{volume}{452}},
  \bibinfo{pages}{854} (\bibinfo{year}{2016}),
  \urlprefix\url{http://dx.doi.org/10.1038/nature06838}.

\bibitem[{\citenamefont{Lazarides
  et~al.}(2014{\natexlab{a}})\citenamefont{Lazarides, Das, and
  Moessner}}]{LDM_PRE}
\bibinfo{author}{\bibfnamefont{A.}~\bibnamefont{Lazarides}},
  \bibinfo{author}{\bibfnamefont{A.}~\bibnamefont{Das}}, \bibnamefont{and}
  \bibinfo{author}{\bibfnamefont{R.}~\bibnamefont{Moessner}},
  \bibinfo{journal}{Phys. Rev. E} \textbf{\bibinfo{volume}{90}},
  \bibinfo{pages}{012110} (\bibinfo{year}{2014}{\natexlab{a}}),
  \urlprefix\url{https://link.aps.org/doi/10.1103/PhysRevE.90.012110}.

\bibitem[{\citenamefont{D'Alessio and Rigol}(2014)}]{Rigol_Infinite_T}
\bibinfo{author}{\bibfnamefont{L.}~\bibnamefont{D'Alessio}} \bibnamefont{and}
  \bibinfo{author}{\bibfnamefont{M.}~\bibnamefont{Rigol}},
  \bibinfo{journal}{Phys. Rev. X} \textbf{\bibinfo{volume}{4}},
  \bibinfo{pages}{041048} (\bibinfo{year}{2014}),
  \urlprefix\url{https://link.aps.org/doi/10.1103/PhysRevX.4.041048}.

\bibitem[{\citenamefont{Basko et~al.}(2006)\citenamefont{Basko, Aleiner, and
  Altshuler}}]{Altshuler_MBL}
\bibinfo{author}{\bibfnamefont{D.}~\bibnamefont{Basko}},
  \bibinfo{author}{\bibfnamefont{I.}~\bibnamefont{Aleiner}}, \bibnamefont{and}
  \bibinfo{author}{\bibfnamefont{B.}~\bibnamefont{Altshuler}},
  \bibinfo{journal}{Annals of Physics} \textbf{\bibinfo{volume}{321}},
  \bibinfo{pages}{1126 } (\bibinfo{year}{2006}), ISSN
  \bibinfo{issn}{0003-4916},
  \urlprefix\url{http://www.sciencedirect.com/science/article/pii/S0003491605002630}.

\bibitem[{\citenamefont{Bardarson et~al.}(2017)\citenamefont{Bardarson,
  Pollmann, Schneider, and Sondhi~(Eds.)}}]{ADP_MBL}
\bibinfo{author}{\bibfnamefont{J.~H.} \bibnamefont{Bardarson}},
  \bibinfo{author}{\bibfnamefont{F.}~\bibnamefont{Pollmann}},
  \bibinfo{author}{\bibfnamefont{H.}~\bibnamefont{Schneider}},
  \bibnamefont{and} \bibinfo{author}{\bibfnamefont{S.~L.}
  \bibnamefont{Sondhi~(Eds.)}}, \emph{\bibinfo{title}{Special Issue: Many-Body
  Localization}}, vol. \bibinfo{volume}{529} (\bibinfo{publisher}{Annelen der
  Physik}, \bibinfo{year}{2017}),
  \urlprefix\url{https://onlinelibrary.wiley.com/toc/15213889/2017/529/7}.

\bibitem[{\citenamefont{Sachdev et~al.}(2002)\citenamefont{Sachdev, Sengupta,
  and Girvin}}]{Kris_Subir_Tilted_Mott}
\bibinfo{author}{\bibfnamefont{S.}~\bibnamefont{Sachdev}},
  \bibinfo{author}{\bibfnamefont{K.}~\bibnamefont{Sengupta}}, \bibnamefont{and}
  \bibinfo{author}{\bibfnamefont{S.~M.} \bibnamefont{Girvin}},
  \bibinfo{journal}{Phys. Rev. B} \textbf{\bibinfo{volume}{66}},
  \bibinfo{pages}{075128} (\bibinfo{year}{2002}),
  \urlprefix\url{https://link.aps.org/doi/10.1103/PhysRevB.66.075128}.

\bibitem[{\citenamefont{{van Nieuwenburg} et~al.}(2019)\citenamefont{{van
  Nieuwenburg}, {Baum}, and {Refael}}}]{2019PNAS..116.9269V}
\bibinfo{author}{\bibfnamefont{E.}~\bibnamefont{{van Nieuwenburg}}},
  \bibinfo{author}{\bibfnamefont{Y.}~\bibnamefont{{Baum}}}, \bibnamefont{and}
  \bibinfo{author}{\bibfnamefont{G.}~\bibnamefont{{Refael}}},
  \bibinfo{journal}{Proceedings of the National Academy of Science}
  \textbf{\bibinfo{volume}{116}}, \bibinfo{pages}{9269} (\bibinfo{year}{2019}),
  \eprint{1808.00471}.

\bibitem[{\citenamefont{Schulz et~al.}(2019)\citenamefont{Schulz, Hooley,
  Moessner, and Pollmann}}]{Moessner_Stark}
\bibinfo{author}{\bibfnamefont{M.}~\bibnamefont{Schulz}},
  \bibinfo{author}{\bibfnamefont{C.~A.} \bibnamefont{Hooley}},
  \bibinfo{author}{\bibfnamefont{R.}~\bibnamefont{Moessner}}, \bibnamefont{and}
  \bibinfo{author}{\bibfnamefont{F.}~\bibnamefont{Pollmann}},
  \bibinfo{journal}{Phys. Rev. Lett.} \textbf{\bibinfo{volume}{122}},
  \bibinfo{pages}{040606} (\bibinfo{year}{2019}),
  \urlprefix\url{https://link.aps.org/doi/10.1103/PhysRevLett.122.040606}.

\bibitem[{\citenamefont{Lazarides et~al.}(2015)\citenamefont{Lazarides, Das,
  and Moessner}}]{FlqMBL1}
\bibinfo{author}{\bibfnamefont{A.}~\bibnamefont{Lazarides}},
  \bibinfo{author}{\bibfnamefont{A.}~\bibnamefont{Das}}, \bibnamefont{and}
  \bibinfo{author}{\bibfnamefont{R.}~\bibnamefont{Moessner}},
  \bibinfo{journal}{Phys. Rev. Lett.} \textbf{\bibinfo{volume}{115}},
  \bibinfo{pages}{030402} (\bibinfo{year}{2015}),
  \urlprefix\url{https://link.aps.org/doi/10.1103/PhysRevLett.115.030402}.

\bibitem[{\citenamefont{Ponte et~al.}(2015)\citenamefont{Ponte,
  Papi\ifmmode~\acute{c}\else \'{c}\fi{}, Huveneers, and Abanin}}]{FlqMBL2}
\bibinfo{author}{\bibfnamefont{P.}~\bibnamefont{Ponte}},
  \bibinfo{author}{\bibfnamefont{Z.}~\bibnamefont{Papi\ifmmode~\acute{c}\else
  \'{c}\fi{}}}, \bibinfo{author}{\bibfnamefont{F.}~\bibnamefont{Huveneers}},
  \bibnamefont{and} \bibinfo{author}{\bibfnamefont{D.~A.}
  \bibnamefont{Abanin}}, \bibinfo{journal}{Phys. Rev. Lett.}
  \textbf{\bibinfo{volume}{114}}, \bibinfo{pages}{140401}
  (\bibinfo{year}{2015}),
  \urlprefix\url{https://link.aps.org/doi/10.1103/PhysRevLett.114.140401}.

\bibitem[{\citenamefont{Mukherjee et~al.}(2019)\citenamefont{Mukherjee, Nandy,
  Sen, Sen, and Sengupta}}]{Bhaskar_Scar}
\bibinfo{author}{\bibfnamefont{B.}~\bibnamefont{Mukherjee}},
  \bibinfo{author}{\bibfnamefont{S.}~\bibnamefont{Nandy}},
  \bibinfo{author}{\bibfnamefont{A.}~\bibnamefont{Sen}},
  \bibinfo{author}{\bibfnamefont{D.}~\bibnamefont{Sen}}, \bibnamefont{and}
  \bibinfo{author}{\bibfnamefont{K.}~\bibnamefont{Sengupta}}
  (\bibinfo{year}{2019}), \eprint{arXiv:1907.08212v2}.

\bibitem[{\citenamefont{von Keyserlingk et~al.}(2016)\citenamefont{von
  Keyserlingk, Khemani, and Sondhi}}]{Vedika_Stability}
\bibinfo{author}{\bibfnamefont{C.~W.} \bibnamefont{von Keyserlingk}},
  \bibinfo{author}{\bibfnamefont{V.}~\bibnamefont{Khemani}}, \bibnamefont{and}
  \bibinfo{author}{\bibfnamefont{S.~L.} \bibnamefont{Sondhi}},
  \bibinfo{journal}{Phys. Rev. B} \textbf{\bibinfo{volume}{94}},
  \bibinfo{pages}{085112} (\bibinfo{year}{2016}),
  \urlprefix\url{https://link.aps.org/doi/10.1103/PhysRevB.94.085112}.

\bibitem[{\citenamefont{Else et~al.}(2016)\citenamefont{Else, Bauer, and
  Nayak}}]{Norm_Chetan_Rev}
\bibinfo{author}{\bibfnamefont{D.~V.} \bibnamefont{Else}},
  \bibinfo{author}{\bibfnamefont{B.}~\bibnamefont{Bauer}}, \bibnamefont{and}
  \bibinfo{author}{\bibfnamefont{C.}~\bibnamefont{Nayak}},
  \bibinfo{journal}{Phys. Rev. Lett.} \textbf{\bibinfo{volume}{117}},
  \bibinfo{pages}{090402} (\bibinfo{year}{2016}),
  \urlprefix\url{https://link.aps.org/doi/10.1103/PhysRevLett.117.090402}.

\bibitem[{\citenamefont{Bernien et~al.}(2017)\citenamefont{Bernien, Schwartz,
  Keesling, Levine, Omran, Pichler, Choi, Zibrov, Endres, Greiner
  et~al.}}]{Scar_Lukin_Nature}
\bibinfo{author}{\bibfnamefont{H.}~\bibnamefont{Bernien}},
  \bibinfo{author}{\bibfnamefont{S.}~\bibnamefont{Schwartz}},
  \bibinfo{author}{\bibfnamefont{A.}~\bibnamefont{Keesling}},
  \bibinfo{author}{\bibfnamefont{H.}~\bibnamefont{Levine}},
  \bibinfo{author}{\bibfnamefont{A.}~\bibnamefont{Omran}},
  \bibinfo{author}{\bibfnamefont{H.}~\bibnamefont{Pichler}},
  \bibinfo{author}{\bibfnamefont{S.}~\bibnamefont{Choi}},
  \bibinfo{author}{\bibfnamefont{A.~S.} \bibnamefont{Zibrov}},
  \bibinfo{author}{\bibfnamefont{M.}~\bibnamefont{Endres}},
  \bibinfo{author}{\bibfnamefont{M.}~\bibnamefont{Greiner}},
  \bibnamefont{et~al.}, \bibinfo{journal}{Nature}
  \textbf{\bibinfo{volume}{551}}, \bibinfo{pages}{579} (\bibinfo{year}{2017}),
  \urlprefix\url{https://doi.org/10.1038/nature24622}.

\bibitem[{\citenamefont{Turner et~al.}(2018{\natexlab{a}})\citenamefont{Turner,
  Michailidis, Abanin, Serbyn, and Papic}}]{Scar_Abanin_NatPhys}
\bibinfo{author}{\bibfnamefont{C.~J.} \bibnamefont{Turner}},
  \bibinfo{author}{\bibfnamefont{A.~A.} \bibnamefont{Michailidis}},
  \bibinfo{author}{\bibfnamefont{D.~A.} \bibnamefont{Abanin}},
  \bibinfo{author}{\bibfnamefont{M.}~\bibnamefont{Serbyn}}, \bibnamefont{and}
  \bibinfo{author}{\bibfnamefont{Z.}~\bibnamefont{Papic}},
  \bibinfo{journal}{Nature Physics} \textbf{\bibinfo{volume}{14}},
  \bibinfo{pages}{745} (\bibinfo{year}{2018}{\natexlab{a}}),
  \urlprefix\url{https://doi.org/10.1038/s41567-018-0137-5}.

\bibitem[{\citenamefont{Shiraishi and Mori}(2017)}]{Scar_Shiraishi-Mori}
\bibinfo{author}{\bibfnamefont{N.}~\bibnamefont{Shiraishi}} \bibnamefont{and}
  \bibinfo{author}{\bibfnamefont{T.}~\bibnamefont{Mori}},
  \bibinfo{journal}{Phys. Rev. Lett.} \textbf{\bibinfo{volume}{119}},
  \bibinfo{pages}{030601} (\bibinfo{year}{2017}),
  \urlprefix\url{https://link.aps.org/doi/10.1103/PhysRevLett.119.030601}.

\bibitem[{\citenamefont{Turner et~al.}(2018{\natexlab{b}})\citenamefont{Turner,
  Michailidis, Abanin, Serbyn, and Papi\ifmmode~\acute{c}\else
  \'{c}\fi{}}}]{Scar_Abanin_PRB}
\bibinfo{author}{\bibfnamefont{C.~J.} \bibnamefont{Turner}},
  \bibinfo{author}{\bibfnamefont{A.~A.} \bibnamefont{Michailidis}},
  \bibinfo{author}{\bibfnamefont{D.~A.} \bibnamefont{Abanin}},
  \bibinfo{author}{\bibfnamefont{M.}~\bibnamefont{Serbyn}}, \bibnamefont{and}
  \bibinfo{author}{\bibfnamefont{Z.}~\bibnamefont{Papi\ifmmode~\acute{c}\else
  \'{c}\fi{}}}, \bibinfo{journal}{Phys. Rev. B} \textbf{\bibinfo{volume}{98}},
  \bibinfo{pages}{155134} (\bibinfo{year}{2018}{\natexlab{b}}),
  \urlprefix\url{https://link.aps.org/doi/10.1103/PhysRevB.98.155134}.

\bibitem[{\citenamefont{Khemani et~al.}(2019)\citenamefont{Khemani, Laumann,
  and Chandran}}]{Scar_Vedika_PRB}
\bibinfo{author}{\bibfnamefont{V.}~\bibnamefont{Khemani}},
  \bibinfo{author}{\bibfnamefont{C.~R.} \bibnamefont{Laumann}},
  \bibnamefont{and} \bibinfo{author}{\bibfnamefont{A.}~\bibnamefont{Chandran}},
  \bibinfo{journal}{Phys. Rev. B} \textbf{\bibinfo{volume}{99}},
  \bibinfo{pages}{161101} (\bibinfo{year}{2019}),
  \urlprefix\url{https://link.aps.org/doi/10.1103/PhysRevB.99.161101}.

\bibitem[{\citenamefont{Ho et~al.}(2019)\citenamefont{Ho, Choi, Pichler, and
  Lukin}}]{Scar_Lukin_PRL}
\bibinfo{author}{\bibfnamefont{W.~W.} \bibnamefont{Ho}},
  \bibinfo{author}{\bibfnamefont{S.}~\bibnamefont{Choi}},
  \bibinfo{author}{\bibfnamefont{H.}~\bibnamefont{Pichler}}, \bibnamefont{and}
  \bibinfo{author}{\bibfnamefont{M.~D.} \bibnamefont{Lukin}},
  \bibinfo{journal}{Phys. Rev. Lett.} \textbf{\bibinfo{volume}{122}},
  \bibinfo{pages}{040603} (\bibinfo{year}{2019}),
  \urlprefix\url{https://link.aps.org/doi/10.1103/PhysRevLett.122.040603}.

\bibitem[{\citenamefont{Choi et~al.}(2019)\citenamefont{Choi, Turner, Pichler,
  Ho, Michailidis, Papi\ifmmode~\acute{c}\else \'{c}\fi{}, Serbyn, Lukin, and
  Abanin}}]{Scar_Serbyn_PRL}
\bibinfo{author}{\bibfnamefont{S.}~\bibnamefont{Choi}},
  \bibinfo{author}{\bibfnamefont{C.~J.} \bibnamefont{Turner}},
  \bibinfo{author}{\bibfnamefont{H.}~\bibnamefont{Pichler}},
  \bibinfo{author}{\bibfnamefont{W.~W.} \bibnamefont{Ho}},
  \bibinfo{author}{\bibfnamefont{A.~A.} \bibnamefont{Michailidis}},
  \bibinfo{author}{\bibfnamefont{Z.}~\bibnamefont{Papi\ifmmode~\acute{c}\else
  \'{c}\fi{}}}, \bibinfo{author}{\bibfnamefont{M.}~\bibnamefont{Serbyn}},
  \bibinfo{author}{\bibfnamefont{M.~D.} \bibnamefont{Lukin}}, \bibnamefont{and}
  \bibinfo{author}{\bibfnamefont{D.~A.} \bibnamefont{Abanin}},
  \bibinfo{journal}{Phys. Rev. Lett.} \textbf{\bibinfo{volume}{122}},
  \bibinfo{pages}{220603} (\bibinfo{year}{2019}),
  \urlprefix\url{https://link.aps.org/doi/10.1103/PhysRevLett.122.220603}.

\bibitem[{\citenamefont{Moudgalya
  et~al.}(2018{\natexlab{a}})\citenamefont{Moudgalya, Regnault, and
  Bernevig}}]{Scar_Bernevig_1}
\bibinfo{author}{\bibfnamefont{S.}~\bibnamefont{Moudgalya}},
  \bibinfo{author}{\bibfnamefont{N.}~\bibnamefont{Regnault}}, \bibnamefont{and}
  \bibinfo{author}{\bibfnamefont{B.~A.} \bibnamefont{Bernevig}},
  \bibinfo{journal}{Phys. Rev. B} \textbf{\bibinfo{volume}{98}},
  \bibinfo{pages}{235156} (\bibinfo{year}{2018}{\natexlab{a}}),
  \urlprefix\url{https://link.aps.org/doi/10.1103/PhysRevB.98.235156}.

\bibitem[{\citenamefont{Khemani and Nandkishore}(2019)}]{Scar_Vedika_Rahul}
\bibinfo{author}{\bibfnamefont{V.}~\bibnamefont{Khemani}} \bibnamefont{and}
  \bibinfo{author}{\bibfnamefont{R.~M.} \bibnamefont{Nandkishore}}
  (\bibinfo{year}{2019}), \eprint{arXiv:1904.04815v2}.

\bibitem[{\citenamefont{Moudgalya
  et~al.}(2018{\natexlab{b}})\citenamefont{Moudgalya, Rachel, Bernevig, and
  Regnault}}]{Scar_Bernevig_2}
\bibinfo{author}{\bibfnamefont{S.}~\bibnamefont{Moudgalya}},
  \bibinfo{author}{\bibfnamefont{S.}~\bibnamefont{Rachel}},
  \bibinfo{author}{\bibfnamefont{B.~A.} \bibnamefont{Bernevig}},
  \bibnamefont{and} \bibinfo{author}{\bibfnamefont{N.}~\bibnamefont{Regnault}},
  \bibinfo{journal}{Phys. Rev. B} \textbf{\bibinfo{volume}{98}},
  \bibinfo{pages}{235155} (\bibinfo{year}{2018}{\natexlab{b}}),
  \urlprefix\url{https://link.aps.org/doi/10.1103/PhysRevB.98.235155}.

\bibitem[{\citenamefont{Mondaini et~al.}(2018)\citenamefont{Mondaini, Mallayya,
  Santos, and Rigol}}]{Rigol_Comment}
\bibinfo{author}{\bibfnamefont{R.}~\bibnamefont{Mondaini}},
  \bibinfo{author}{\bibfnamefont{K.}~\bibnamefont{Mallayya}},
  \bibinfo{author}{\bibfnamefont{L.~F.} \bibnamefont{Santos}},
  \bibnamefont{and} \bibinfo{author}{\bibfnamefont{M.}~\bibnamefont{Rigol}},
  \bibinfo{journal}{Phys. Rev. Lett.} \textbf{\bibinfo{volume}{121}},
  \bibinfo{pages}{038901} (\bibinfo{year}{2018}),
  \urlprefix\url{https://link.aps.org/doi/10.1103/PhysRevLett.121.038901}.

\bibitem[{\citenamefont{D'Alessio and Polkovnikov}(2013)}]{Luca_Polku}
\bibinfo{author}{\bibfnamefont{L.}~\bibnamefont{D'Alessio}} \bibnamefont{and}
  \bibinfo{author}{\bibfnamefont{A.}~\bibnamefont{Polkovnikov}},
  \bibinfo{journal}{Annals of Physics} \textbf{\bibinfo{volume}{333}},
  \bibinfo{pages}{19} (\bibinfo{year}{2013}).

\bibitem[{\citenamefont{Bukov et~al.}(2016)\citenamefont{Bukov, Heyl, Huse, and
  Polkovnikov}}]{Bukov_Polku_Huse}
\bibinfo{author}{\bibfnamefont{M.}~\bibnamefont{Bukov}},
  \bibinfo{author}{\bibfnamefont{M.}~\bibnamefont{Heyl}},
  \bibinfo{author}{\bibfnamefont{D.~A.} \bibnamefont{Huse}}, \bibnamefont{and}
  \bibinfo{author}{\bibfnamefont{A.}~\bibnamefont{Polkovnikov}},
  \bibinfo{journal}{Phys. Rev. B} \textbf{\bibinfo{volume}{93}},
  \bibinfo{pages}{155132} (\bibinfo{year}{2016}),
  \urlprefix\url{https://link.aps.org/doi/10.1103/PhysRevB.93.155132}.

\bibitem[{\citenamefont{Bukov et~al.}(2015)\citenamefont{Bukov, D'Alessio, and
  Polkovnikov}}]{Anatoli_Rev}
\bibinfo{author}{\bibfnamefont{M.}~\bibnamefont{Bukov}},
  \bibinfo{author}{\bibfnamefont{L.}~\bibnamefont{D'Alessio}},
  \bibnamefont{and}
  \bibinfo{author}{\bibfnamefont{A.}~\bibnamefont{Polkovnikov}},
  \bibinfo{journal}{Advances in Physics} \textbf{\bibinfo{volume}{64}},
  \bibinfo{pages}{139} (\bibinfo{year}{2015}).

\bibitem[{\citenamefont{Agarwala and Sen}(2017)}]{Adhip_Diptiman}
\bibinfo{author}{\bibfnamefont{A.}~\bibnamefont{Agarwala}} \bibnamefont{and}
  \bibinfo{author}{\bibfnamefont{D.}~\bibnamefont{Sen}},
  \bibinfo{journal}{Phys. Rev. B} \textbf{\bibinfo{volume}{95}},
  \bibinfo{pages}{014305} (\bibinfo{year}{2017}),
  \urlprefix\url{https://link.aps.org/doi/10.1103/PhysRevB.95.014305}.

\bibitem[{\citenamefont{Dasgupta et~al.}(2015)\citenamefont{Dasgupta,
  Bhattacharya, and Dutta}}]{Sayak_Utsa_Amit}
\bibinfo{author}{\bibfnamefont{S.}~\bibnamefont{Dasgupta}},
  \bibinfo{author}{\bibfnamefont{U.}~\bibnamefont{Bhattacharya}},
  \bibnamefont{and} \bibinfo{author}{\bibfnamefont{A.}~\bibnamefont{Dutta}},
  \bibinfo{journal}{Phys. Rev. E} \textbf{\bibinfo{volume}{91}},
  \bibinfo{pages}{052129} (\bibinfo{year}{2015}),
  \urlprefix\url{https://link.aps.org/doi/10.1103/PhysRevE.91.052129}.

\bibitem[{\citenamefont{Bordia et~al.}(2017)\citenamefont{Bordia, L\"{u}schen,
  Schneider, Knap, and Bloch}}]{Bordia_Knap_Bloch}
\bibinfo{author}{\bibfnamefont{P.}~\bibnamefont{Bordia}},
  \bibinfo{author}{\bibfnamefont{H.}~\bibnamefont{L\"{u}schen}},
  \bibinfo{author}{\bibfnamefont{U.}~\bibnamefont{Schneider}},
  \bibinfo{author}{\bibfnamefont{M.}~\bibnamefont{Knap}}, \bibnamefont{and}
  \bibinfo{author}{\bibfnamefont{I.}~\bibnamefont{Bloch}},
  \bibinfo{journal}{Nat. Phys.} \textbf{\bibinfo{volume}{13}},
  \bibinfo{pages}{460} (\bibinfo{year}{2017}),
  \urlprefix\url{https://www.nature.com/articles/nphys4020?WT.feed_name=subjects_physics}.

\bibitem[{\citenamefont{Pal et~al.}(2018)\citenamefont{Pal, Nishad, Mahesh, and
  Sreejith}}]{Sreejith_Mahesh}
\bibinfo{author}{\bibfnamefont{S.}~\bibnamefont{Pal}},
  \bibinfo{author}{\bibfnamefont{N.}~\bibnamefont{Nishad}},
  \bibinfo{author}{\bibfnamefont{T.~S.} \bibnamefont{Mahesh}},
  \bibnamefont{and} \bibinfo{author}{\bibfnamefont{G.~J.}
  \bibnamefont{Sreejith}}, \bibinfo{journal}{Phys. Rev. Lett.}
  \textbf{\bibinfo{volume}{120}}, \bibinfo{pages}{180602}
  (\bibinfo{year}{2018}),
  \urlprefix\url{https://link.aps.org/doi/10.1103/PhysRevLett.120.180602}.

\bibitem[{\citenamefont{Qin and Hofstetter}(2018)}]{Qin_Hofstetter}
\bibinfo{author}{\bibfnamefont{T.}~\bibnamefont{Qin}} \bibnamefont{and}
  \bibinfo{author}{\bibfnamefont{W.}~\bibnamefont{Hofstetter}},
  \bibinfo{journal}{Phys. Rev. B} \textbf{\bibinfo{volume}{97}},
  \bibinfo{pages}{125115} (\bibinfo{year}{2018}),
  \urlprefix\url{https://link.aps.org/doi/10.1103/PhysRevB.97.125115}.

\bibitem[{\citenamefont{Seetharam
  et~al.}(2018{\natexlab{a}})\citenamefont{Seetharam, Titum, Kolodrubetz, and
  Refael}}]{gil-fss}
\bibinfo{author}{\bibfnamefont{K.}~\bibnamefont{Seetharam}},
  \bibinfo{author}{\bibfnamefont{P.}~\bibnamefont{Titum}},
  \bibinfo{author}{\bibfnamefont{M.}~\bibnamefont{Kolodrubetz}},
  \bibnamefont{and} \bibinfo{author}{\bibfnamefont{G.}~\bibnamefont{Refael}},
  \bibinfo{journal}{Phys. Rev. B} \textbf{\bibinfo{volume}{97}},
  \bibinfo{pages}{014311} (\bibinfo{year}{2018}{\natexlab{a}}),
  \urlprefix\url{https://link.aps.org/doi/10.1103/PhysRevB.97.014311}.

\bibitem[{\citenamefont{Prosen}(1998)}]{Prosen_prl_98}
\bibinfo{author}{\bibfnamefont{T.}~\bibnamefont{Prosen}},
  \bibinfo{journal}{Phys. Rev. Lett.} \textbf{\bibinfo{volume}{80}},
  \bibinfo{pages}{1808} (\bibinfo{year}{1998}),
  \urlprefix\url{https://link.aps.org/doi/10.1103/PhysRevLett.80.1808}.

\bibitem[{\citenamefont{Prosen}(2002)}]{Prosen_Tilted_NoHeat}
\bibinfo{author}{\bibfnamefont{T.}~\bibnamefont{Prosen}},
  \bibinfo{journal}{Phys. Rev. E} \textbf{\bibinfo{volume}{65}},
  \bibinfo{pages}{036208} (\bibinfo{year}{2002}),
  \urlprefix\url{https://link.aps.org/doi/10.1103/PhysRevE.65.036208}.

\bibitem[{\citenamefont{Luitz et~al.}(2018)\citenamefont{Luitz, Lazarides, and
  Bar~Lev}}]{AL_DL_PRB}
\bibinfo{author}{\bibfnamefont{D.~J.} \bibnamefont{Luitz}},
  \bibinfo{author}{\bibfnamefont{A.}~\bibnamefont{Lazarides}},
  \bibnamefont{and} \bibinfo{author}{\bibfnamefont{Y.}~\bibnamefont{Bar~Lev}},
  \bibinfo{journal}{Phys. Rev. B} \textbf{\bibinfo{volume}{97}},
  \bibinfo{pages}{020303} (\bibinfo{year}{2018}),
  \urlprefix\url{https://link.aps.org/doi/10.1103/PhysRevB.97.020303}.

\bibitem[{\citenamefont{Haldar et~al.}(2018)\citenamefont{Haldar, Moessner, and
  Das}}]{Onset}
\bibinfo{author}{\bibfnamefont{A.}~\bibnamefont{Haldar}},
  \bibinfo{author}{\bibfnamefont{R.}~\bibnamefont{Moessner}}, \bibnamefont{and}
  \bibinfo{author}{\bibfnamefont{A.}~\bibnamefont{Das}},
  \bibinfo{journal}{Phys. Rev. B} \textbf{\bibinfo{volume}{97}},
  \bibinfo{pages}{245122} (\bibinfo{year}{2018}),
  \urlprefix\url{https://link.aps.org/doi/10.1103/PhysRevB.97.245122}.

\bibitem[{\citenamefont{Das}(2010)}]{AD-DMF}
\bibinfo{author}{\bibfnamefont{A.}~\bibnamefont{Das}}, \bibinfo{journal}{Phys.
  Rev. B} \textbf{\bibinfo{volume}{82}}, \bibinfo{pages}{172402}
  (\bibinfo{year}{2010}),
  \urlprefix\url{http://link.aps.org/doi/10.1103/PhysRevB.82.172402}.

\bibitem[{\citenamefont{Bhattacharyya et~al.}(2012)\citenamefont{Bhattacharyya,
  Das, and Dasgupta}}]{SB_AD_SDG}
\bibinfo{author}{\bibfnamefont{S.}~\bibnamefont{Bhattacharyya}},
  \bibinfo{author}{\bibfnamefont{A.}~\bibnamefont{Das}}, \bibnamefont{and}
  \bibinfo{author}{\bibfnamefont{S.}~\bibnamefont{Dasgupta}},
  \bibinfo{journal}{Phys. Rev. B} \textbf{\bibinfo{volume}{86}},
  \bibinfo{pages}{054410} (\bibinfo{year}{2012}),
  \urlprefix\url{http://link.aps.org/doi/10.1103/PhysRevB.86.054410}.

\bibitem[{\citenamefont{Hegde et~al.}(2014)\citenamefont{Hegde, Katiyar,
  Mahesh, and Das}}]{Mahesh_Freezing}
\bibinfo{author}{\bibfnamefont{S.~S.} \bibnamefont{Hegde}},
  \bibinfo{author}{\bibfnamefont{H.}~\bibnamefont{Katiyar}},
  \bibinfo{author}{\bibfnamefont{T.~S.} \bibnamefont{Mahesh}},
  \bibnamefont{and} \bibinfo{author}{\bibfnamefont{A.}~\bibnamefont{Das}},
  \bibinfo{journal}{Phys. Rev. B} \textbf{\bibinfo{volume}{90}},
  \bibinfo{pages}{174407} (\bibinfo{year}{2014}),
  \urlprefix\url{https://link.aps.org/doi/10.1103/PhysRevB.90.174407}.

\bibitem[{\citenamefont{Mondal et~al.}(2012)\citenamefont{Mondal, Pekker, and
  Sengupta}}]{Kris-Periodic}
\bibinfo{author}{\bibfnamefont{S.}~\bibnamefont{Mondal}},
  \bibinfo{author}{\bibfnamefont{D.}~\bibnamefont{Pekker}}, \bibnamefont{and}
  \bibinfo{author}{\bibfnamefont{K.}~\bibnamefont{Sengupta}},
  \bibinfo{journal}{EPL} \textbf{\bibinfo{volume}{100}}, \bibinfo{pages}{60007}
  (\bibinfo{year}{2012}),
  \urlprefix\url{http://dx.doi.org/10.1209/0295-5075/100/60007}.

\bibitem[{\citenamefont{Russomanno et~al.}(2013)\citenamefont{Russomanno,
  Silva, and Santoro}}]{Russomanno_JStatMech}
\bibinfo{author}{\bibfnamefont{A.}~\bibnamefont{Russomanno}},
  \bibinfo{author}{\bibfnamefont{A.}~\bibnamefont{Silva}}, \bibnamefont{and}
  \bibinfo{author}{\bibfnamefont{G.~E.} \bibnamefont{Santoro}},
  \bibinfo{journal}{J. Stat. Mech.} \textbf{\bibinfo{volume}{2013}},
  \bibinfo{pages}{P09012} (\bibinfo{year}{2013}).

\bibitem[{\citenamefont{Lazarides
  et~al.}(2014{\natexlab{b}})\citenamefont{Lazarides, Das, and Moessner}}]{PGE}
\bibinfo{author}{\bibfnamefont{A.}~\bibnamefont{Lazarides}},
  \bibinfo{author}{\bibfnamefont{A.}~\bibnamefont{Das}}, \bibnamefont{and}
  \bibinfo{author}{\bibfnamefont{R.}~\bibnamefont{Moessner}},
  \bibinfo{journal}{Phys. Rev. Letts.} \textbf{\bibinfo{volume}{112}},
  \bibinfo{pages}{150401} (\bibinfo{year}{2014}{\natexlab{b}}).

\bibitem[{\citenamefont{Seetharam
  et~al.}(2018{\natexlab{b}})\citenamefont{Seetharam, Titum, Kolodrubetz, and
  Refael}}]{Kolodrubetz}
\bibinfo{author}{\bibfnamefont{K.}~\bibnamefont{Seetharam}},
  \bibinfo{author}{\bibfnamefont{P.}~\bibnamefont{Titum}},
  \bibinfo{author}{\bibfnamefont{M.}~\bibnamefont{Kolodrubetz}},
  \bibnamefont{and} \bibinfo{author}{\bibfnamefont{G.}~\bibnamefont{Refael}},
  \bibinfo{journal}{Phys. Rev. B} \textbf{\bibinfo{volume}{97}},
  \bibinfo{pages}{014311} (\bibinfo{year}{2018}{\natexlab{b}}),
  \urlprefix\url{https://link.aps.org/doi/10.1103/PhysRevB.97.014311}.

\bibitem[{\citenamefont{Rodriguez-Vega
  et~al.}(2018)\citenamefont{Rodriguez-Vega, Lentz, and Seradjeh}}]{Babak_1}
\bibinfo{author}{\bibfnamefont{M.}~\bibnamefont{Rodriguez-Vega}},
  \bibinfo{author}{\bibfnamefont{M.}~\bibnamefont{Lentz}}, \bibnamefont{and}
  \bibinfo{author}{\bibfnamefont{B.}~\bibnamefont{Seradjeh}},
  \bibinfo{journal}{New Journal of Physics} \textbf{\bibinfo{volume}{20}},
  \bibinfo{pages}{093022} (\bibinfo{year}{2018}),
  \urlprefix\url{https://doi.org/10.1088%2F1367-2630%2Faade37}.

\bibitem[{\citenamefont{Vogl et~al.}(2019{\natexlab{a}})\citenamefont{Vogl,
  Laurell, Barr, and Fiete}}]{Babak_2}
\bibinfo{author}{\bibfnamefont{M.}~\bibnamefont{Vogl}},
  \bibinfo{author}{\bibfnamefont{P.}~\bibnamefont{Laurell}},
  \bibinfo{author}{\bibfnamefont{A.~D.} \bibnamefont{Barr}}, \bibnamefont{and}
  \bibinfo{author}{\bibfnamefont{G.~A.} \bibnamefont{Fiete}},
  \bibinfo{journal}{Phys. Rev. X} \textbf{\bibinfo{volume}{9}},
  \bibinfo{pages}{021037} (\bibinfo{year}{2019}{\natexlab{a}}),
  \urlprefix\url{https://link.aps.org/doi/10.1103/PhysRevX.9.021037}.

\bibitem[{\citenamefont{Vogl et~al.}(2019{\natexlab{b}})\citenamefont{Vogl,
  Rodriguez-Vega, and Fiete}}]{Babak_3}
\bibinfo{author}{\bibfnamefont{M.}~\bibnamefont{Vogl}},
  \bibinfo{author}{\bibfnamefont{M.}~\bibnamefont{Rodriguez-Vega}},
  \bibnamefont{and} \bibinfo{author}{\bibfnamefont{G.~A.} \bibnamefont{Fiete}}
  (\bibinfo{year}{2019}{\natexlab{b}}), \eprint{arXiv:1909.04263}.

\bibitem[{\citenamefont{Srednicki}(1994{\natexlab{b}})}]{Srednicki_DE}
\bibinfo{author}{\bibfnamefont{M.}~\bibnamefont{Srednicki}},
  \bibinfo{journal}{arXiv:cond-mat/9410046v2}
  (\bibinfo{year}{1994}{\natexlab{b}}).

\bibitem[{\citenamefont{Rigol et~al.}(2007)\citenamefont{Rigol, Dunjko,
  Yurovsky, and Olshanii}}]{Rigol_GGE_1}
\bibinfo{author}{\bibfnamefont{M.}~\bibnamefont{Rigol}},
  \bibinfo{author}{\bibfnamefont{V.}~\bibnamefont{Dunjko}},
  \bibinfo{author}{\bibfnamefont{V.}~\bibnamefont{Yurovsky}}, \bibnamefont{and}
  \bibinfo{author}{\bibfnamefont{M.}~\bibnamefont{Olshanii}},
  \bibinfo{journal}{Phys. Rev. Lett.} \textbf{\bibinfo{volume}{98}},
  \bibinfo{pages}{050405} (\bibinfo{year}{2007}),
  \urlprefix\url{https://link.aps.org/doi/10.1103/PhysRevLett.98.050405}.

\bibitem[{\citenamefont{Cassidy et~al.}(2011)\citenamefont{Cassidy, Clark, and
  Rigol}}]{Rigol_GGE_2}
\bibinfo{author}{\bibfnamefont{A.~C.} \bibnamefont{Cassidy}},
  \bibinfo{author}{\bibfnamefont{C.~W.} \bibnamefont{Clark}}, \bibnamefont{and}
  \bibinfo{author}{\bibfnamefont{M.}~\bibnamefont{Rigol}},
  \bibinfo{journal}{Phys. Rev. Lett.} \textbf{\bibinfo{volume}{106}},
  \bibinfo{pages}{140405} (\bibinfo{year}{2011}),
  \urlprefix\url{https://link.aps.org/doi/10.1103/PhysRevLett.106.140405}.

\bibitem[{\citenamefont{Reimann}(2008)}]{Reimann}
\bibinfo{author}{\bibfnamefont{P.}~\bibnamefont{Reimann}},
  \bibinfo{journal}{Phys. Rev. Lett.} \textbf{\bibinfo{volume}{101}},
  \bibinfo{pages}{190403} (\bibinfo{year}{2008}),
  \urlprefix\url{https://link.aps.org/doi/10.1103/PhysRevLett.101.190403}.

\bibitem[{\citenamefont{Eckardt and Anisimovas}(2015)}]{Andre_Anisimovas_Rev}
\bibinfo{author}{\bibfnamefont{A.}~\bibnamefont{Eckardt}} \bibnamefont{and}
  \bibinfo{author}{\bibfnamefont{E.}~\bibnamefont{Anisimovas}},
  \bibinfo{journal}{New Journal of Physics} \textbf{\bibinfo{volume}{17}},
  \bibinfo{pages}{093039} (\bibinfo{year}{2015}),
  \urlprefix\url{https://doi.org/10.1088%2F1367-2630%2F17%2F9%2F093039}.

\bibitem[{\citenamefont{Soori and Sen}(2010)}]{soori}
\bibinfo{author}{\bibfnamefont{A.}~\bibnamefont{Soori}} \bibnamefont{and}
  \bibinfo{author}{\bibfnamefont{D.}~\bibnamefont{Sen}},
  \bibinfo{journal}{Phys. Rev. B} \textbf{\bibinfo{volume}{82}},
  \bibinfo{pages}{115432} (\bibinfo{year}{2010}),
  \urlprefix\url{https://link.aps.org/doi/10.1103/PhysRevB.82.115432}.

\bibitem[{\citenamefont{Zhang et~al.}(2017)\citenamefont{Zhang, Hess,
  Kuprianidis, Becker, Lee, Smith, Pagano, Potirniche, Potter, Vishwanath
  et~al.}}]{Monroe}
\bibinfo{author}{\bibfnamefont{J.}~\bibnamefont{Zhang}},
  \bibinfo{author}{\bibfnamefont{P.~W.} \bibnamefont{Hess}},
  \bibinfo{author}{\bibfnamefont{A.}~\bibnamefont{Kuprianidis}},
  \bibinfo{author}{\bibfnamefont{P.}~\bibnamefont{Becker}},
  \bibinfo{author}{\bibfnamefont{A.}~\bibnamefont{Lee}},
  \bibinfo{author}{\bibfnamefont{J.}~\bibnamefont{Smith}},
  \bibinfo{author}{\bibfnamefont{G.}~\bibnamefont{Pagano}},
  \bibinfo{author}{\bibfnamefont{I.-D.} \bibnamefont{Potirniche}},
  \bibinfo{author}{\bibfnamefont{A.~C.} \bibnamefont{Potter}},
  \bibinfo{author}{\bibfnamefont{A.}~\bibnamefont{Vishwanath}},
  \bibnamefont{et~al.}, \bibinfo{journal}{Nature}
  \textbf{\bibinfo{volume}{543}}, \bibinfo{pages}{217} (\bibinfo{year}{2017}).

\bibitem[{\citenamefont{{Rovny} et~al.}(2018)\citenamefont{{Rovny}, {Blum}, and
  {Barrett}}}]{Rovny}
\bibinfo{author}{\bibfnamefont{J.}~\bibnamefont{{Rovny}}},
  \bibinfo{author}{\bibfnamefont{R.~L.} \bibnamefont{{Blum}}},
  \bibnamefont{and} \bibinfo{author}{\bibfnamefont{S.~E.}
  \bibnamefont{{Barrett}}}, \bibinfo{journal}{Phys. Rev. Lett.}
  \textbf{\bibinfo{volume}{120}}, \bibinfo{eid}{180603} (\bibinfo{year}{2018}).

\bibitem[{\citenamefont{Das and Moessner}(2012)}]{AD-RM}
\bibinfo{author}{\bibfnamefont{A.}~\bibnamefont{Das}} \bibnamefont{and}
  \bibinfo{author}{\bibfnamefont{R.}~\bibnamefont{Moessner}}
  (\bibinfo{year}{2012}), \eprint{1208.0217}.

\bibitem[{\citenamefont{Bukov}(2016)}]{Bukov_Thesis}
\bibinfo{author}{\bibfnamefont{M.}~\bibnamefont{Bukov}},
  \bibinfo{journal}{SciPost Thesis Link} p.~\bibinfo{pages}{32}
  (\bibinfo{year}{2016}), \urlprefix\url{https://scipost.org/theses/32/}.

\bibitem[{\citenamefont{Weinberg and Bukov}(2017)}]{Quspin_D1}
\bibinfo{author}{\bibfnamefont{P.}~\bibnamefont{Weinberg}} \bibnamefont{and}
  \bibinfo{author}{\bibfnamefont{M.}~\bibnamefont{Bukov}},
  \bibinfo{journal}{SciPost Phys.} \textbf{\bibinfo{volume}{2}},
  \bibinfo{pages}{003} (\bibinfo{year}{2017}),
  \urlprefix\url{https://scipost.org/10.21468/SciPostPhys.2.1.003}.

\bibitem[{\citenamefont{Weinberg and Bukov}(2019)}]{Quspin_D2}
\bibinfo{author}{\bibfnamefont{P.}~\bibnamefont{Weinberg}} \bibnamefont{and}
  \bibinfo{author}{\bibfnamefont{M.}~\bibnamefont{Bukov}},
  \bibinfo{journal}{SciPost Phys.} \textbf{\bibinfo{volume}{7}},
  \bibinfo{pages}{20} (\bibinfo{year}{2019}),
  \urlprefix\url{https://scipost.org/10.21468/SciPostPhys.7.2.020}.

\bibitem[{\citenamefont{Luiz et~al.}(2019)\citenamefont{Luiz, Moessner, Sondhi,
  and Khemani}}]{LMSK}
\bibinfo{author}{\bibfnamefont{D.~J.} \bibnamefont{Luiz}},
  \bibinfo{author}{\bibfnamefont{R.}~\bibnamefont{Moessner}},
  \bibinfo{author}{\bibfnamefont{S.~L.} \bibnamefont{Sondhi}},
  \bibnamefont{and} \bibinfo{author}{\bibfnamefont{V.}~\bibnamefont{Khemani}}
  (\bibinfo{year}{2019}), \eprint{arXiv:1908.10371}.

\end{thebibliography}

\end{document}